\documentclass[times]{weauth}
\usepackage{graphicx}
\usepackage{amsmath}
\usepackage{subfigure}
\usepackage{caption}
\usepackage{array,url}
\usepackage{blindtext}
\usepackage{epstopdf}
\usepackage{multirow}
\usepackage{ar}
\usepackage{moreverb,epsfig,ar}
\usepackage{lineno}

\usepackage{tabu}
\graphicspath{{images/}}

\usepackage{nomencl}
\makenomenclature

\newcommand{\norm}[1]{\left\lVert#1\right\rVert}

\def\volumeyear{2018}

\begin{document}

\title{Systematic design methodology for development and flight testing of a variable pitch quadrotor biplane VTOL UAV for payload delivery}

\author{Vishnu S. Chipade\footnote{Graduate Student}, Abhishek \footnote{Assistant Professor, corresponding author (abhish@iitk.ac.in)}, Mangal Kothari \footnote{Assistant Professor} and Rushikesh R. Chaudhari\footnote{Graduate Student}}
\address{Department of Aerospace Engineering \\Indian Institute of Technology Kanpur, Kanpur, UP, India - 208016}

\maketitle

\section*{Abstract}
  This paper discusses the conceptual design and proof-of-concept flight demonstration of a novel variable pitch quadrotor biplane Unmanned Aerial Vehicle concept for payload delivery. The proposed design combines vertical takeoff and landing (VTOL),  precise hover capabilities of a quadrotor helicopter and high range, endurance and high forward cruise speed characteristics of a fixed wing aircraft. The proposed UAV is designed for a mission requirement of carrying and delivering 6 kg payload to a destination at 16 km from the point of origin. First, the design of  proprotors is carried out using a physics based modified Blade Element Momentum Theory (BEMT) analysis, which is validated using experimental data generated for the purpose. Proprotors have conflicting requirement for optimal hover and forward flight performance. Next, the biplane wings are designed using simple lifting line theory. The airframe design is followed by power plant selection and transmission design. Finally, weight estimation is carried out to complete the design process. The proprotor design with 24$^\circ$ preset angle and -24$^\circ$ twist is designed based on 70\% weightage to forward flight and 30\% weightage to hovering flight conditions. The operating RPM of the proprotors is reduced from 3200 during hover to 2000 during forward flight to ensure optimal performance during cruise flight. The estimated power consumption during forward flight mode is 64\% less than that required for hover, establishing the benefit of this hybrid concept. A proof-of-concept scaled prototype is fabricated using commercial-off-the-shelf parts. A PID controller is developed and implemented on the PixHawk board to enable stable hovering flight and attitude tracking.

\subsection*{Keywords}
Variable Pitch, Quadrotor Tailsitter UAV, UAV Design, Blade Element Theory, Payload Delivery

\section*{Nomenclature}
\begin{tabbing}
XXX \= \kill
$A$ \> \indent \qquad \qquad Rotor disk area \\ 
$\AR,\;\AR_w$ \> \indent \qquad \qquad Aspect ratio of rotor and wing respectively\\
$a_t, d_t$ \> \indent \qquad \qquad Addendum and dedendum of gear respectively\\
$b_t$ \> \indent \qquad \qquad Gear face width\\
$b_w$ \> \indent \qquad \qquad Span of the wing\\
$B$ \> \indent \qquad \qquad Bevel gear\\
$c, c_w$ \> \indent \qquad \qquad Chord of the rotor and wing respectively \\
$C_l,\; C_d$ \> \indent \qquad \qquad 2D Coefficient of lift and drag respectively \\
$C_{D_0}$ \> \indent \qquad \qquad Parasitic drag coefficient of the wing\\
$C_{L_{cr}}, \;C_{L_{max}}$ \> \indent \qquad \qquad Lift coefficient during cruise and maximum lift coefficient\\
$C_T, \;C_P$ \> \indent \qquad \qquad Coefficient of thrust and power respectively\\
$C_{Th}$ \> \indent \qquad \qquad Coefficient of thrust during hover\\
$D$ \> \indent \qquad \qquad Drag\\
$E$ \> \indent \qquad \qquad Engine gear\\
$\textbf{E},\; \textbf{E}_d$ \> \indent \qquad \qquad Vector representing actual Euler angles and desired Euler angles respectively\\
$\textbf{e}_a,\; \textbf{e}_p$ \> \indent \qquad \qquad Error vector for Euler angles and position respectively\\
$F$ \> \indent \qquad \qquad Prandtl's tip loss factor\\
$FM$ \> \indent \qquad \qquad Figure of merit\\
$F_t$ \> \indent \qquad \qquad Load transmitted by gears\\
$g$ \> \indent \qquad \qquad Acceleration due to gravity\\
$K$ \> \indent \qquad \qquad Induced drag factor\\
$K_f,\; K_m$ \> \indent \qquad \qquad Factor for stress concentration effect and load distribution respectively\\
$K_F$ \> \indent \qquad \qquad Rotor force constant\\
$\textbf{$\textbf{K}_p^a$,  $\textbf{K}_i^a$, $\textbf{K}_d^a$}$ \> \indent \qquad \qquad PID gains for attitude control\\ 
$\textbf{$\textbf{K}_p^p$, $\textbf{K}_i^p$, $\textbf{K}_d^p$}$ \> \indent \qquad \qquad PID gains for position control\\ 
$K_o,\; K_v$ \> \indent \qquad \qquad Factor for loading and dynamic loading respectively\\
$l, m, n$ \> \indent \qquad \qquad Total Moments about body fixed x, y, z axes\\
$L$ \> \indent \qquad \qquad Lift \\
$M$ \> \indent \qquad \qquad Mass of the vehicle\\
$\textbf{M}$ \> \indent \qquad \qquad Vector of moments acting on the vehicle\\
$M1, M2$ \> \indent \qquad \qquad Middle gears\\
$N_b$ \> \indent \qquad \qquad Number of blades \\
$P, \; P_{ind}$ \> \indent \qquad \qquad Total power and induced power respectively\\
$P_d$ \> \indent \qquad \qquad Pitch circle diameter\\
$PL$ \> \indent \qquad \qquad Power loading\\
$r$ \> \indent \qquad \qquad Non-dimensional radial position, = y/R \\
$\textbf{r},\;\textbf{r}_d$ \> \indent \qquad \qquad Actual position and desired position vector of the vehicle\\
$R$ \> \indent \qquad \qquad Rotor blade radius \\
$S1, \; S2$ \> \indent \qquad \qquad Shaft gears\\
$S_w$ \> \indent \qquad \qquad Wing area\\
$T$ \> \indent \qquad \qquad Thrust\\
$T_1, \; T_2$ \> \indent \qquad \qquad Gear teeth ratio for gear 1 and 2\\
$TR$ \> \indent \qquad \qquad Taper ratio\\
$u_i, w_i$ \> \indent \qquad \qquad Induced swirl and  axial velocity respectively  \\
$U_T, U_P$ \> \indent \qquad \qquad Sectional in plane and out of plane velocity respectively  \\
$U$ \> \indent \qquad \qquad Resultant Velocity\\
$v_i$ \> \indent \qquad \qquad Resultant induced velocity\\
$V_\infty, V_{cr}, V_{stall}$ \> \indent \qquad \qquad Free stream velocity, cruise velocity and stall velocity respectively\\
$W$ \> \indent \qquad \qquad Weight \\
$(x, \;y, \; z)$ \> \indent \qquad \qquad Position of the vehicle\\
$(x_d, \;y_d, \; z_d)$ \> \indent \qquad \qquad Desired position of the vehicle\\
$y_r$ \> \indent \qquad \qquad Distance along blade from rotational axis\\
$Y$ \> \indent \qquad \qquad Lewis factor\\
$\alpha$ \> \indent \qquad \qquad Angle of attack\\
$\beta$ \> \indent \qquad \qquad Ratio of span of biplane to that of monoplane\\
$\gamma$ \> \indent \qquad \qquad $tan^{-1}(C_d/C_l)$\\
$\eta_P$ \> \indent \qquad \qquad Propeller efficiency\\
$\theta$ \> \indent \qquad \qquad Blade pitch angle \\
$\lambda, \lambda_i$ \> \indent \qquad \qquad Out of plane velocity ratio and induced velocity ratio respectively\\
$\mu$ \> \indent \qquad \qquad Advance ratio \\
$\xi, \xi_i$ \> \indent \qquad \qquad In plane plane velocity ratio and induced velocity ratio respectively\\
$\rho$ \> \indent \qquad \qquad Air density\\
$\sigma$ \> \indent \qquad \qquad solidity =$\frac{N_b c}{\pi R}$\\
$\sigma_b$ \> \indent \qquad \qquad Maximum bending stress\\
$\Phi$ \> \indent \qquad \qquad Inflow angle\\
$(\phi, \;\theta, \; \psi)$ \> \indent \qquad \qquad Euler angle representing attitude of the quadrotor\\
$(\phi_d, \;\theta_d, \; \psi_d)$ \> \indent \qquad \qquad Desired Euler angle representing desired attitude of the quadrotor\\
$\phi_0$ \> \indent \qquad \qquad Pressure angle for gear \\
$\Omega$ \> \indent \qquad \qquad Rotor angular velocity (rad/sec) \\
AGMA \>\indent \qquad \qquad American Gear Manufacturers Association\\
VTOL \> \indent \qquad \qquad Vertical Take-off Landing\\
UAV \> \indent \qquad \qquad Unmanned Aerial Vehicle\\
\end{tabbing}
\section{Introduction}
With increased popularity of internet shopping, e-commerce sector faces a daunting task of delivering huge volumes of packages on time. The technological revolution in the area of Unmanned Aerial System has provided the sector a plausible solution that can not only deliver the packages on time but also to remote areas with difficult terrains. Amazon, Google, Alibaba and other e-commerce giants are already working on developing their own Unmanned Aerial Vehicle (UAV) solution with small payload capacity to cater to the ``last mile'', as it can augment their profit by cutting labor cost. Apart from e-commerce sector, emergency health-care sector is also showing tremendous interest in UAVs with automatic payload delivery system, which can be used to deliver life saving drugs or blood over remote areas. Autonomous payload delivery requires the UAV to do precise hover at low altitude above ground and fly long distances efficiently, both of which have conflicting requirements. The rotary wing UAVs offer the ability to hover and vertical take-off and landing (VTOL), but are known to have less efficiency and speed than the fixed wing UAVs. Fixed wing UAVs have been used widely because of their simple design, high endurance, high range and high speed capability but are incapable of VTOL. Therefore, there is a need for hybrid design of UAVs that combine VTOL capabilities with high speed, range and endurance.

The common hybrid VTOL concepts studied in literature include  monoplane wing tail-sitter~\cite{tailsit1,tailsit2}, tiltrotor~\cite{tiltrotor1,tiltrotor2}, compound helicopter, quad tiltrotor~\cite{quadtiltrotor}, quad-tiltwing~\cite{quadtiltwing} concepts. The monoplane tail sitter concept has the simplest design, but suffers from poor control authority during hover due to use of one propeller. The transition from hover to forward flight and back is quite challenging in these vehicles. Mimicking the V-22 Osprey tiltrotor design in small sized UAVs is cumbersome as it requires use of swashplate for controlling the pitch angle of each of the rotors making it mechanically complicated and therefore is only suitable for larger sized UAVs such as that of~\cite{tiltrotor1}. The fixed pitch propeller based designs as that of~\cite{tiltrotor2} are easy to build but suffer from poor stability and control authority in pitch mode. Lift and thrust compounding results in high empty weight and therefore has not been attempted in 1--2 kg class small UAVs. The quad tiltrotor design has good control authority during hover and transition and its flight control is simpler than that for monoplane tail-sitter. But, the rotor wing interaction in hover and the aerodynamic interaction between fore and aft wings result in loss of efficiency. The interaction losses during hover for quad tiltrotor is addressed in the quad-tiltwing design of~\cite{quadtiltwing}.  It has four rotors that are mounted on the four wings. The wings, together with the rotors, are tilted between vertical and horizontal configurations to accomplish vertical and horizontal flights. It hovers like a quadrotor and a flies like a tandem wing airplane in forward flight mode. However, the transition from hover to forward flight and back to hover offers significant challenges due to the flow separation on the wing operating at very angle of attack at the beginning of the transition. The convertiplane design of~\cite{abhishek} offers a better compromise with part of the wing under the downwash of rotor tilting with it to eliminate hover losses. Also, since significant portion of the wing remains straight, it facilitates transition.

Google X, a division of Google, developed and carried out field trials of a tailsitter aircraft with four fixed pitch propellers offset from the wing~\cite{Stewart} to enable smooth transition from hover to forward flight and back. The objective was to develop a UAV solution for ``last mile'' delivery. After extensive testing the design was deemed unsuitable at maintaining its attitude during windy conditions due to lack of control power as the four propellers were quite close to each other and therefore had poor control authority.  Amazon's solution has been brute force and simple. Instead of focusing on efficiency, the focus is on simplicity to ensure that a working solution is achieved in least time. The design of Amazon Prime Air package delivery UAV prototype is a battery driven vehicle with four sets of coaxial contra-rotating propellers for vertical takeoff, hover and landing and tandem wings with pusher propeller for forward flight. The four coaxial propellers operate like that of a quadrotor with X-8 configuration to provide control during hover and transition and are possibly turned off during forward flight increasing the empty weight of the vehicle.

A more efficient solution has been developed at University of Maryland, which is a quadrotor biplane MAV~\cite{Hrishikeshavan1,Hrishikeshavan2, Hrishikeshavan3}. The design tries to overcome the mechanical complexity associated with tilting of rotors and wings in tiltrotor or tiltwing type hybrid concepts by making use of the maneuverability of the vehicle for transition. It consists of counter-rotating propellers arranged in a conventional quadrotor configuration, to which are attached two wings in biplane configuration. This design is expected to offer significantly higher control authority and can successfully transition from hover to forward flight and back to hover. This design offers better overall performance compared to some of the other options discussed above as it uses same set of propellers for hover and forward flight. However, being battery powered compromises its industrial usability. 

The novel design of a quadrotor biplane Vertical Takeoff and Landing (VTOL) UAV proposed and designed in this paper aims at providing solution to several of the challenges discussed above. In this design, a variable pitch H-shaped quadrotor design is combined with two fixed wings attached to the parallel limbs of the quadrotor in  biplane configuration similar to that of~\cite{Hrishikeshavan1}. The key differences between the two designs are: 1) the use of variable pitch rotors instead of fixed pitch ones, which enables use of internal combustion (IC) engine as power plant, 2) significantly higher endurance, in comparison to battery powered solutions, due to use of fuel engine and 3) use of mechanical transmission to transfer power from engine to the rotors. The vehicle can take-off and land vertically and hover in helicopter mode before transitioning to fixed wing mode to fly with good fuel economy like a fixed wing UAV. The present design is based on the IC engine powered quadrotor work carried out by Abhishek et al.~\cite{ram}. 

While, the use of variable pitch propeller and mechanical transmission adds complexity to a simple battery powered earlier design. The use of variable pitch rotors instead of fixed pitch ones offer the following advantages:
\begin{enumerate}
\item Higher control authority and control bandwidth than fixed pitch propeller due to higher rate of change of thrust and ability to generate negative thrust~\cite{Cutler}. This would improve maneuverability and gust rejection capability. 
\item The ability to use variable pitch enables the use of fuel engines whereby increasing the endurance dramatically. Gasoline has significantly higher energy density as compared to current Lithium Polymer batteries which are commonly used in UAVs.
\item Different rotor RPMs can be chosen for the rotor during hover and forward flight modes to optimize the rotor power consumption during hover and forward flight.
\item Use of fuel engine reduces turnaround time between flights as refueling is faster than charging batteries. 
\end{enumerate}
  
Design of rotary wing UAVs is well established~\cite{heli}. However, the design methodology followed for hybrid UAV systems is not fully understood. This paper describes the design, development and flight testing of this novel quadrotor biplane VTOL UAV which has not been studied in literature. The objective is to systematically carry out design of a novel UAV concept, followed by fabrication and flight testing of a scaled proof-of-concept prototype. In section 2, detailed description of the vehicle is given followed by a discussion of the design methodology in section 3. For this a physics based modified Blade Element Momentum Theory (BEMT) simulation is developed and validated using experimental data generated for the purpose.
To initiate the vehicle design process, first, the proprotor is carefully designed using the above BEMT analysis as discussed in section 4. Section 5 details the wing design.  The conceptual design of the UAV proprotor and wing is followed by selection of power plant described in section 6. The transmission mechanism required to transfer power from the engine to the rotors is discussed in section 7. Preliminary structural analysis is presented in section 8 followed by weight and center of gravity (cg) analysis in section 9. The avionics and telemetry system used are described in section 10. In section 11, the proof-of-concept prototype integration and flight testing is discussed and finally the paper is concluded in section 12.

\section{Vehicle configuration}

\begin{figure}[h]
  \centering
  \subfigure[Hover]{
    \label{Vehicle:hover}
    \includegraphics[width=0.47\columnwidth]{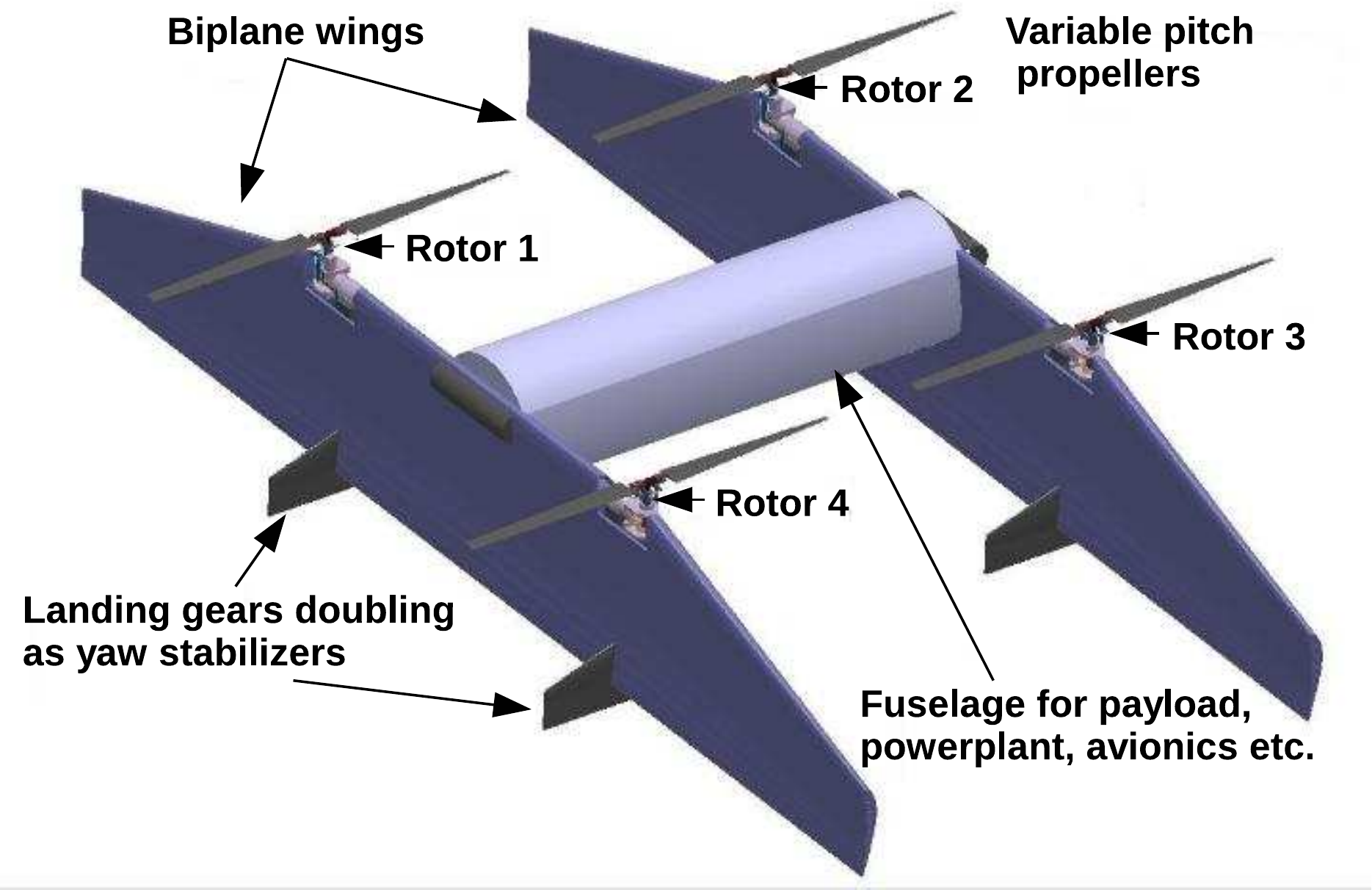}} \\
  \subfigure[Forward flight]{
    \label{Vehicle:forward}
      \includegraphics[width=0.47\columnwidth]{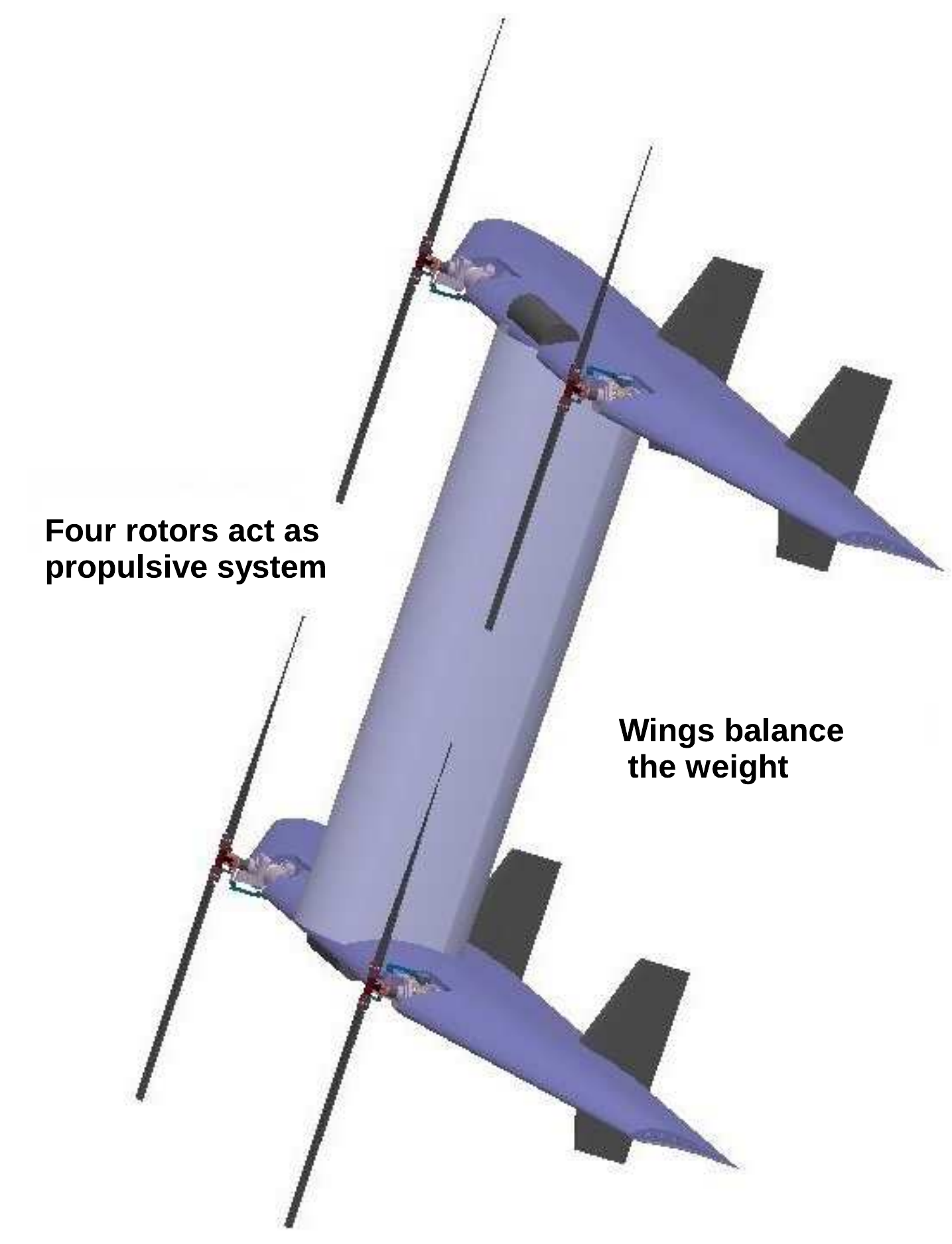}}
    \caption{Hover and forward flight modes of the quadrotor tailsitter UAV prototype}
\label{fig:schematic}
\end{figure}

The proposed design, a quadrotor biplane tailsitter concept, is a novel tailsitter concept comprising of four variable pitch rotors arranged in a quadrotor like arrangement and driven through a mechanical transmission using a single power plant with two fixed wings (in bi-plane configuration) attached to parallel limbs of the quadrotor as shown in Fig.~\ref{fig:schematic}. During entire duration of the flight, the wings remain parallel to the propeller axis and hence do not result in any download penalty due to rotor downwash hitting the rotors, commonly experienced by tilrotor type configurations.

The payload can be mounted inside the fuselage of the vehicle which has been designed for this purpose. After gaining safe altitude in hover mode, the rotors would then maneuver and change its attitude from hover mode to forward flight (airplane) mode by generating differential thrust resulting in pitching moment about the center of gravity of the vehicle, required to transition the vehicle to fixed wing mode (forward flight mode) as shown in Fig. \ref{Vehicle:forward}. The VTOL vehicle being designed, attempts to selectively incorporate the advantages offered by both the fixed wing and quadrotor type of vehicles, as it can take off, land vertically and hover at a point which is necessary requirement to deliver the payload safely to any destination.  During the forward flight mode, it can fly efficiently without any dead weight and attain high speeds to cover large distances required for the last mile delivery.

\subsection{Control strategy}
The primary control of various motions (three translational, roll and pitch motions) in the design is achieved by changing the thrust and torque of different rotors in various combinations during the hover as well as forward flight modes. In the hover mode, the flight control is identical to the control of quadrotor helicopter with fixed pitch, however the mechanisms of thrust variation are different for the two. The change in thrust for the present design is achieved by changing the blade pitch angle. The control of yawing motion and the mechanism involved is significantly different as discussed below.

\begin{figure}[h]
\centering
\includegraphics[scale=0.6]{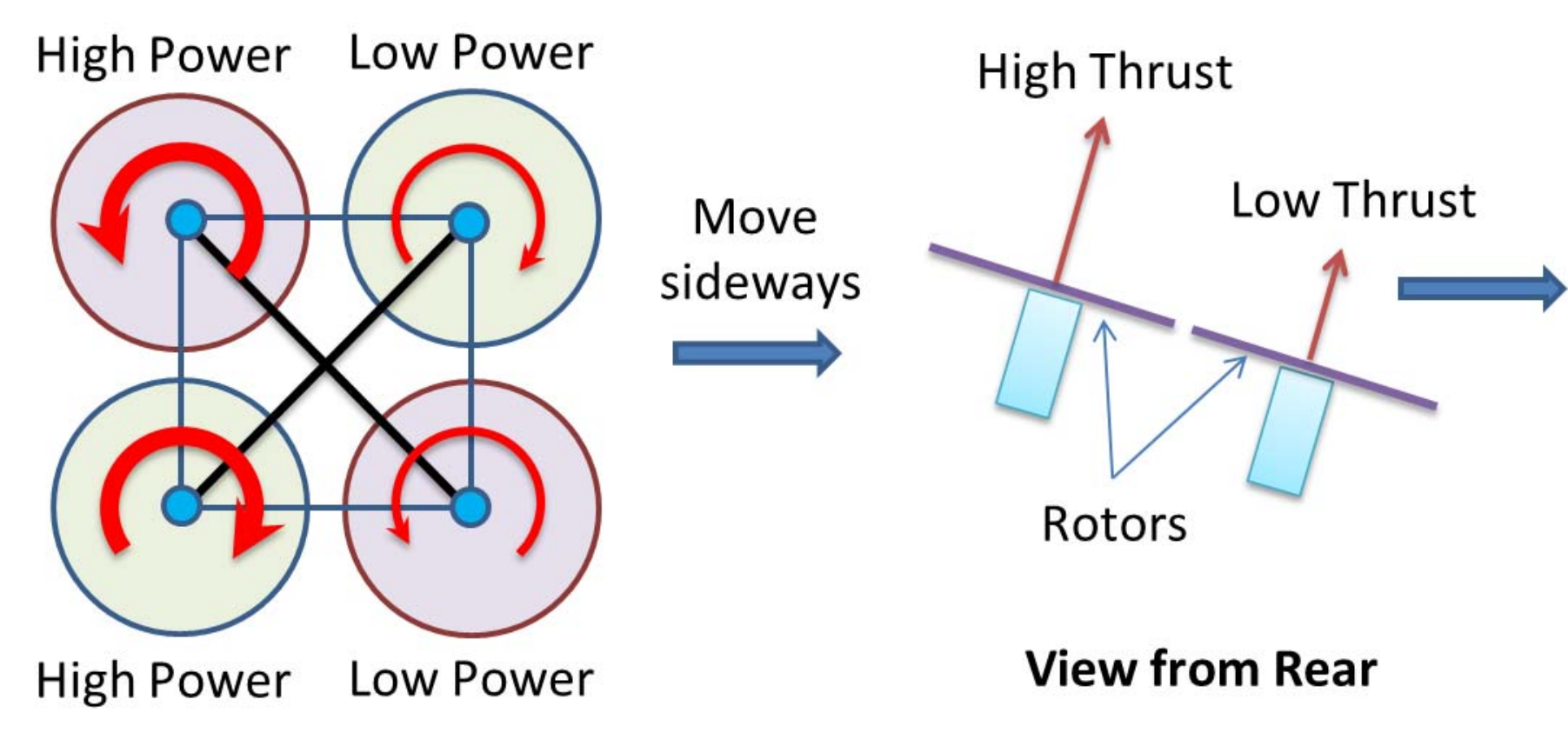}
\caption{Control strategy for sideways motion}
\label{Side_control}
\end{figure}

\begin{figure}
\centering
\includegraphics[scale=0.6]{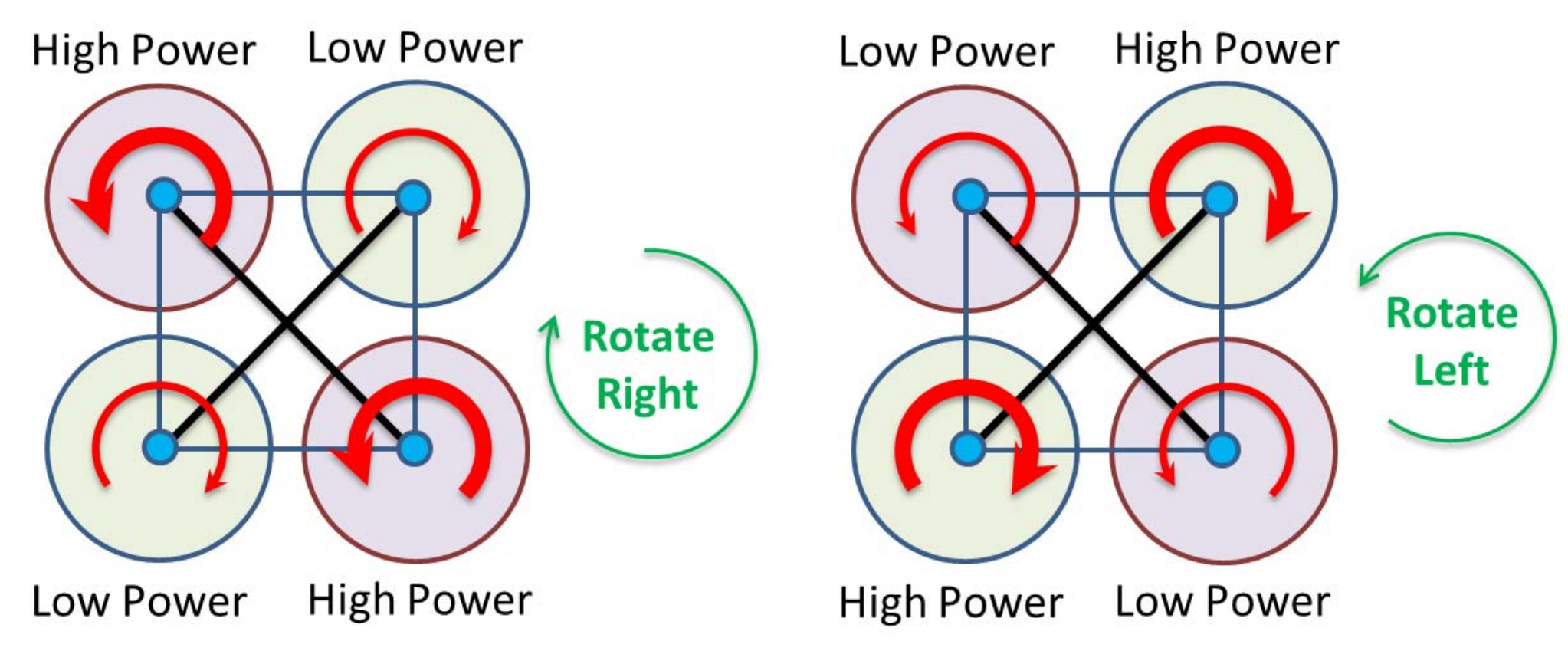}
\caption{Control strategy for yaw motion}
\label{Yaw_control}
\end{figure}

The up/down motion is easily controlled by collectively increasing or decreasing the blade pitch angle for all the rotors simultaneously. Sideways flight can be achieved as explained in Fig. \ref{Side_control}. For example, to move right, the collective inputs of the left rotors are increased and the inputs to right rotors are decreased by same amount, and thereby increasing the thrust of the two left rotors which lifts the left side up and generates a net thrust component to the right. In similar way, the vehicle can be moved to left. By the same principle, increase in collective of the two rear rotors would result in forward flight and increase in collective of the two front rotors would result in backward motion. The yaw control is less intuitive. The method of generating yawing moment is identical to that used for coaxial and tandem helicopters. It is known as ``differential collective''. In this, the collective pitch of the two diagonal rotors rotating in the same direction is increased and that of the other two is decreased by same amount. The change of collective pitch results in increase of lift and drag experienced by both these rotors compared to the other two rotors. This increase in lift and drag results an increase in profile and induced torque components of these rotors compared to the other two rotors. This increment in the torque value results in yawing motion of the quadrotor as explained in Fig. \ref{Yaw_control}. It is important, as explained earlier, to note that because of the diagonal arrangement, this operation has no effect on translational motion.

The roll, pitch and yaw motions in forward flight mode can be controlled by using the same set of controls that are used for flight control during hovering flight. It should be noted that, after the vehicle maneuvers into the forward flight mode, the yaw and roll degree of freedom during hovering flight becomes the roll and yaw degree of freedom respectively. The pitch motion during hover remains the pitch motion during forward flight. And the flight control for various motions can be achieved as explained earlier. In addition, the pitching moment contribution from the wings can be compensated by differential thrust generation by the top and bottom rotors, during forward flight. 

The rotor RPM in forward flight mode is reduced from that in the hover mode as the thrust requirements are significantly reduced in forward flight. The entire weight of the vehicle is supported using the thrust from the four rotors during hover, hence the rotors need to operate at a higher RPM as the thrust requirement from each rotor is higher. However, during the forward flight, thrust is required only to overcome the drag of the vehicle which is significantly smaller than the weight of the vehicle. Therefore, the rotor RPM is reduced in forward flight mode to save power. This observation would be elucidated later in text.   

Since, the vehicle doesn't have a tail and sits on the biplane wings during take-off and landing, four landing gears shaped like small vertical fins are attached at the trailing edge of the wings as shown in Fig.~\ref{fig:schematic}. These vertical fins provide rigid support for the vehicle to land on the ground and also act as vertical fin during forward flight to augment the yaw damping of the vehicle.

\section{Approach} 
 
The design methodology for hybrid UAVs is established by carrying out the design of a hybrid tailsitter UAV with the following mission objectives:
\begin{itemize}
  \item Payload: 6 kg
  \item Range: 32 km
  \item Cruising height: 500 m above sea level 
  \item Cruising speed: 20 m/s
  \item Compact vehicle
\end{itemize}
These mission objectives have been decided based on the request for proposal of 32nd American Helicopter Society student design competition~\cite{ahs2014} for a payload delivery UAV. 

\begin{figure}[h]
	\centering
	\includegraphics[scale=0.65]{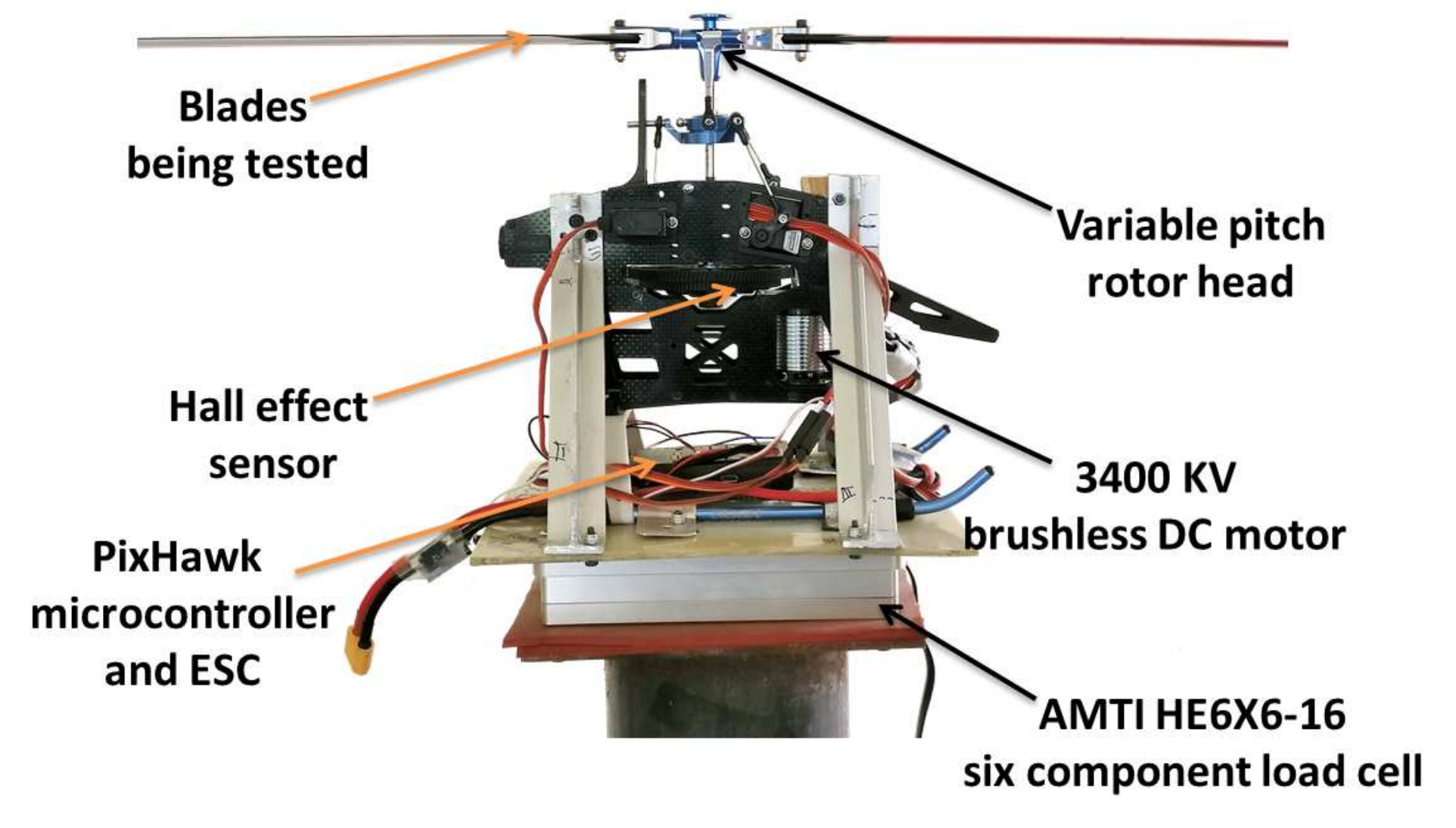}
	\caption{Experimental setup used for thrust and torque measurements}
	\label{fig:Exp_setup}
\end{figure}

The complete design and development of the tailsitter vehicle is carried out in various steps. First, the preliminary sizing of the proprotors is carried out. Essentially, proprotors need to perform optimally in both hover as well as forward flight modes. The proprotor sizing and design is carried out by using a modified blade element momentum theory (BEMT) specifically formulated for propellers and proprotors~\cite{Stahlhut}. The analysis developed for this purpose is used to study the effect of varying different design parameters such as blade taper, aspect ratio (solidity), twist, radius and tip speed on the rotor performance during hover as well as forward flight. This modified BEMT differs from the conventional BEMT~\cite{Leishman} in the following respects: 1) It doesn't have the small inflow angle assumption, 2) It includes the in plane or swirl velocity component, 3) Prandtl's tip loss function is modified according to large inflow angles. The predictions from the above modified BEMT analysis is first validated using measured thrust and torque data for a variable pitch rotor. A hover test stand with variable pitch rotor head is setup for this purpose and is shown in Fig.~\ref{fig:Exp_setup}. The experimental setup consists of a six component load cell mounted on hover test stand with a variable pitch rotor system mounted on top of it. 

Next, the airfoil selection is carried out for the proprotor system. Once, the airfoil is chosen, the proprotors are sized for radius, chord, twist, taper, number of blades, tip speed etc. based on the performance requirements for both hover and forward flight. The preliminary calculations are carried out based on a nominal weight of approximately 16 kg, which is later on updated as the actual weight of the final design is arrived through detailed weight estimation using CAD drawing. 

Next, the aerodynamic design of the biplane wing is carried out to meet the established performance indices. This involves wing loading selection, airfoil selection, planform sizing, material selection. After the rotor and wings have been sized the estimated performance indices allow for the selection of the off-the-shelf power plant. Once, the power plant is selected, the transmission system and the variable pitch mechanisms are designed in detail. After the aerodynamic design is completed, preliminary structural analysis is carried out using off the shelf FEM tools to analyze the factor of safety available with the aircraft. The weight estimation and centre of gravity locations are also found out by using CAD software. The estimated weight is used to iterate over the entire design process till a convergence is achieved. Once the design process is completed, the fabrication of the prototype is carried out for a scaled model using appropriate material described later in text. 

Finally, a simple proportional-integral-derivative (PID) based attitude controller is developed for the variable pitch quad-rotor system and used to stabilize the vehicle in hovering flight. It is implemented on a PixHawk autopilot board to demonstrate stable hovering flight for the scaled prototype. The control design for transition and forward flight would be carried out in future work.

\subsection{Modified Blade Element Theory (BET)}
The modified blade element momentum theory (BEMT), described in this section, has been specifically formulated for propellers and proprotors by Stahlhut~\cite{Stahlhut}. It is used to analyze the effect of various rotor design parameters such as taper, aspect ratio (solidity for constant radius), twist, radius and tip speed on its performance as a helicopter rotor and propeller. The modified BEMT differs from the conventional BEMT~\cite{Leishman} in the following respects: 1) the small inflow angle assumption is not made, 2) in-plane or swirl velocity component is included in the definition of inflow angle, and 3) Prandtl's tip loss function is modified for large inflow angles.

\begin{figure}[h]
	\centering
	\includegraphics[scale=0.6]{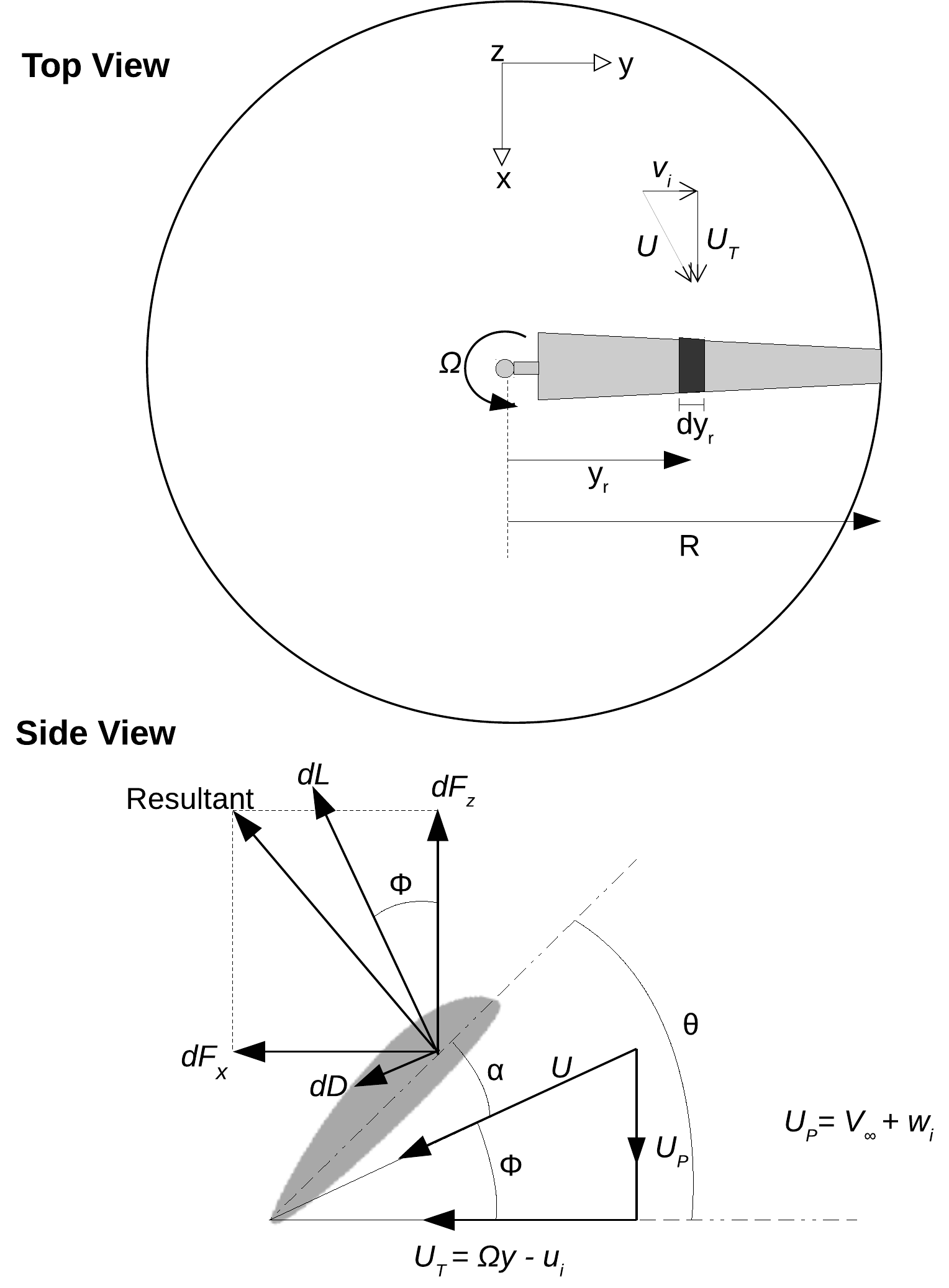}
	\caption{Velocity components that the blade sections are experiencing}
     \label{Vel_comp}
\end{figure}

Figure~\ref{Vel_comp} shows the schematic of a rotor blade and an element therein. The inflow angle $\Phi $, at each blade section, depends on both the $w_{i} $ and $ u_{i} $ components of the induced velocity. For a proprotor in high advance ratio forward flight, the inflow angles can be large, often exceeding $45^{\circ}$ near the blade tips. Therefore, the lift vector on the blade sections may induce flow velocities in the in-plane direction that may be greater than the induced inflow component, therefore, both the thrust and torque component are considered for estimating the induced flow-field.

Using the blade element theory, the infinitesimal thrust coefficient generated by a blade element is given by:

\begin{equation}
 {\rm d}C_T=\frac{{\rm d}T}{\rho A(\Omega R)^2} =\frac{N_b ({\rm d}L cos(\Phi)-{\rm d}D sin(\Phi))} {\rho A(\Omega R)^2}
\end{equation}

\begin{equation}
{\rm d}C_T=\frac{N_b (\frac{1}{2} \rho U^2 c)(C_l  cos⁡(\Phi)-C_d sin(\Phi)){\rm d}y_r}{\rho A(\Omega R)^2}
\end{equation}

\begin{equation}
{\rm d}C_T=\frac{N_b ( \frac{1}{2} \rho c  \sqrt{U_T^2+U_P^2 })(C_l U_T-C_d U_P ){\rm d}(\frac{y_r}{R} )  }{\rho A(\Omega R)^2}
\end{equation}

\begin{equation}
{\rm d}C_T= \frac{1}{2} \sigma \sqrt{\xi^2+\lambda^2 }(C_l \xi-C_d \lambda ){\rm d}r
\end{equation}

where inflow ratio $\lambda=\frac{(V_{\infty}+w_i)}{\Omega R}$ and azimuthal flow ratio $\xi=\frac{(\Omega y_r-u_{i})}{\Omega R}$
 
Similarly,
\begin{equation}
{\rm d}C_P= \frac{{\rm d}P}{\rho A(\Omega R)^3} = \frac{N_b ({\rm d}L sin(\Phi)+{\rm d}D cos(\Phi))\Omega y_r} {\rho A(\Omega R)^3}
\end{equation}

 \begin{equation}
 {\rm d}C_P= \frac{1}{2} \sigma \sqrt{\xi^2+\lambda^2 })(C_l \lambda+C_d \xi )r{\rm d}r
 \end{equation}

The elemental thrust and power coefficients estimated from Momentum theory for a rotor annulus are given by:
\begin{equation}
{\rm d}C_T= 4|\lambda|\lambda_i r{\rm d}r
\end{equation}
\begin{equation}
{\rm d}C_P= 4|\lambda|\xi_i r^2{\rm d}r
\end{equation}

\subsubsection{Tip loss effect}

It is necessary to account for tip-loss effects because only some of the induced flow arises from momentum conservation, the remaining part is because of the tip vortices. Goldstein \cite{Goldstein} established a method for tip-loss by using the velocity potential of a series of helical vortex sheets in the rotor wake. A simplified version of the Goldstein result was first developed by Prandtl, who approximated the helical surface as a series of two-dimensional planar sheets that convect at the slipstream velocity, which is a more practical mathematical realization of Goldstein's approach \cite{Bradley}. 

Prandtl's tip loss factor is given by:
\begin{equation}
F=\frac{2}{\pi}cos^{-1}[exp⁡(\frac{\frac{N_b}{2}(r-1)}{2 rsin(\Phi)})]
\end{equation}

This tip loss factor is incorporated in the momentum theory using Kutta-Joukowski theorem
\begin{equation}\label{eqn:Thrust1}
{\rm d}C_T = 4F|\lambda|\lambda_i r{\rm d}r
\end{equation}
\begin{equation} \label{eqn:Power1}
{\rm d}C_P = 4F|\lambda|\xi_i r^2{\rm d}r
\end{equation}

Because of the inviscid flow assumption, equation \ref{eqn:Thrust1} is only strictly valid when $\Phi=0^{0}$ and equation \ref{eqn:Power1} is only valid when  $\Phi=90^{0}$. At $\Phi=90^{0}$  for $dC_T$ and $\Phi=0^{0}$ for $dC_P$, the tip vortices do not contribute to the induced velocities $w_{i}$ and $u_i$. So, in order for these result to match with the expected physical behaviour, the decrease in thrust and power after tip loss i.e $ dC_T=4(1-F)|\lambda|\lambda_i r{\rm d}r $ and $ dC_P=4(1-F)|\lambda|\xi_i r^2{\rm d}r $ should gradually decrease to zero as the angles decrease between the tip vortex axes and the blade force vectors. For incorporating this, the theory is modified as follows.
\begin{equation}\label{eqn:Thrust2}
{\rm d}C_T = 4K_T|\lambda|\lambda_i r{\rm d}r
\end{equation}
\begin{equation} \label{eqn:Power2}
{\rm d}C_P = 4K_P|\lambda|\xi_i r^2{\rm d}r
\end{equation}
Where $K_T=1-(1-F)cos(\Phi)$ and $K_P=1-(1-F)  sin⁡(\Phi)$

Now to solve for the unknown inflow and swirl components, the thrust and power expressions from blade element and momentum theory are equated. 
\begin{equation}\label{eqn:Thrust3}
{\rm d}C_T= 4K_T|\lambda|\lambda_i r{\rm d}r =  \frac{1}{2} \sigma \sqrt{\xi^2+\lambda^2 })(C_l \xi-C_d \lambda ){\rm d}r
\end{equation}
\begin{equation} \label{eqn:Power3}
{\rm d}C_P= 4K_P|\lambda|\xi_i r^2{\rm d}r = \frac{1}{2} \sigma \sqrt{\xi^2+\lambda^2 })(C_l \lambda+C_d \xi )r{\rm d}r 
\end{equation}

These two equations can be solved simultaneously for $\lambda$ and $\xi$ using fixed point iteration or any other method. But a different approach, which combines these two equations into a single equation, as shown in \cite{Winarto}, and solves for inflow using bracketed bisection method, is used here because of the guaranteed solution that this method gives.

 Replacing $\lambda = \frac{U sin(\Phi)}{\Omega R}$ and $\xi = \frac{U cos(\Phi)}{\Omega R}$ and defining $tan⁡(\gamma) = C_d/C_l$ and putting in equation \ref{eqn:Thrust3}, we get,
 
 \begin{equation}
 \frac{1}{2} \sigma U^2 C_l  sec(\gamma) (cos\gamma  cos\Phi-sin\gamma  sin\Phi )= 4K_T U sin(|\Phi |) w_i r 
 \end{equation}
 
 We know $w_i=U sin(\Phi-V_\infty)$ 
 \begin{equation}
\frac{σU^2 C_l  sec(\gamma)  cos⁡(\Phi+\gamma)}{8K_T U sin(|\Phi|) r }=U sin(\Phi)-V_\infty
 \end{equation}
 Let us define, 	
\begin{equation}
B_1 (\Phi)=\frac{V_\infty}{U}=sin(\Phi)  - \frac{\sigma C_l  sec(\gamma)  cos⁡(\Phi+\gamma)}{8K_T  sin⁡(|\Phi|) r}
\end{equation}
Similarly, using equation \ref{eqn:Power3} we get,
\begin{equation}
B_2 (\Phi)=\frac{\Omega y}{U}=cos(\Phi)  + \frac{\sigma C_l  sec(\gamma)  sin⁡(\Phi+\gamma)}{8K_P  sin(⁡|\Phi|) r}
\end{equation} 
These two equations can be combined to get a single equation 
\begin{equation}
g(\Phi)=[B_1 (\Phi)\Omega y- B_2 (\Phi) V_\infty ]  sin\Phi=0 
\end{equation}
\begin{equation}\label{eqn:Transcend}
g(\Phi)=[\Omega y sin\Phi-V_\infty  cos\Phi ]  sin\Phi - sgn(\Phi) \frac{\sigma C_l  sec\gamma}{r}  [\frac{V_\infty}{K_P}   sin(\Phi+\gamma)  +\frac{\Omega y}{K_T}  cos⁡(\Phi+\gamma) ]  
\end{equation}

Equation \ref{eqn:Transcend} has $\Phi$ as the only unknown. It can be solved by using bisection method by having proper range for solution of inflow angle $\Phi$. 

\subsubsection{Experimental validation of modified BEMT}
Modified BEMT is validated with experimental results for an untwisted rectangular rotor with variable pitch and a fixed pitch twisted propeller. The validated analysis is then used for performance predictions of various rotor configurations for design study. 

\begin{figure}[h]
	\centering
	\subfigure[Thrust validation]{\includegraphics[scale=0.51]{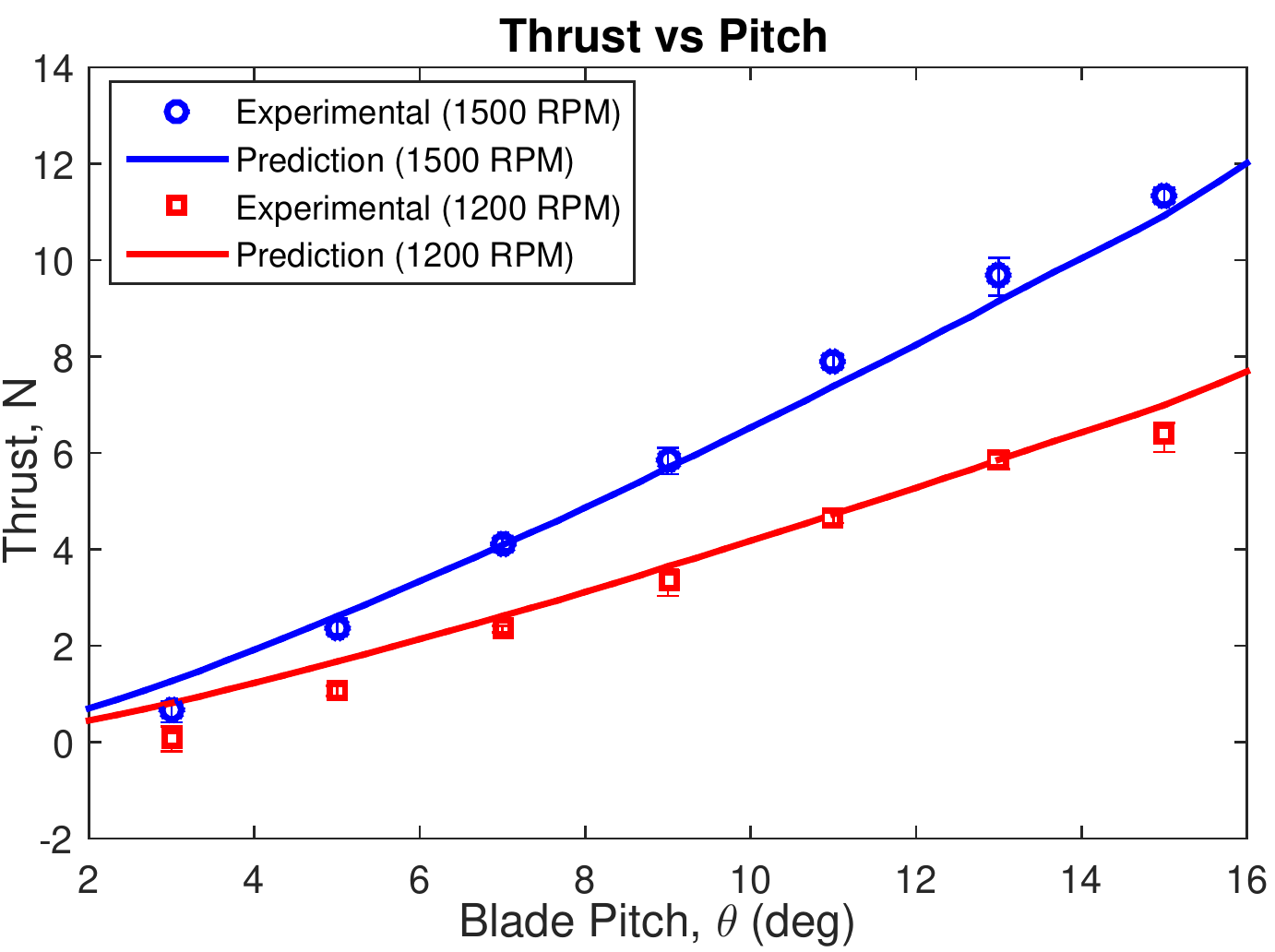}}
	\subfigure[Power  validation]{\includegraphics[scale=.51]{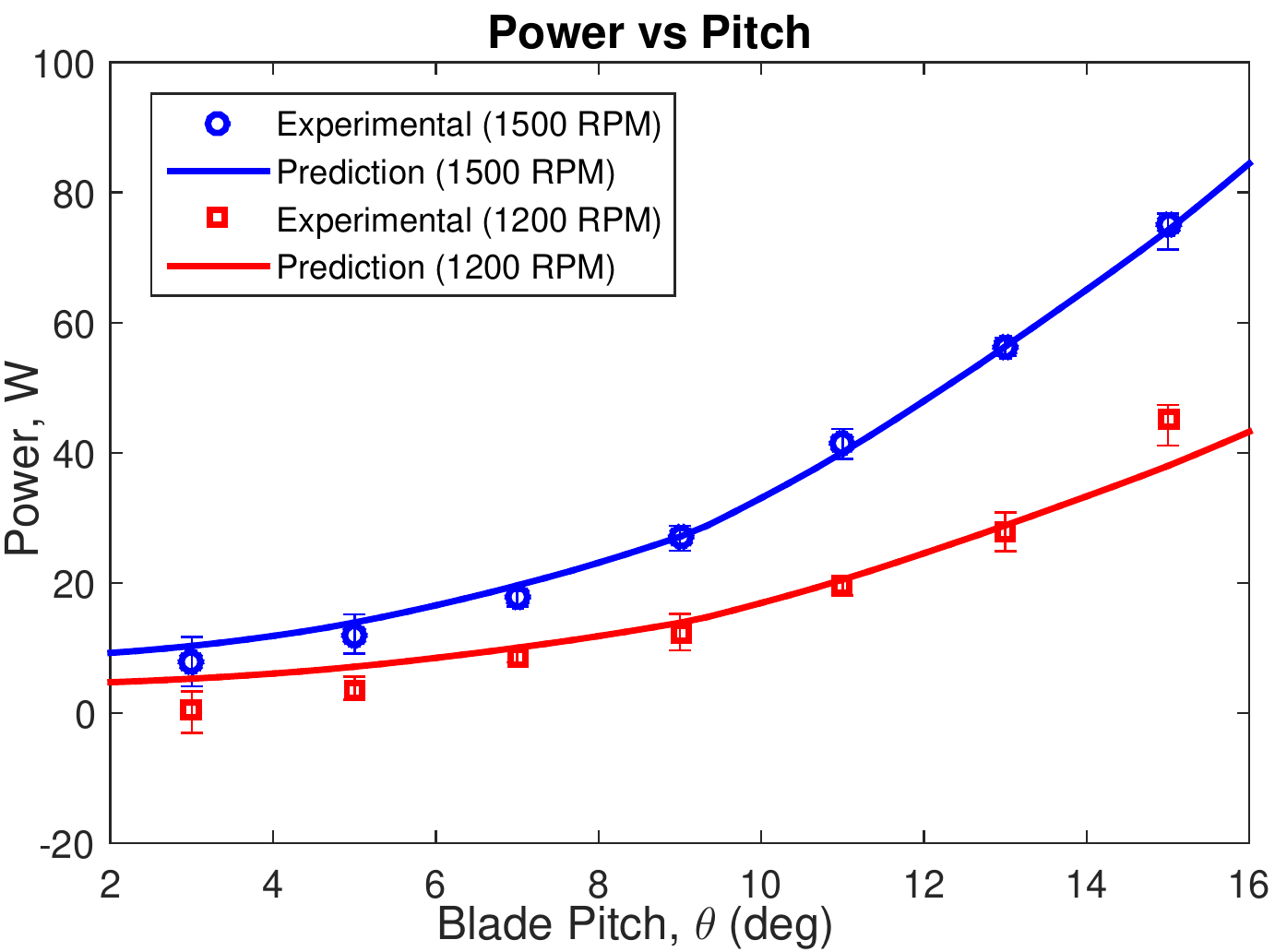}}	
	\caption{Experimental validation of present analysis for untwisted helicopter rotor blade with variable collective pitch input}
	\label{fig:Exp_Valid_rot}
\end{figure}

\begin{figure}[h]
	\centering
	\subfigure[Thrust validation]{\includegraphics[scale=0.51]{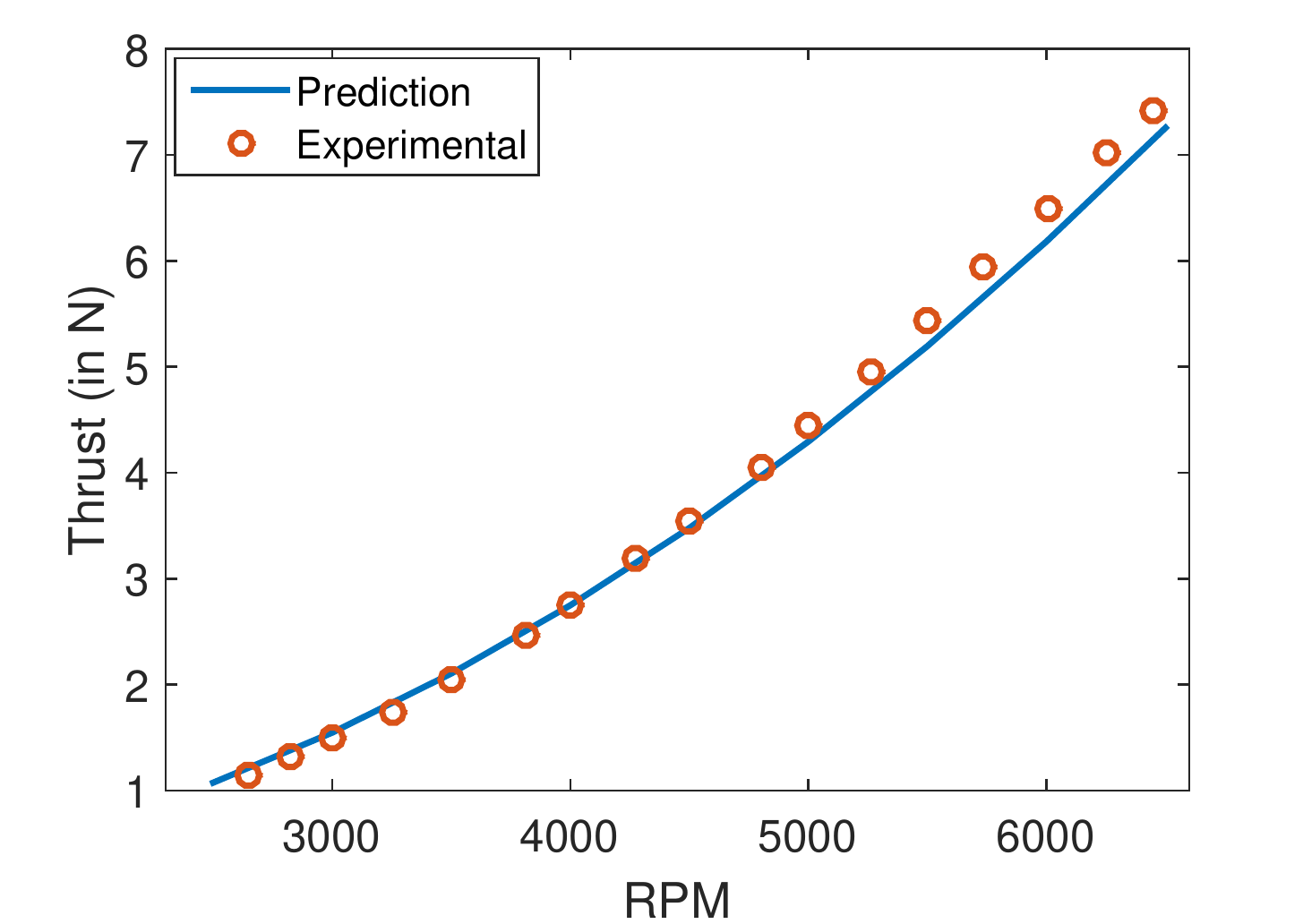}}
	\subfigure[Power  validation]{\includegraphics[scale=.51]{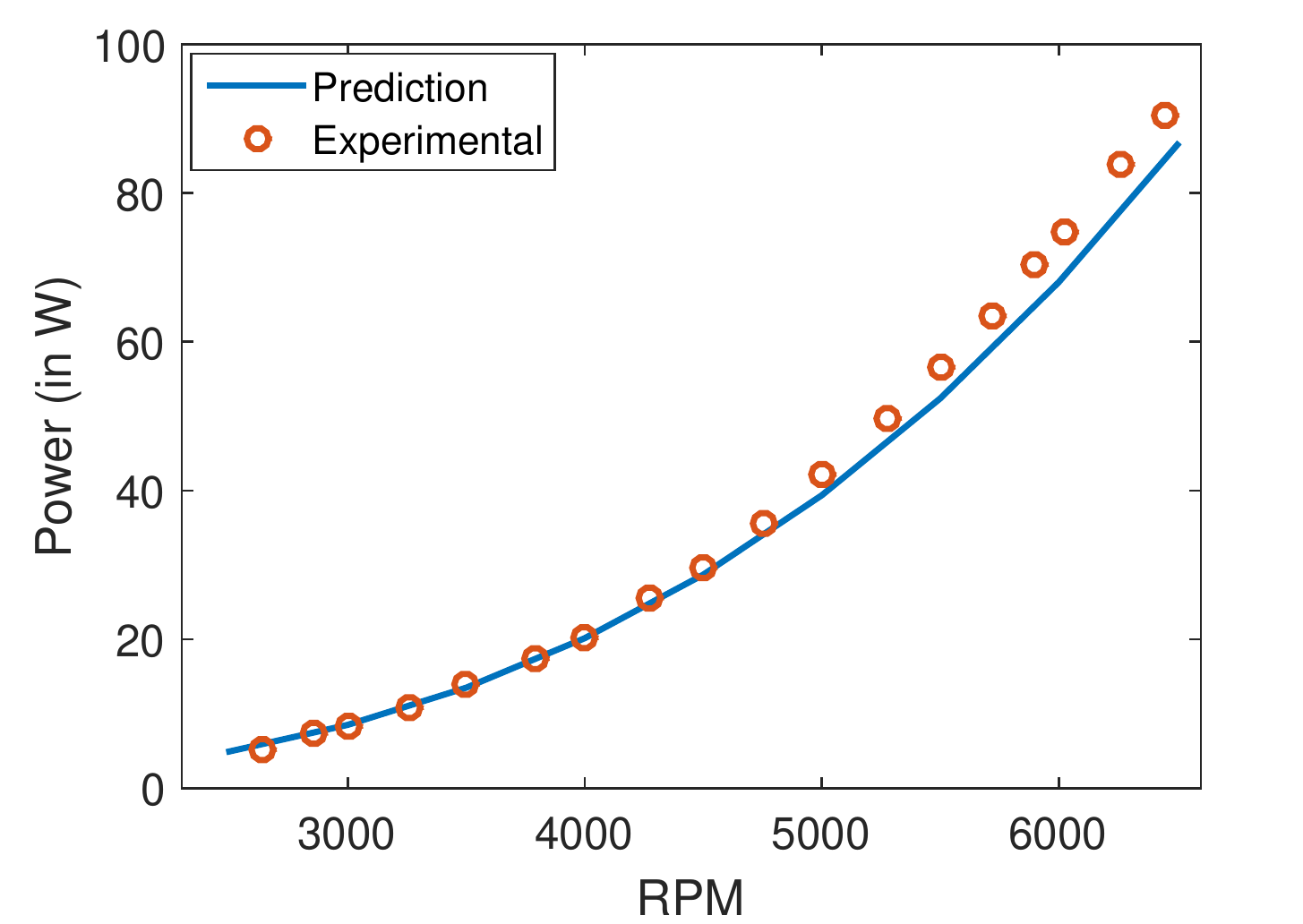}}	
	\caption{Experimental Validation of present analysis for a twisted fixed pitch propeller}
	\label{fig:Exp_Valid_prop}
\end{figure}

Aerodynamic performance characteristics of a rectangular blade (radius = 0.3325 m, chord = 0.32 cm, max t/c=12\%) with symmetric airfoil (NACA0012) is measured experimentally and also predicted using BEMT. The experimental and predicted values of thrust and power, using BEMT, for different blade pitch angles for two different rotor RPMs are plotted in Fig. \ref{fig:Exp_Valid_rot}. The predictions from the current analysis show satisfactory correlation with the experimental results. The modified BEMT analysis is also validated for a fixed pitch propeller tested by Shastry~\cite{shastry}. The pitch and chord variation along the radius for the propeller (with radius of 14 cm)  are approximated by equation \ref{eqn:pitch_chord}. 
\begin{equation} \label{eqn:pitch_chord}
\begin{aligned}
\theta(y) &= 0.0009159 y^5 - 0.04202 y^4 + 0.742 y^3 - 6.151 y^2 + 21.24 y + 0.1216\\
c(y) &= -0.001067*y^3 - 0.02944*y^2 + 0.6259*y + 0.5392
\end{aligned}
\end{equation} 
The experimental thrust and power data for the propeller at different rotational speeds is taken from~\cite{shastry}.  As shown in Fig. \ref{fig:Exp_Valid_prop}, the thrust and power predicted from the current modified BEMT analysis show excellent correlation with the experimental results. Therefore, the modified BEMT analysis developed can be used for the preliminary design of the proprotor with confidence.

\section{Rotor design}
The design of proprotor is carried out with the objective of extracting optimal performance in both hover and forward flight modes. In hover, the proprotors provide thrust to support the weight of the vehicle (plus any airframe download) and for this, generous blade area and higher tip speed is desirable. In cruise / forward flight mode, the thrust generated by the rotors has to overcome the aircraft drag only which is a fraction of the vehicle weight. To ensure good propulsive efficiency, the profile losses need to be reduced, which hints at the use of smaller blade area and lower tip speed in cruise mode. This is in contradiction with the requirements for hover. In this section, the influence of various rotor design parameters on hover efficiency and propulsive efficiency is assessed. In the rest of this paper, hover efficiency is assessed either by power loading which is given as $PL=\frac{\text{Thrust}}{\text{Power}}$ or by figure of merit $FM=\frac{\text{ideal power}}{\text{actual power}}$ and propulsive efficiency of the rotor is compared by using propeller efficiency defined as $\eta_p = {\frac{C_T \mu}{C_P}}$. Typically the payload ratio of rotary wing vehicles is around 3-4. For a payload of 6 kg, the assumption of useful load fraction of 0.3 results in approximate gross take-off weight of 20 kg. This requires that each proprotor should produce at least 50 N thrust.

\begin{figure}[h]
	\centering
	\includegraphics[scale=0.65]{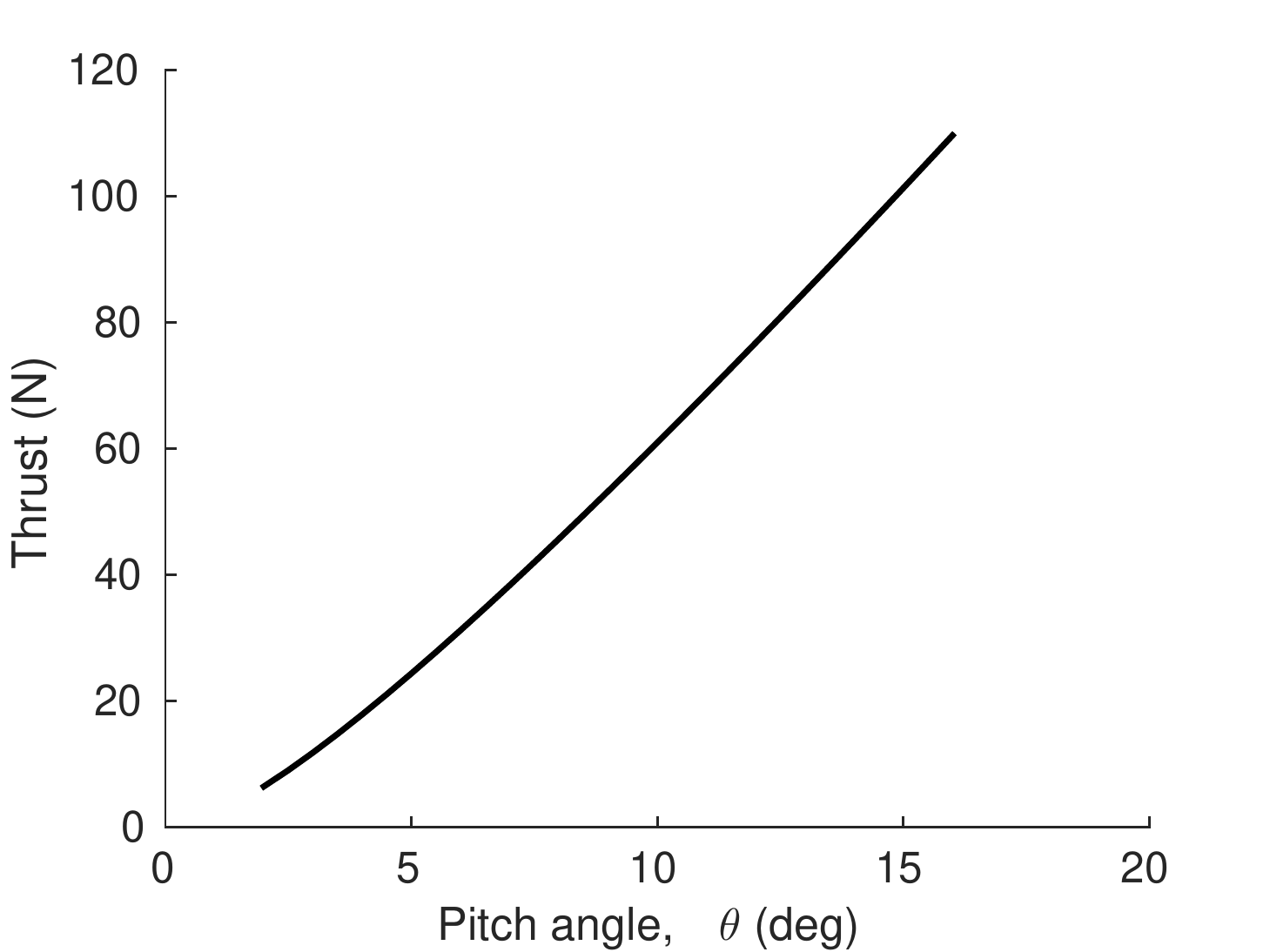}
	\caption{Predicted thrust vs. blade pitch angle for baseline rotor $ \{ \theta_{tw}=0^{\circ}, \theta_{pre}=0^{\circ}, R=.42 m, RPM=3200, \AR=10 \} $}
	\label{thrust_pred}
\end{figure}

\subsection{Baseline configuration selection}
The baseline rotor configuration is chosen to initiate the design process. A disc loading of 90 $N/m^2$ (1.88 lb/ft$^2$) is chosen, which has been reported to be reasonable for small UAVs~\cite{vikram}. This results in a radius of approximately 0.42 m for the desired thrust of 50 N. The tip speed is chosen to be 140 m/s which corresponds to Mach number of 0.412 and rotor RPM of 3200. The variation of predicted thrust, using the BEMT analysis validated earlier, for a two bladed rotor with rectangular untwisted blade (selected as baseline) with radius 0.42 m and aspect ratio 10 (chord = 0.042 m) is shown in Fig.~\ref{thrust_pred}. The blade airfoil is assumed to be symmetric. It can be observed that the baseline rotor selected is capable of generating 50 N thrust at a pitch angle of 8.5$^\circ$ justifying the choices made. The number of blades is kept at two which is known to be reasonable for small rotors with low disc loading. Typical to prop-rotor design, the rotor blades are given a preset angle at the point of its attachment to the hub. This is done to allow for high negative linear twist used in most prop-rotors to prevent the blades from stall under forward flight condition. To begin with the preset angle ($\theta_{pre}$) is set to $0^\circ$  for the untwisted baseline blades. Therefore, for a given collective pitch input of $\theta_0$, the blade pitch angle at any spanwise location ($r$) is given as: $\theta = \theta_0 + \theta_{pre} + \theta_{tw} r$. It is common practise to use the blade preset angle equal to the magnitude of the negative linear twist angle of the blade. This ensures that the blade pitch angle applied directly defines the pitch angle seen by the blade tip.

\subsection{Airfoil selection}
Airfoil with high $C_l/C_d$, high $C_{l\alpha}$, low $C_{m\alpha}$, and high stall angle are desirable for the proprotor, as such an airfoil would minimize the power requirement of the proprotor. There are several airfoils designed for propellers such as Onera Series HOR 07, 12, 20; EPPLER series E850 to E858; HS1 family HS1404, 30, HS1606, 20 etc.~\cite{UIUC} which give satisfactory propulsive efficiency for the propellers. Similarly, airfoils designed for helicopter blade such as EPPLER 360, 361; NPL 9615, 26, 27, 60; Onera series OA206, 09, 12, 13; NASA RC 08,08(N)1, 10-64C, 10(N)1; Sikorsky SC 1094r8, 1095, 1095r8; Boeing-Vertol series VR12 to VR15 etc.~\cite{UIUC} are known to perform well for hovering and edgewise flight of helicopter rotors. For proprotors, airfoils optimized both for hovering as well as high forward speed are needed. Generally, proprotors have multiple airfoils along its span. Because of the inherent high twist that is present in proprotors (reason for which will be discussed later), the root sections operate at very high pitch angles causing the root sections to stall. Therefore, use of thicker airfoil sections with higher stall angles near the proprotor root would maximize hover efficiency. Near the tip, significantly thinner airfoil sections are preferred to reduce the profile power loss, while maximizing figure of merit would require relatively thicker airfoil. The authors in \cite{Stahlhut} have reported that for medium-lift tiltrotor, optimizing airfoil section for forward flight has adverse effects on the hover performance. However, optimizing airfoil sections for weighted compromise between hover and forward flight results in only marginal gain in comparison to the results obtained for purely hover based optimization. Compromised optimization becomes a concern only at high altitudes where higher $C_{l_{max}}$ values are needed to prevent stall near the blade tip. The kind of UAV which is being designed here is not meant to fly at very high altitudes. Therefore, airfoil design and optimization for maximum hover efficiency can give better overall performance without really going into weighted compromised optimization. At the preliminary design stage, which is the focus of the current paper, detailed airfoil optimization is not considered. Instead, a popular helicopter airfoil SC1095, the experimental data for which are available in public domain~\cite{davis,bousman}, is used for proprotor design and analysis.

\subsection{Aspect ratio}
\begin{figure}
	\centering
	\begin{minipage}{0.49\linewidth}
	\includegraphics[scale=0.5]{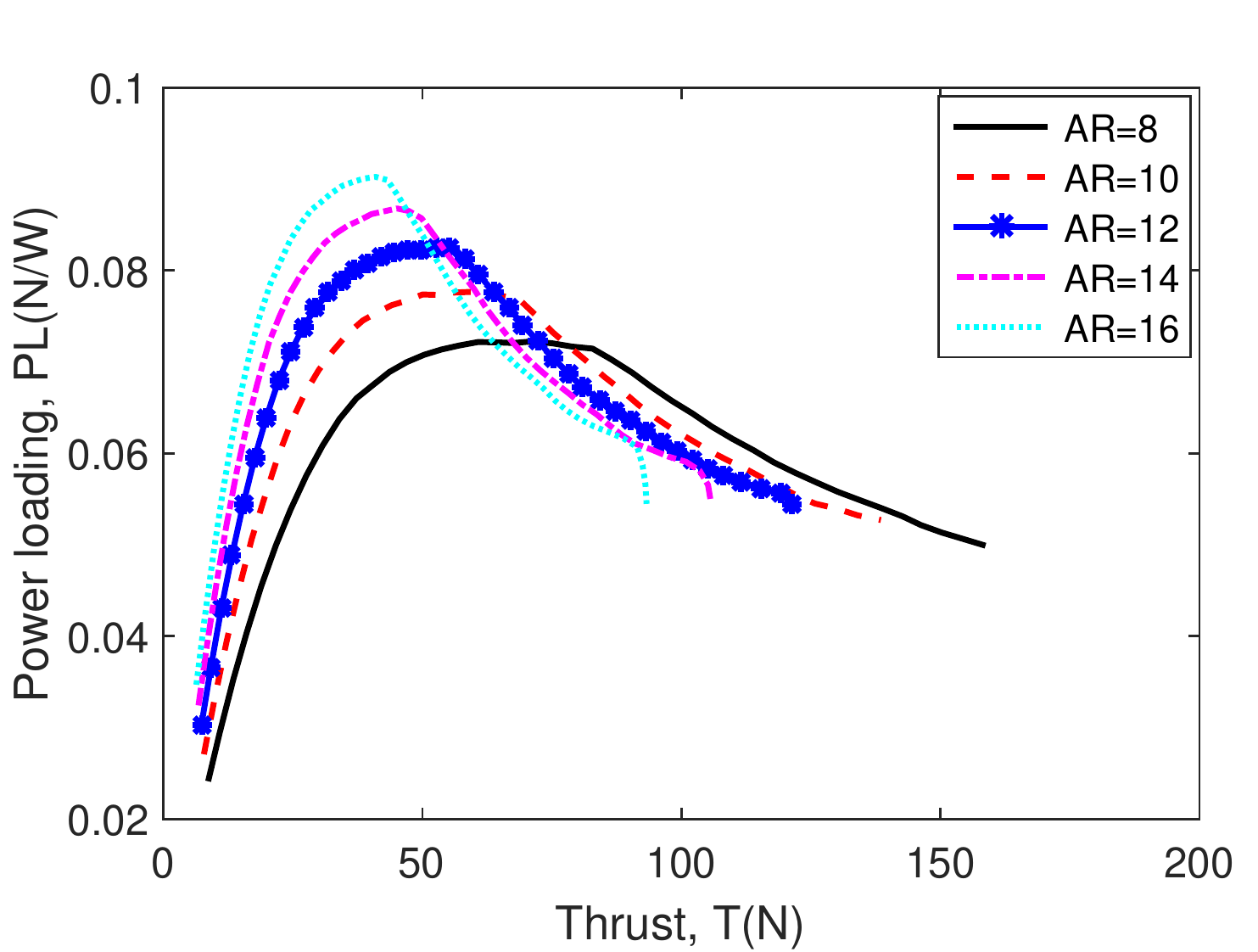}
	\caption{Effect of aspect ratio ($\AR$) on power loading $ \{ \theta_{tw}=0^{\circ}, \theta_{pre}=0^{\circ}, R=.42 m$, $RPM=3200$, no taper \} }
	\label{fig:Effect_AR}
    \end{minipage}	
   \begin{minipage}{0.49\linewidth}
	\includegraphics[scale=0.5]{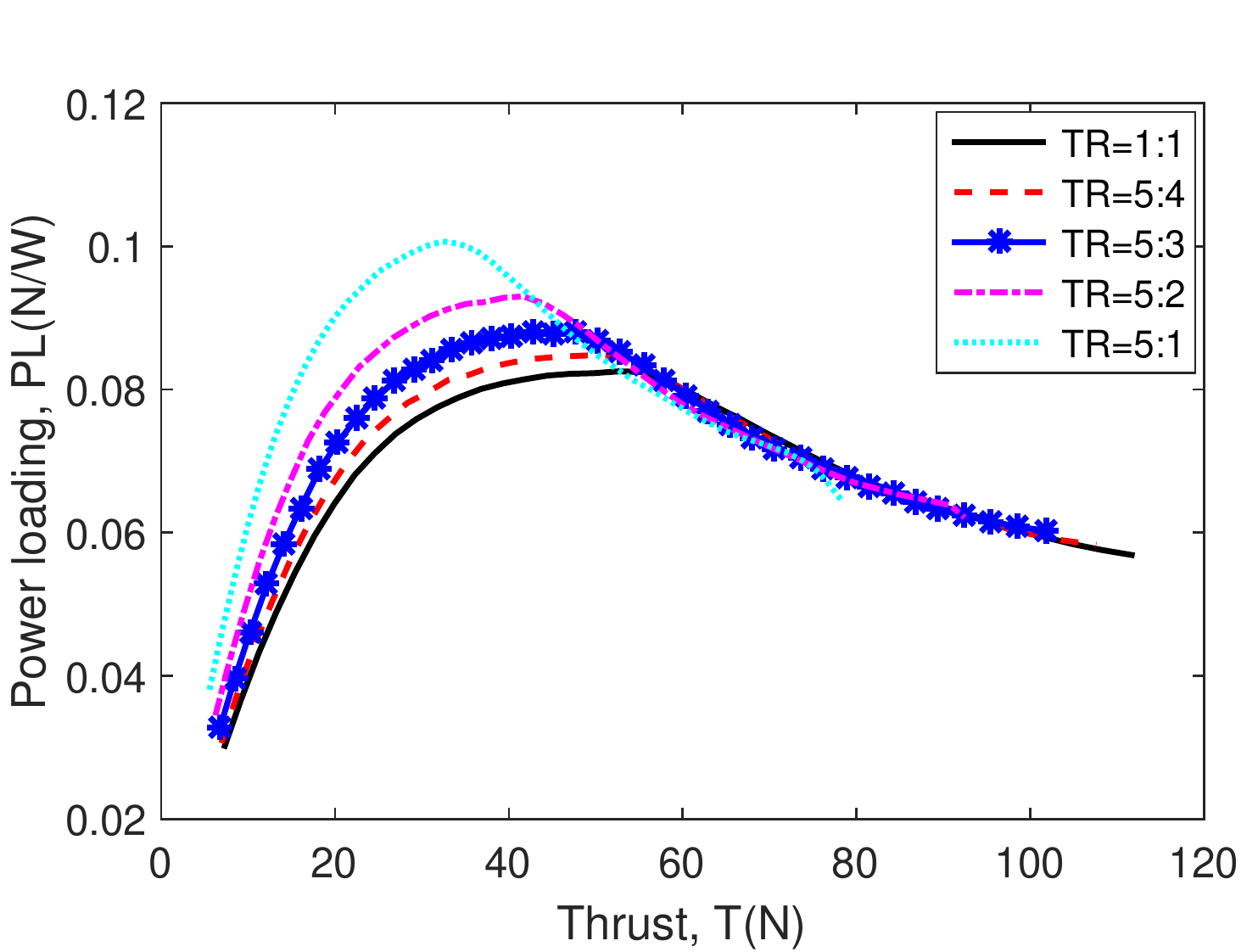}
	\caption{Effect of taper ratio (TR) on power loading  $\{ \theta_{tw}=0^{\circ}, \theta_{pre}=0^{\circ}, R=.42 m, RPM=3200, \AR=12 \}$}
	\label{fig:Effect_TR}
   \end{minipage}
	
\end{figure}

The effect of changing the aspect ratio of the blade is studied by changing the chord of the blade while keeping the radius constant. Using the modified BEMT analysis, the effect of varying aspect ratio (i.e. rotor solidity for constant radius) during hover is studied as shown in Fig.~\ref{fig:Effect_AR}. As expected, an increase in aspect ratio $(\AR)$ results in increase in power loading $(PL=\text{Thrust/Power})$ for hover condition. However, reduction of chord length reduces blade area, thereby decreasing the maximum thrust generated by the rotor. For the desired thrust requirement of more than 50 N, $\AR$ of 12 appears to be an acceptable compromise between highest power loading and maximum thrust output. Therefore, aspect ratio of 12 is chosen for the current blade design.

\subsection{Blade taper}
It is known that use of some planform taper for the rotor blade has beneficial effect on hovering rotor performance. This is due to the fact that taper enables the rotor to operate close to maximum $C_l/C_d$~\cite{Leishman}. Therefore, linear taper is incorporated in the BEMT analysis to study its effect on performance of the rotor by using taper ratio, which is defined as $TR=\frac{\text{root chord}}{\text{tip chord}}$. Effect of varying the blade taper ratio, while keeping radius and other parameters same, is shown in Fig.~\ref{fig:Effect_TR}. As observed, the power loading $(PL)$ required for hovering flight increases with increasing taper ratio in the range of 20 to 60 N thrust output. However, the maximum thrust output also decreases with increase in taper, due to decrease in the blade area. During the hovering flight condition, the linear taper of 5:3 seems to be a good compromise for generating high thrust (100 N) and moderately high power loading. Hence, the linear taper of 5:3 is chosen for the current design. 


\subsection{Blade twist}
\begin{figure}[h]
	\centering
	\subfigure[Power loading vs. thrust $\{\theta_{pre}=-\theta_{tw} , R=.42 m, RPM=3200, \AR=12, TR=5:3 \}$]{\label{fig:plvsth_twist}
          \includegraphics[width=0.47\columnwidth]{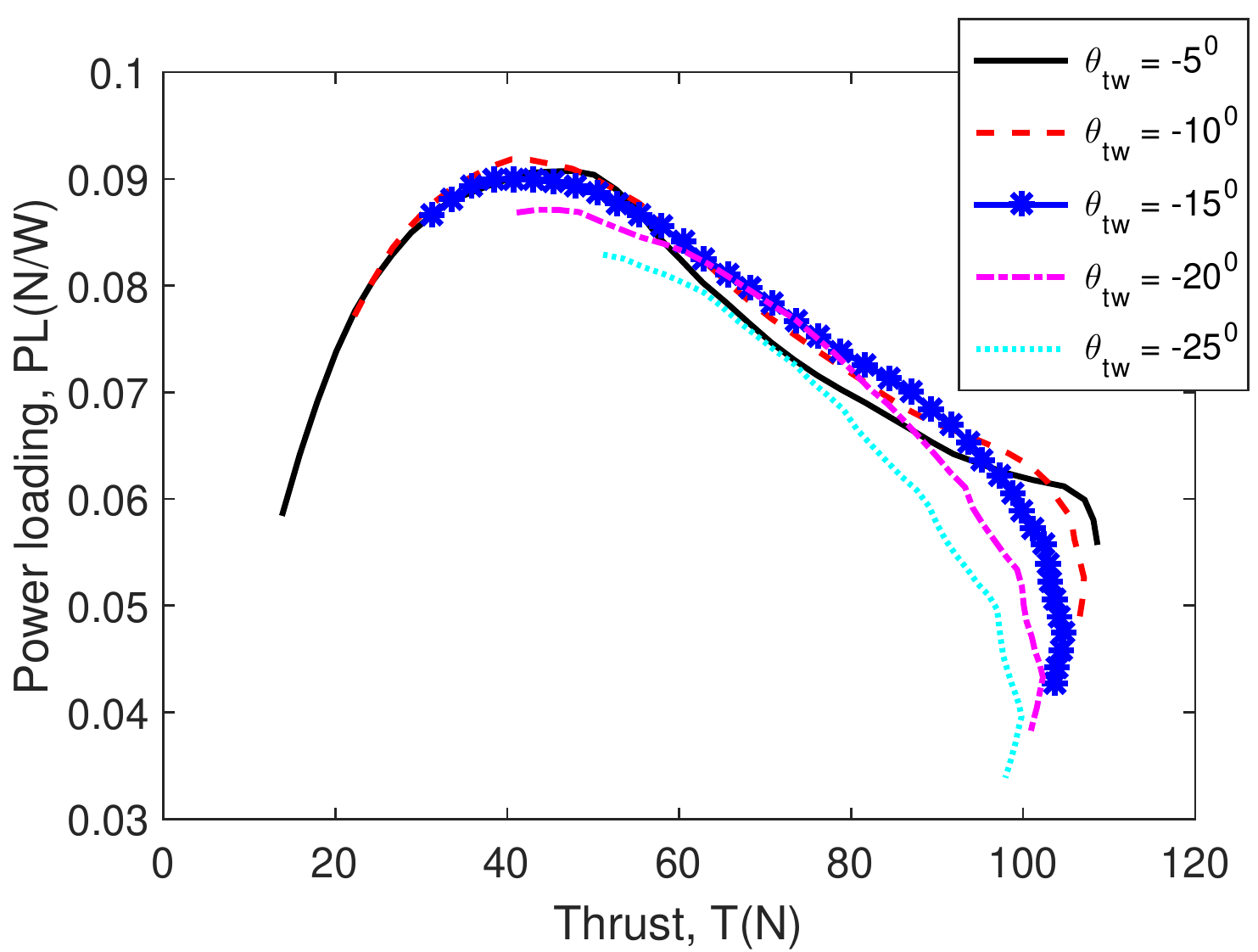}}
	\subfigure[Propeller efficiency vs. forward speed $\{ \theta_{pre}=-\theta_{tw}, R=.42 m, RPM=3200, \AR=12, TR=5:3, \theta_0=5^{\circ} \}$]{\label{fig:prop_eff_twist}
          \includegraphics[width=0.47\columnwidth]{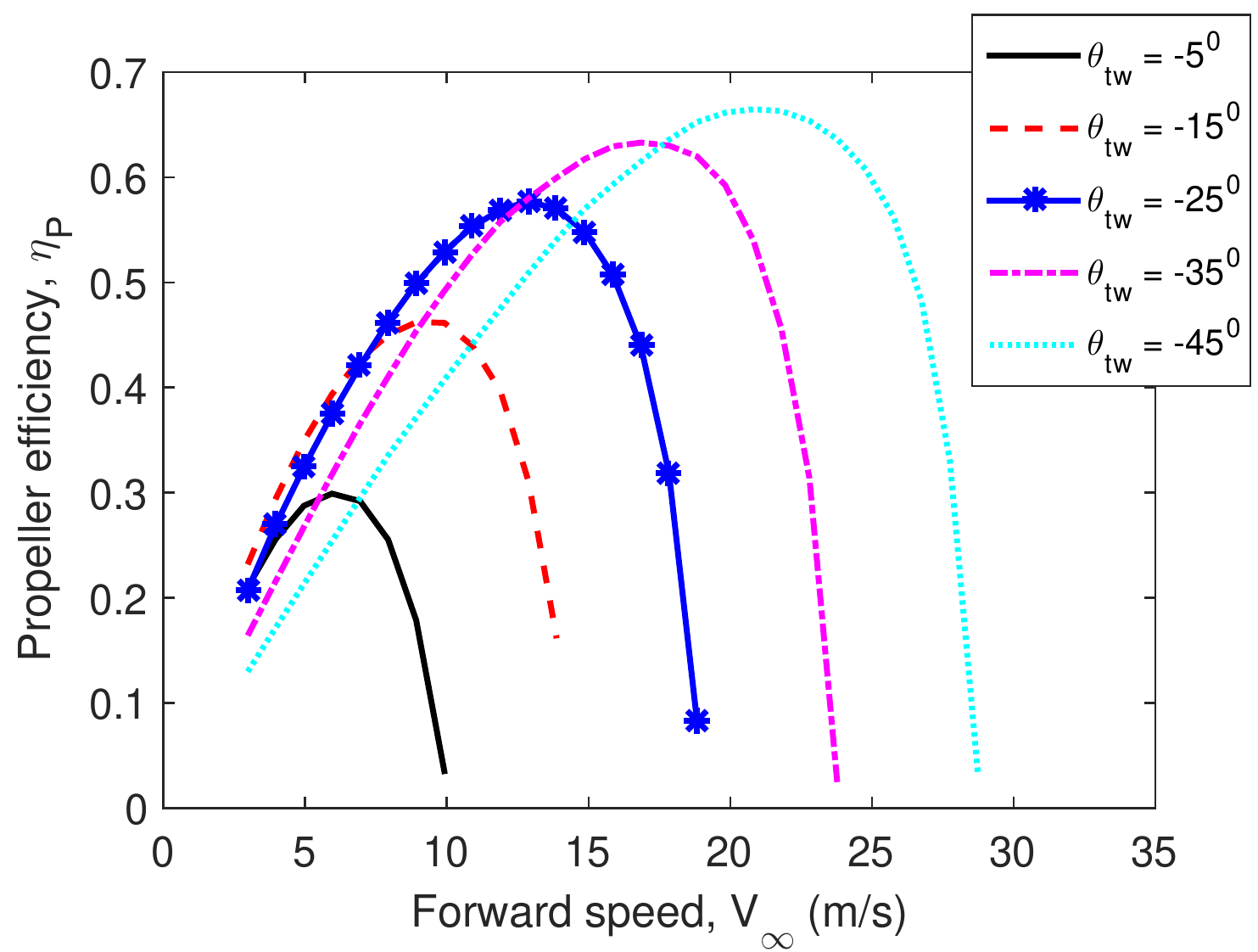}}	\\
	\subfigure[Propeller Efficiency vs. forward speed $\{ \theta_{pre}=-\theta_{tw}, R=.42 m, RPM=3200, \AR=12, TR=5:3, \theta_0=10^{\circ} \}$]{\label{fig:prop_eff_twist1}
          \includegraphics[width=0.47\columnwidth]{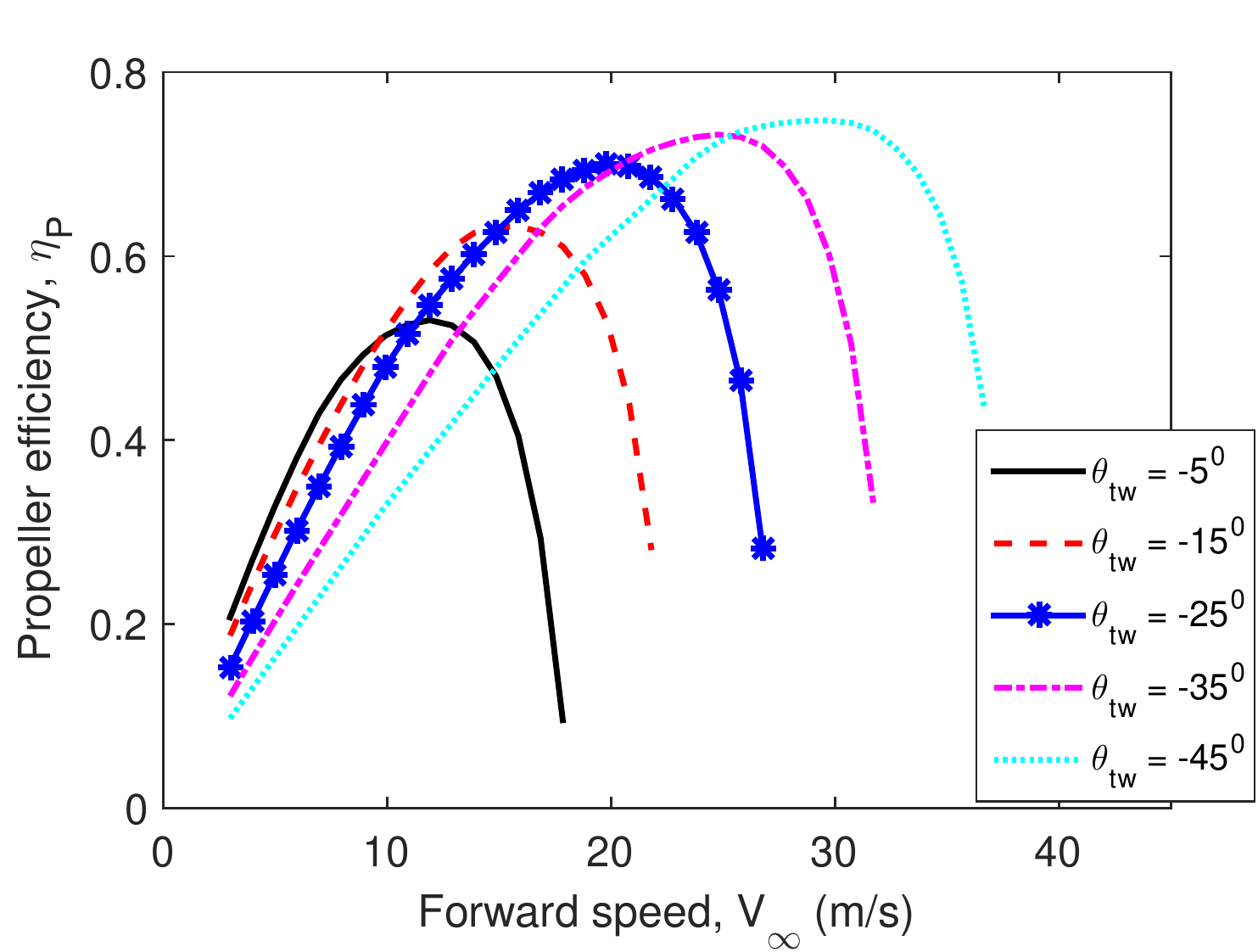}}
	\caption{Effect of twist on hover and forward flight performance of the prop-rotor}
	\label{fig:Effect_Twist}
\end{figure}

The rotor blade twist is a critical parameter for improving the proprotor performance during hover as well as forward flight condition.  The effect of varying twist on hover and forward fight performance is studied carefully to bracket the range of values suitable for hover and forward flight. The final value of twist is determined by optimizing the performance within the bracketed range. As discussed earlier, the preset angle at the blade root is set to be equal to the absolute value of the negative rate of twist to ensure that the tip of the blade has same pitch angle for all cases. The variation of predicted power loading with thrust for different values of blade twist is shown in Fig.~\ref{fig:plvsth_twist}.  The hover performance, as measured by power loading for the desired thrust of more than 50 N, starts to degrade beyond the twist angle of $-15^{\circ}$. Therefore, twist rate greater than $-15^{\circ}$ is less desirable for achieving good hover efficiency. This is due to the fact that with higher twist the blade has to operate at higher pitch input to generate same value of thrust which causes the inboard region to stall early, which increases the profile power resulting in a decrease in power loading.

On the other hand, during forward flight, the proprotor experiences very high inflow angles near the root and the inflow angle decreases along the span. For this reason, a blade with high value of negative blade twist is preferred so that almost every section of the blade operates at optimum lift coefficient. This results in redistribution of the lift and reduces the induced power. The variation of the efficiency of the proprotor is shown in Fig. \ref{fig:prop_eff_twist}. The  optimum propeller efficiency and corresponding speed increase with increase in twist implying that the vehicle can fly faster and with higher efficiency with highly twisted blades. This observation is confirmed for two different blade pitch inputs of $5^{\circ}$ and $10^{\circ}$ in Figs.~\ref{fig:prop_eff_twist} and~\ref{fig:prop_eff_twist1}.  It is further observed that at a given twist, an increase in blade pitch input also increases the maximum efficiency and the corresponding speed.  Having higher twist also allows optimal performance for wider range of speeds around the desired cruise speed and hence higher twist angle is desirable during forward flight. In summary, a twist between $-10^\circ$ to $-15^\circ$ is desirable for best hover performance while twist between $-35^\circ$ to $-45^\circ$ is desirable for higher propeller efficiency. Because of the conflicting requirements, the blade twist is chosen as a compromise between hover and forward efficiency by manual optimization (discussed later) in which more weightage is given to efficiency in forward flight because the vehicle would be spending a significant amount of its mission time in forward flight mode.

\subsection{Rotational speed (RPM)}
\begin{figure}
	\centering
	\includegraphics[scale=.5]{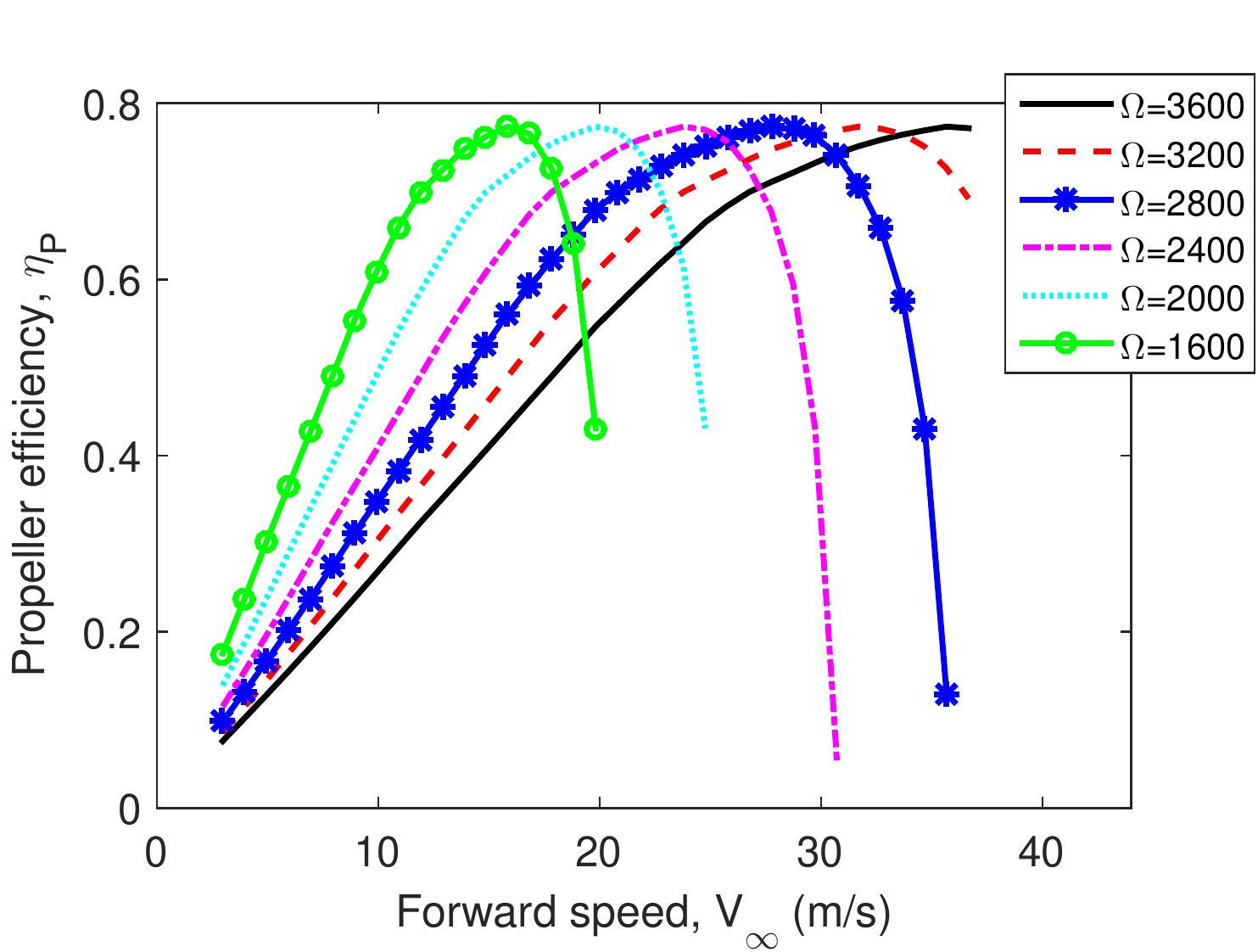}
	\caption{Effect of RPM on the propeller efficiency $ \{\theta_{tw}=-30^{\circ}, \theta_{pre}=30^{\circ}, R=.42 m, TR=5:3, \AR=12, \theta_0=16^{\circ}\} $}
	\label{fig:Effect_RPM}
\end{figure}

For a given total power available, the rotor torque is inversely proportional to the rotational speed being used. Hence, high rotor speed is desirable as it reduces the weight penalty associated with a heavier transmission system required for higher torque requirements. However, with increase in rotational speed, profile power increases due to compressibility effect and rotor noise also increase.  During forward flight, the thrust required to overcome vehicle drag is less than that required to balance the weight of the vehicle during hovering flight. Therefore, the power required dueing forward flight is significantly less, and it may be advantageous to operate the proprotor at reduced RPM during forward flight. Figure~\ref{fig:Effect_RPM} shows the effect of changing RPM on propeller efficiency during forward flight mode. It was established in the previous section that the effect of twist variation for optimal performance lies between $-10^\circ$ to $-40^\circ$, therefore a nominal value of $-30^\circ$ twist and preset angle of $30^\circ$ is assumed for the comparison. It is observed that, for the forward speeds in the viscinity of cruise speed $(V_{cr})$ of 20 m/s, the proprotor has highest efficiency for 2000 RPM. The propeller efficiency at 3200 RPM is 0.6079 and that at 2000 RPM is 0.7733 which gives approximately 27\% increase in efficiency due to reduction in proprotor RPM.  Therefore, the operating rotor RPM during forward flight mode is set to 2000 to ensure high propeller efficiency. This is possibly due to reduction in profile power with reduced RPM. 

\subsection{Optimization}

In the previous sections, the aspect ratio, taper ratio and the RPM for the two operational modes of the UAV being designed have been selected. It was noted that the blade twist requirements for the hover and forward flight modes are contradictory in nature. In the current section, manual optimization is carried out to maximize a simple performance cost function defined as $\text{cost}=0.3 FM+0.7\eta_p$, for the aircraft by varying the blade twist and rotor radius. The cost function defined uses non-dimensional parameters figure of merit (FM) as an index of hover efficiency and propeller efficiency ($\eta_p$) as an index of forward flight efficiency. Based on the typical mission envisioned for the UAV, 30\% weightage is been given to efficient hovering capability and 70\% weightage is given to forward flight performance. This is based on the assumption that the objective of door-to-door package delivery would require the vehicle to spend significant amount of flight time (70\% to 80\%) in forward flight mode and approximately 20\% to 30\% of time in hovering mode to perform precise delivery of payload. It should be noted that figure of merit has its usual meaning which is defined as the ratio of ideal power for the proprotor in hover obtained from momentum theory and the actual power estimated using the modified BEMT. 

\begin{figure}
	\centering
	\subfigure[Figure of merit (FM)]{\label{fig:fmvar}
	\includegraphics[width=0.47\columnwidth]{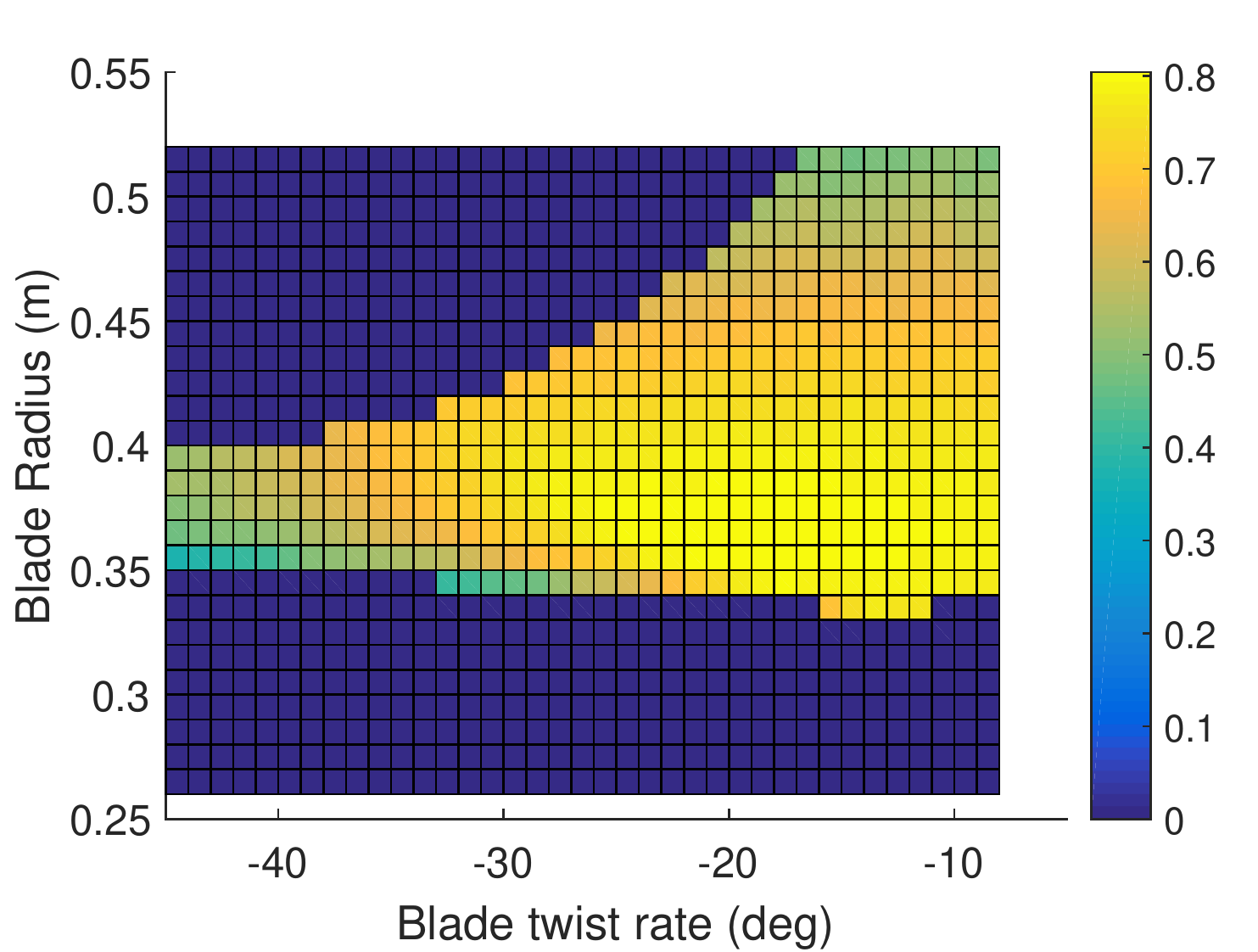}}
	\subfigure[Propeller efficiency ($\eta_{p}$ for cruise velocity)]{\label{fig:propeff}
	\includegraphics[width=0.47\columnwidth]{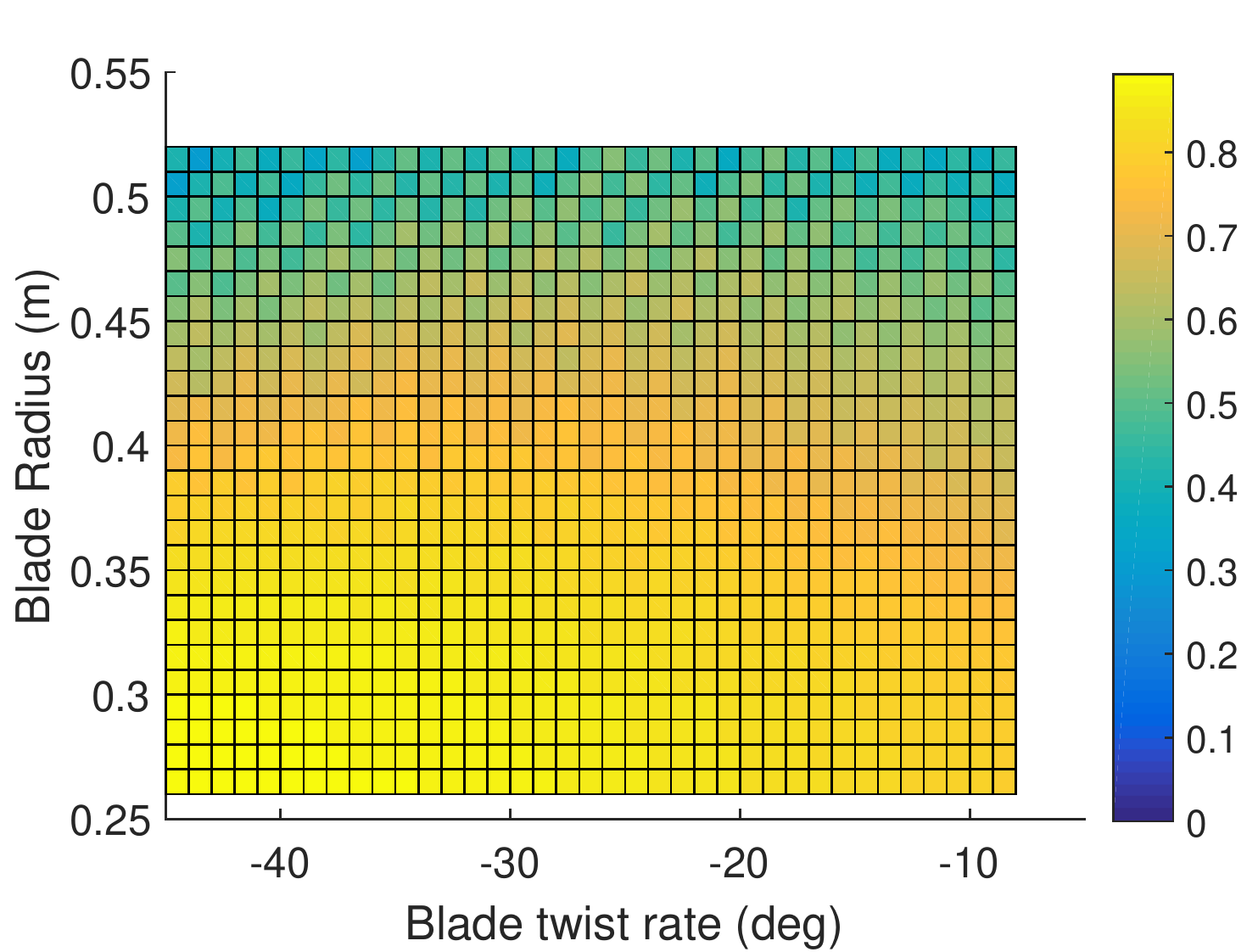}}\\
	\subfigure[Combined performance cost function]{\label{fig:costfunvar}
	\includegraphics[width=0.47\columnwidth]{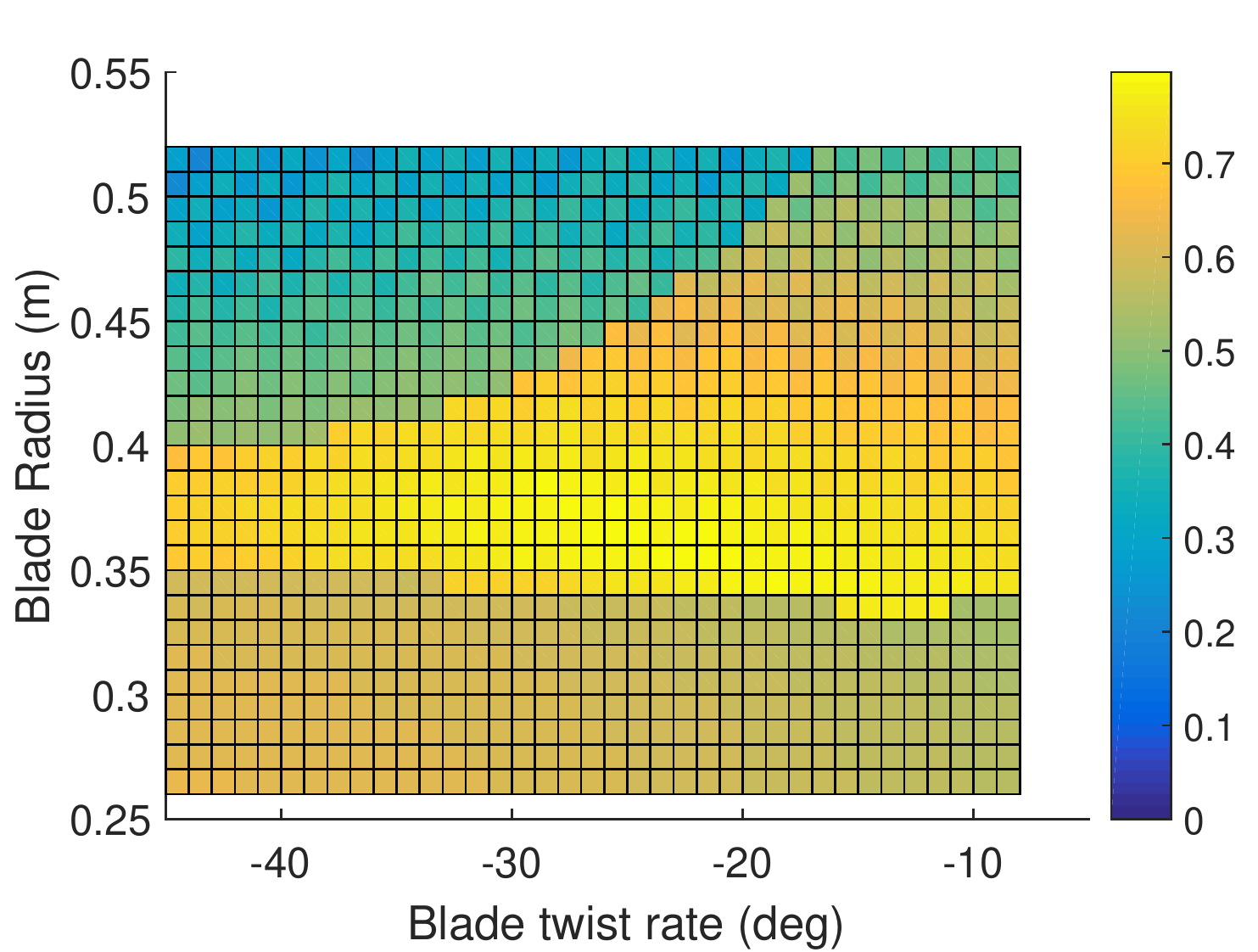}}
	\caption{Manual optimization of UAV performance cost function for radius and twist}
	\label{fig:Optimize}
\end{figure}

To optimize the performance, the blade twist angle is varied from $-45^\circ$ to $-8^\circ$ (with an interval of $1^\circ$) and rotor radius is varied from 0.26 m to 0.53 m (with an interval of 0.01 m) and the figure of merit, propeller efficiency and combined performance cost function are plotted as shown in Fig.~\ref{fig:Optimize}. Figures~\ref{fig:fmvar} and~\ref{fig:propeff} reiterate the contradictory nature of the requirements for optimal hover and forward flight performance. Moderate radius and low twist angle seems to give best hover efficiency, while low rotor radius and high twist are desirable for high propeller efficiency. The combined cost function variation is shown in Fig.~\ref{fig:costfunvar} which is a weighted combination of Figs.~\ref{fig:fmvar} and~\ref{fig:propeff}. Based on these calculations, an optimal radius of 0.38 m and blade twist of $-24^{\circ}$ are chosen which maximizes the cost function for the range of speeds considered. Table \ref{tab:Rotor_param} shows the finalized design parameters for the proprotor obtained by optimizing the defiend cost function.

\begin{table}
	\centering
	\caption{Proprotor design parameters}
	\label{tab:Rotor_param}
		\begin{tabular}{ c @{\hskip .3in}  c}  
		\hline \hline
		\textbf{Parameter} & \textbf{Value}\\
		\hline\hline
		Aspect ratio & 12\\
		Blade taper ratio & 5:3\\
		Blade radius & 0.38 m\\
		Root chord & 0.0475 m\\
		Tip chord & 0.0285 m\\
		Blade twist & $-24^{\circ}$\\
		Blade preset angle & $24^{\circ}$\\
		Rotor RPM (hover) & 3200 \\
		Rotor RPM (forward flight) & 2000 \\
	    \hline
		\hline		
		\end{tabular}
\end{table}

\section{Wing design}
As discussed in section II, the proposed vehicle has two wings in biplane configuration. In this section, the detailed design of the wing is discussed along with the justification for selecting the biplane configuration over conventional monoplane design.

\subsection{Wing configuration selection}
\begin{figure}[h]
	\centering
	\includegraphics[scale=0.65]{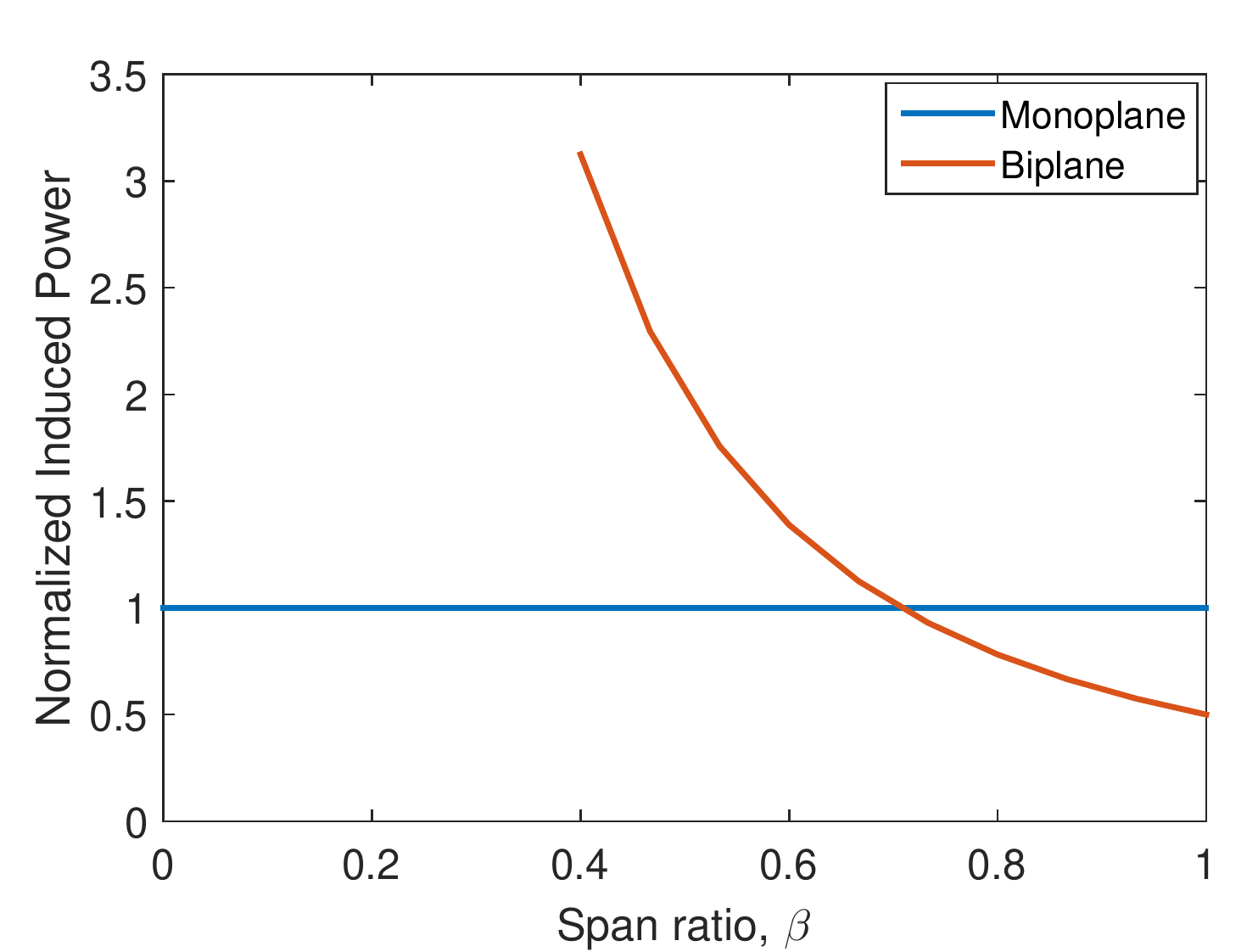}
	\caption{Comparison of power for monoplane and biplane with same total wing area}
	\label{fig:Bi_Mono}
\end{figure}

The mission profile of the VTOL UAV being designed requires it to deliver packages in urban canyons with cluttered space. Therefore, the compactness is a desirable feature in the vehicle design. It also ensures easy stowage in warehouses. It is easy to notice that for a given chord, monoplane configuration needs larger span when compared to the biplane design for same wing area. This can be assessed using a simplified analysis based on classical lifting line theory with elliptic lift distribution~\cite{Anderson}. Consider a monoplane and a biplane, having rectangular planform, with same total wing area and span ratio, $\beta$, defined as  $\beta=\frac{\textup{span of biplane}}{\textup{span of monoplane}}$. For a given $\beta$ and same total wing area, the span and chord of monoplane ($b_{w_m},\; c_{w_m}$) and biplane ($b_{w_b},\; c_{w_b}$) are related as follows
\begin{equation}
\begin{aligned}
b_{w_b}=\beta b_{w_m}\\
c_{w_b}= \frac{c_{w_m}}{2 \beta}
\end{aligned}
\end{equation}
Induced power for monoplane and biplane is calculated using the following equation,
\begin{equation}
P_{ind}=\frac{1}{2}\rho V_{cr}^3 S_{w} K  C_{L_{cr}}^2
\end{equation} 
where $K=\frac{1}{\pi (\AR_w) e}$ and $\AR_w=\frac{b_w}{c_w}$. For simplicity, it is assumed that wings operate independent of each other in biplane configuration. Figure~\ref{fig:Bi_Mono} shows normalized induced power variation for different span ratios defined above. Normalized induced power is the ratio of induced power for biplane to that for monoplane. From this simple analysis, it is observed that in biplane configuration the span can be reduced by up to 30\% of the span in monoplane design with the added benefit of requiring less induced power compared to monoplane case. It is known that biplane configuration makes the vehicle structurally more robust and agile~\cite{Raymer}. In the current design, the biplane wing also serves the purpose of landing gear. Because of all these benefits, biplane configuration is chosen. For the preliminary design, it is assumed that both the wings behave independent of each other and each wing contributes towards supporting the half of the vehicle weight. Although in practice because of interference between the two wings, the reduction in induced power, in case of biplane, is less than predicted theoretically. For the same span and same area, the biplane should require 50\% less induced power than monoplane. But, due to mutual interference between the wings the reduction in power is around 30\% instead of 50\%~\cite{Raymer}. As shown in~\cite{Bogdanowicz}, biplane wings with separation greater than 1.5 times the chord generates approximately 90\% of the lift produced by the monoplane with the same total wing area. Based on this, in the current design approach, wing parameters are initially chosen corresponding to monoplane configuration and the span ratio of 0.8 (see Fig.~\ref{fig:Bi_Mono}) is used to obtain the corresponding span for biplane design to match the induced power of the monoplane configuration. Once, the span is decided, other wing parameters for biplane configuration are obtained to meet the requirement for wing lift. 

\subsection{Wing loading}
\begin{figure}[h]
	\centering
	\includegraphics[scale=0.65]{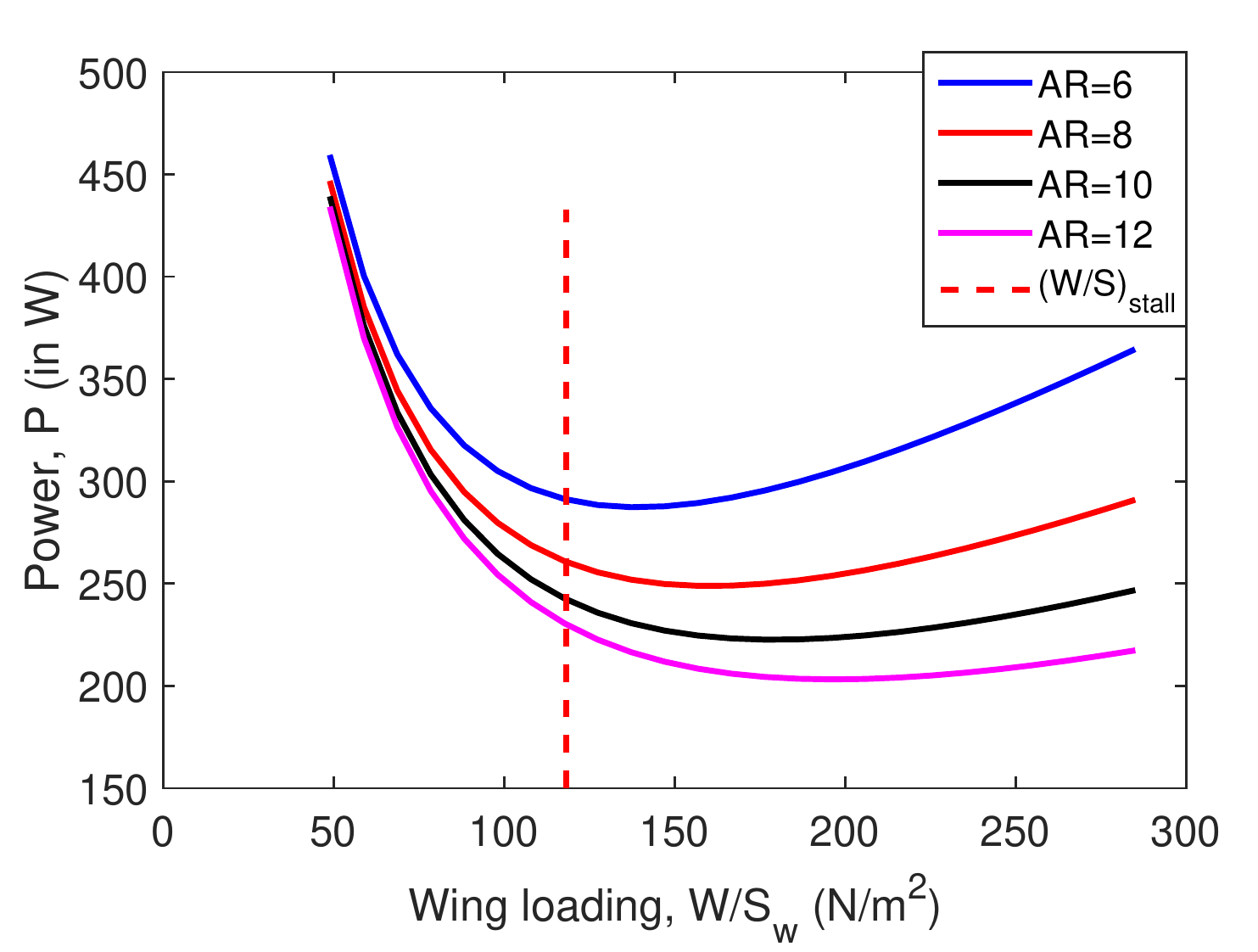}
	\caption{Power variation with wing loading}
	\label{fig:Power_Wing_load}
\end{figure}

Wing loading $\left ( \frac{W}{S_w} \right )$, defined as the ratio of weight of the vehicle divided by the total wing area, affects stall speed, cruise performance and gust tolerance of the vehicle. Since the proposed vehicle can takeoff and land vertically, the constraints on the wing loading enforced by takeoff and landing distances are not relevant. To choose wing loading, power variation with wing loading is plotted for different aspect ratios as shown in Fig.~\ref{fig:Power_Wing_load}. The power is calculated based on the simplified expression shown in Eq.~\ref{eqn:Wing_power}. The calculations are performed based on the following typical values for fixed wing UAVs of this category taken from~\cite{Landolfo}: drag coefficient, $C_{D0} $ of 0.025; stall speed, $V_{stall}$ of 12 m/s; oswald efficiency, $e$ of 0.8; maximum lift coefficient, $C_{L_{max}}$ of 1.5.
\begin{equation}\label{eqn:Wing_power}
P=\frac{1}{2}\rho V_{cr}^3 C_{D_0}\frac{W}{(\frac{W}{S_w})}+2K W \frac{(\frac{W}{S_w})}{\rho V_{cr}}
\end{equation}
\begin{equation}\label{eqn:Wing_loading}
\left (\frac{W}{S_w} \right )_{stall}=\frac{1}{2}\rho V_{stall}^2 C_{L_{max}}
\end{equation}

The critical wing loading corresponding to stall speed, $\left (\frac{W}{S_w} \right )_{stall}$, calculated using Eq.~\ref{eqn:Wing_loading}, is also shown in Fig.~\ref{fig:Power_Wing_load}. This value comes out to be around 122 N/$m^2$. With larger aspect ratios, the optimum wing loading increases and the power requirement reduces. But, this also increases stall speed and the weight because of the extra structural strength required to support the corresponding high aspect ratio wings.  Typically the wing loadings of the UAVs of this class, such as ScanEagle, expedition II, Sky Crossbow etc., range from 100 to 170 $N/m^2$. Considering all these factors, wing loading of 130 $N/m^2$ is chosen.

\subsection{Airfoil selection}
\begin{figure}
	\centering
	\subfigure[Lift coefficient with angle of attack]{\includegraphics[width=0.47\columnwidth]{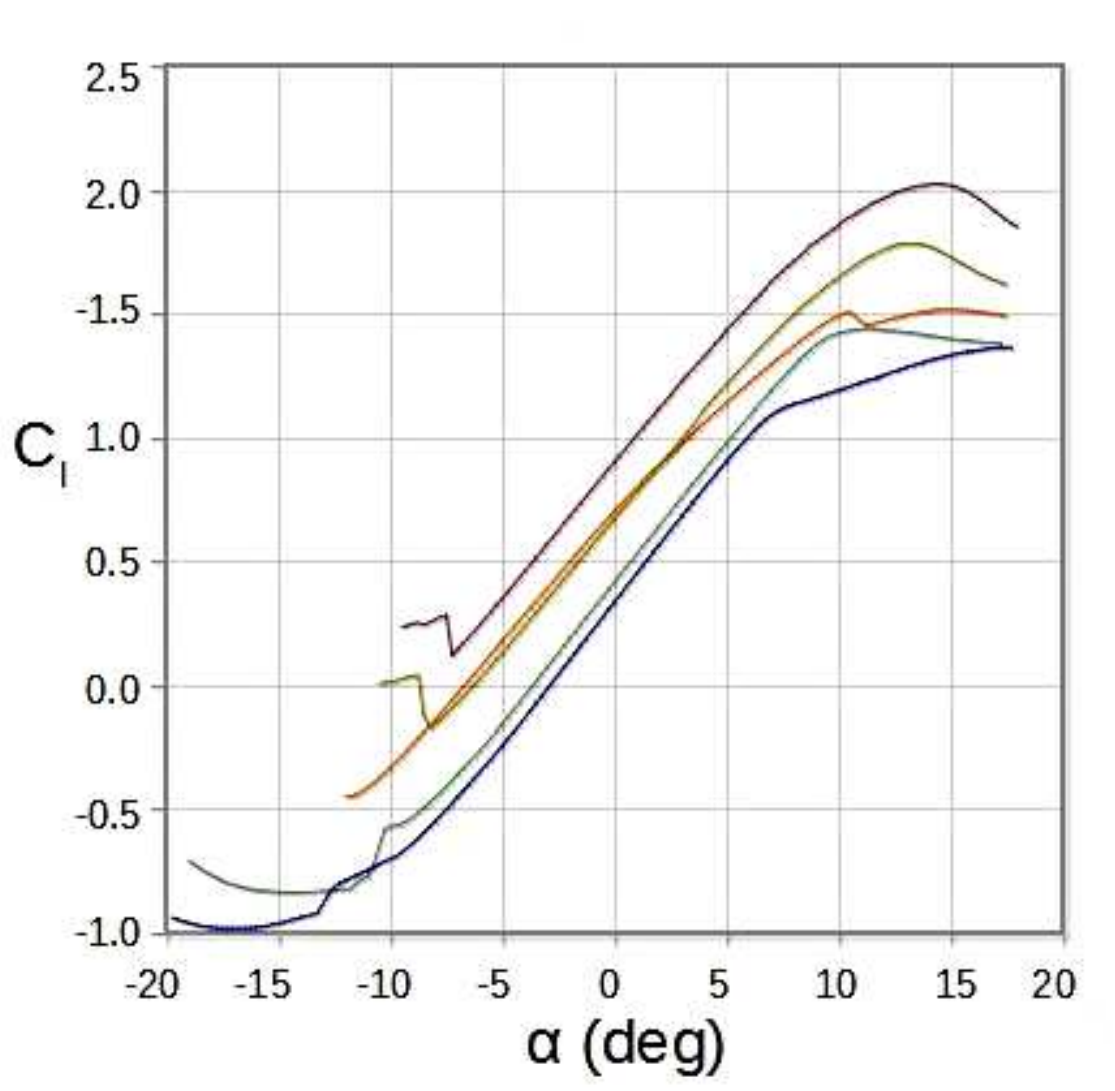}}
	\subfigure[Lift coefficient with drag coefficient]{\includegraphics[width=0.47\columnwidth]{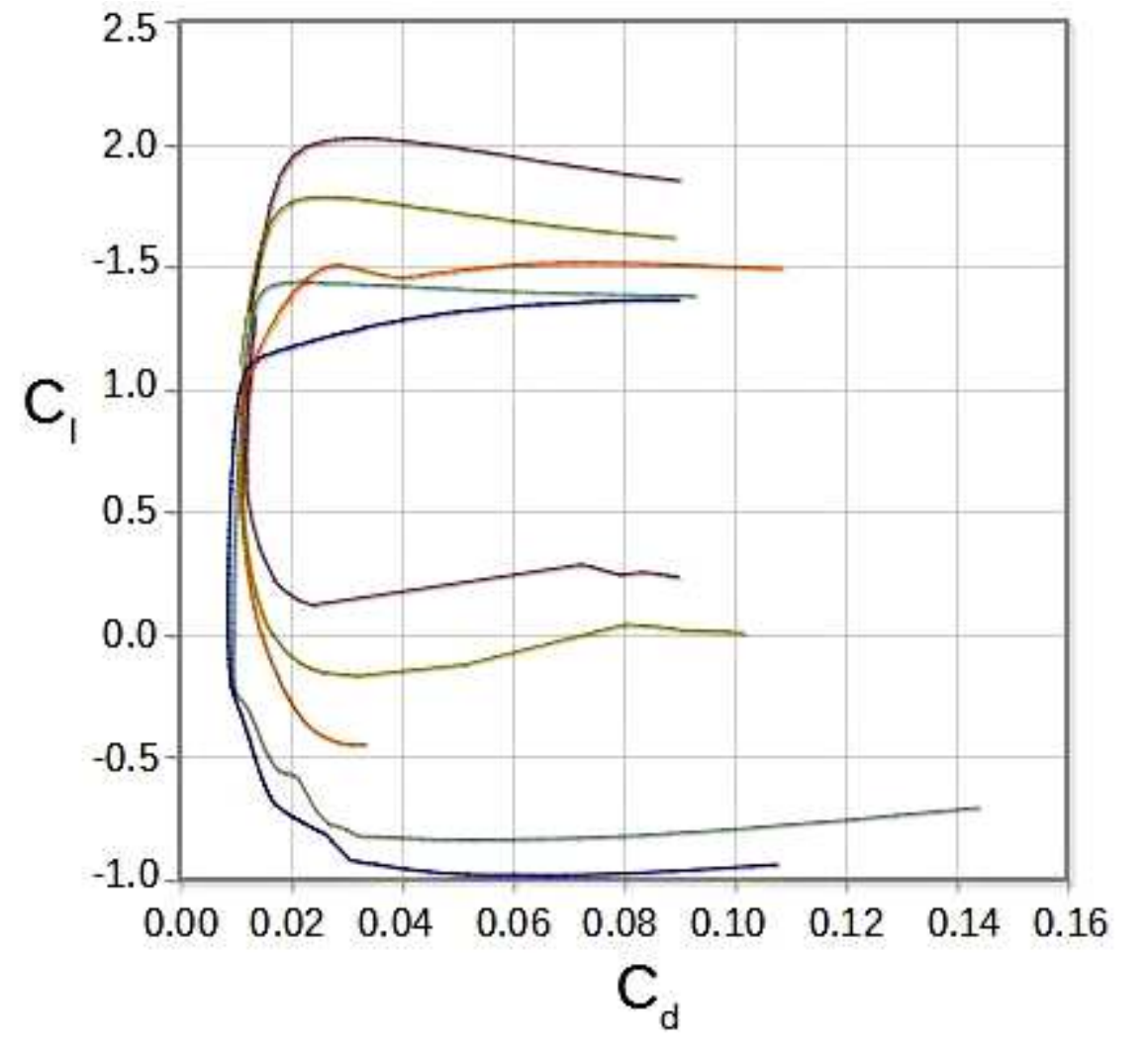}} \\
	\subfigure[Pitching moment coefficient with angle of attack]{\includegraphics[width=0.55\columnwidth]{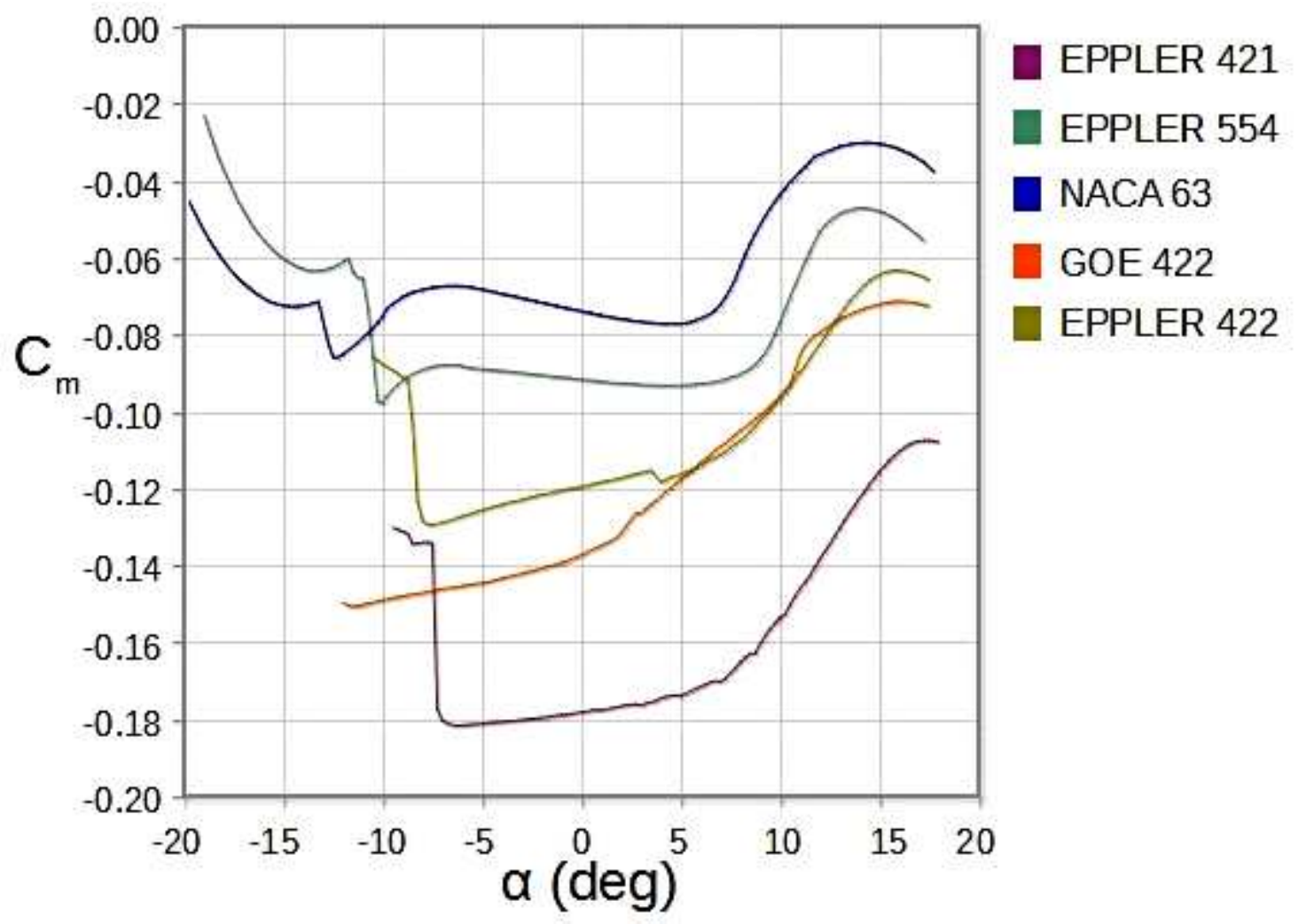}}
	\caption{Comparison of airfoil parameters for different airfoils considered for the wing design}
	\label{fig:airfoil_compare}
\end{figure}

Airfoils with moderate or high camber generate very high lift and typically have high lift to drag ratio. But, high lift comes with high nose down pitching moment \cite{Raymer,Ming}. For the preliminary analysis, 2D airfoil data available from \cite{Airfoil_tools} is used for comparing different airfoils. Figure \ref{fig:airfoil_compare} (a), (b) and (c) show airfoil data($C_l$ vs. $C_d$,  $ C_l$ vs. $\alpha$,  $C_m$ vs. $\alpha$) for the following airfoils commonly used in small UAVs: EPPLER 421, 554, 422; NACA 63; GOE422. EPPLER 421 gives higher lift and lift to drag ratio and also has very high nose-down (negative) pitching moment about quarter chord. The present design is without a tail and the pitching moment from the wing would have to be compensated by use of differential thrust between top and bottom rotor pairs. EPPLER 422 airfoil has second highest lift coefficient among all the airfoils compared and has third lowest nose-down pitching moment and hence is chosen as the primary airfoil for the biplane wings.  

\subsection{Wing planform}
The variation in airfoil characteristics along the span of the wing is largely influenced by wing planform characteristics and so is the performance of the vehicle. An elliptical wing planform has the best lift distribution along the wing but it is difficult to manufacture. A tapered wing has better aerodynamic and structural loads  distribution than a rectangular wing. If properly designed, its performance can approach the lift distribution of elliptical wing.  However, the small chord near the tip results in lower Reynolds number and can cause stall~\cite{Raymer,Ming}. Rectangular planform wings do not suffer from this deficiency but have higher drag and worse loads distribution. In the current design the proprotors are located at the leading edge of the wing, the wing section up to the location of the proprotors is kept rectangular for giving enough space for the transmission system that transfers the power to the rotors. The region of the wing outboard of the rotors is tapered as discussed below. Various aspects of the wing geometry are discussed below.

\subsubsection{Aspect ratio}
A wing with a high aspect ratio has tips farther apart than an equal area wing with a low aspect ratio. Therefore, the amount of the wing affected by the tip vortex is less for a high aspect ratio and hence experience less loss of lift and increase of drag due to tip effects as compare to a low-aspect-ratio wing of equal area \cite{Raymer}. But with large aspect ratio, the span of the wing will also be larger making vehicle bigger in size. Typically aspect ratio of 5-8 is used for small aircrafts of this weight class. Therefore, an aspect ratio of 6 is chosen.
\subsubsection{Taper}
The taper ratio ($\lambda_w$) has significant effect on spanwise lift distribution. The spanwise location of centre of pressure shifts toward the root with decreased taper ratio thereby decreasing the bending moment about the root. Also the chord at the root increases with decreasing taper ratio for the same wing area providing more structural strength. But highly tapered wing is more prone to tip stall because of the thinner airfoil present at the tip due to decreased chord. Taper ratio of 0.45 is chosen because it gives fairly elliptic lift distribution and good stall characteristics \cite{Torenbeek}. 
\subsubsection{Sweep}
Sweep delays the shock waves and accompanying aerodynamic drag rise because of the compressibility effects near the speed of sound, improving performance \cite{Raymer,Torenbeek}. But the vehicle is being designed to fly at low subsonic speeds. Therefore, there is no need for wing sweep.

\subsection{Other considerations for wing design}
Performance of the biplane largely depends on the gap and stagger between the two wings. Because of the quadrotor configuration attached to both the wings, placing wings at different locations along the wing chordwise direction is not possible, suggesting for zero stagger. Increasing the gap between the wings reduces the interference between the two wings, thereby, making them operate independent of each other and improve the overall aerodynamic efficiency \cite{Landolfo, Maqsood, Altman}. As discussed in \cite{Bogdanowicz}, typically a gap of greater than 1.5 times the chord can give almost 90\% of the lift produced by the monoplane with the same total wing area. As discussed earlier, the radius of the rotor blade is 0.38 m so the rotors has to be kept at least 0.76 m apart. For reducing rotor interference, the rotors are kept 1 m apart, thereby constraining the wing separation to be 1 m. Wing gap of 1 m is well above 2.5 times the root chord of the wing, which is 0.39 m, and therefore it is expected that there would be very less interference effect.

Wing dihedral though imparts roll stability but takes away a part of lift and hence has worse aerodynamic performance compared to the wing without dihedral. Winglets significantly reduce drag but only if properly designed. However, their working is still not clearly understood for this scale and a poor design may increase drag instead of reducing it. So neither dihedral nor winglets is considered for the current design. To reduce weight of the wing, it will be made of rohacell foam coated with a thin layer of kevlar composite. Spar made with aluminum alloy will be used to augment the stiffness of the wing. Foam being lighter will help in reducing wing's weight while coating of kevlar composite layer will add stiffness and strength to the wing.

After having decided wing loading, aspect ratio, and taper ratio etc., dimensions of the wing for monoplane configuration are determined. These parameters are scaled, as discussed above, to obtain parameters for a wing in biplane configuration and are shown in Table \ref{tab:Wing_param} 
\begin{table}
	\centering
	\caption{Wing parameters for single wing of the biplane configuration}
	\label{tab:Wing_param}
	\begin{tabular}{c  c}  
		\hline \hline
		\textbf{Parameter} & \textbf{Value}\\
		\hline\hline
		Wing area & 0.754 $m^2$\\
		Wing Span & 2.29 m\\
		Aspect ratio & 6.9\\
		Taper & 0.45 \\
		Root chord & 0.39 m\\
		Tip chord & 0.176 m\\
		Airfoil & EPPLER 422 \\
		\hline
		\hline
		
	\end{tabular}
\end{table}


\section{Power plant}
With the design parameters chosen above, the vehicle require power of approximately  1.93 hp in hover and 0.7 hp during forward flight. Thus, the forward flight mode consumes 64\% less power than the hover mode highlighing the advantage of the proposed configuration.  To account for mechanical losses and aerodynamic losses that the theory might not have incorporated, extra 10\% is added, meaning around 2.13 hp power is needed during hover. To improve endurance, as discussed earlier, Gasoline run 105HZ helicopter engine of OS is selected as the power source of vehicle. Besides, being small in size and weight, 105HZ can provide power required by vehicle to hover and also to execute other heavy maneuvers if needed. The specifications of the engine are given in table \ref{tab:Engine_Specs}.

\begin{table}
	\centering
	\caption{Specifications of 105HZ helicopter engine}
	\label{tab:Engine_Specs}
	\begin{tabular}{ c  c}  
		\hline\hline
		\textbf{Parameter} & \textbf{Value}\\
		Displacement & 1.048 $in^3$ (17.17 $cm^3$) \\
		Bore & 1.442 in (36.63 mm)\\
		Stroke & 1.024 in (26 mm)\\
		Practical rpm & 2000-16,500 \\
		Output & 3.75 hp @ 15000 rpm\\
		Engine weight & 596 gm\\
		\hline \hline
		
	\end{tabular}
\end{table}

\section{Transmission and pitch change mechanism}
\begin{figure}
 	\centering
 	\subfigure[Transmission Mechanism]{\includegraphics[scale=0.6]{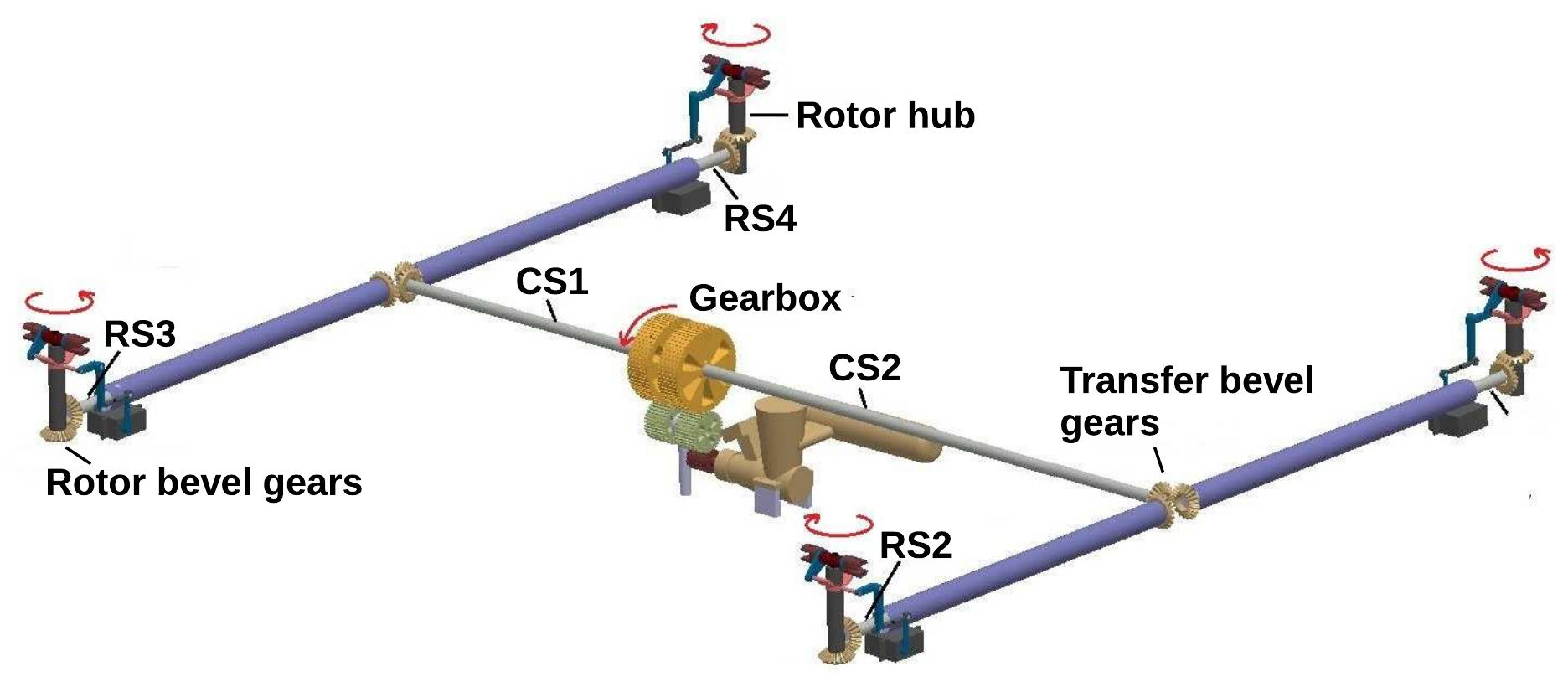}}\\
    \subfigure[Gears in Gearbox]{\includegraphics[scale=.35]{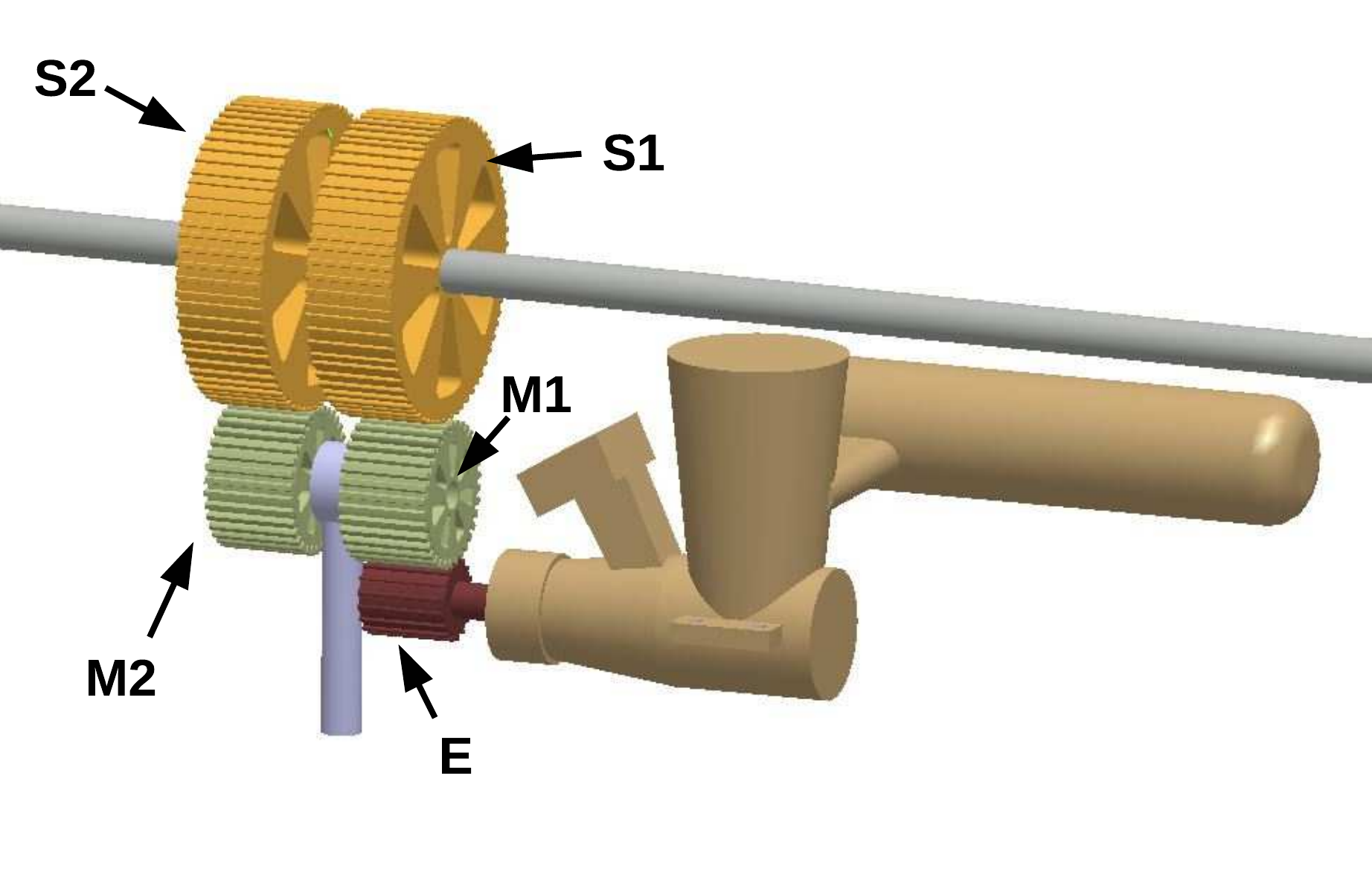}}
 	\subfigure[Transfer bevel gears]{\includegraphics[scale=.35]{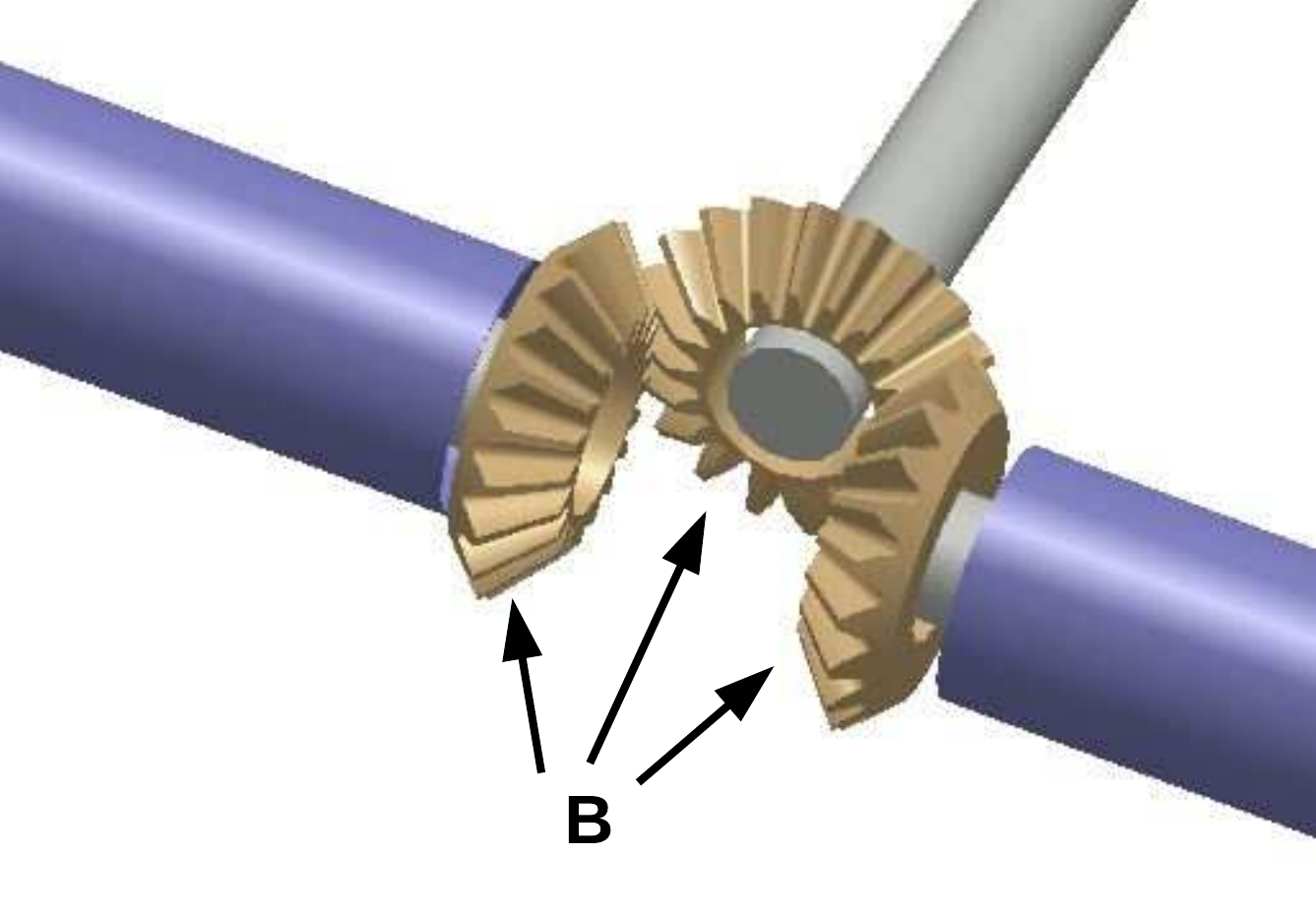}}\\
 	\subfigure[Rotor bevel gears]{\includegraphics[scale=.35]{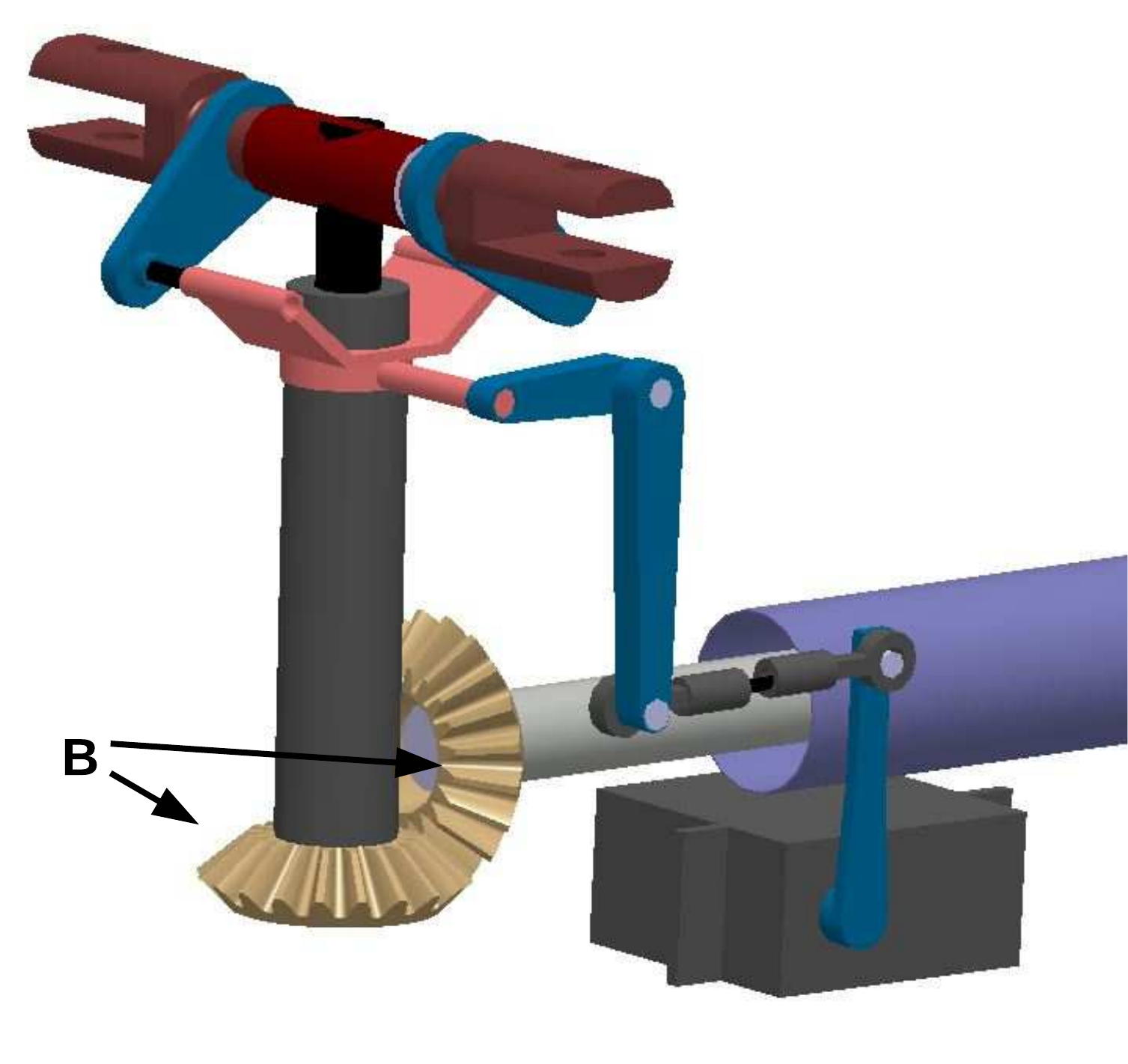}}
 	\subfigure[Rotor bevel gears]{\includegraphics[scale=.35]{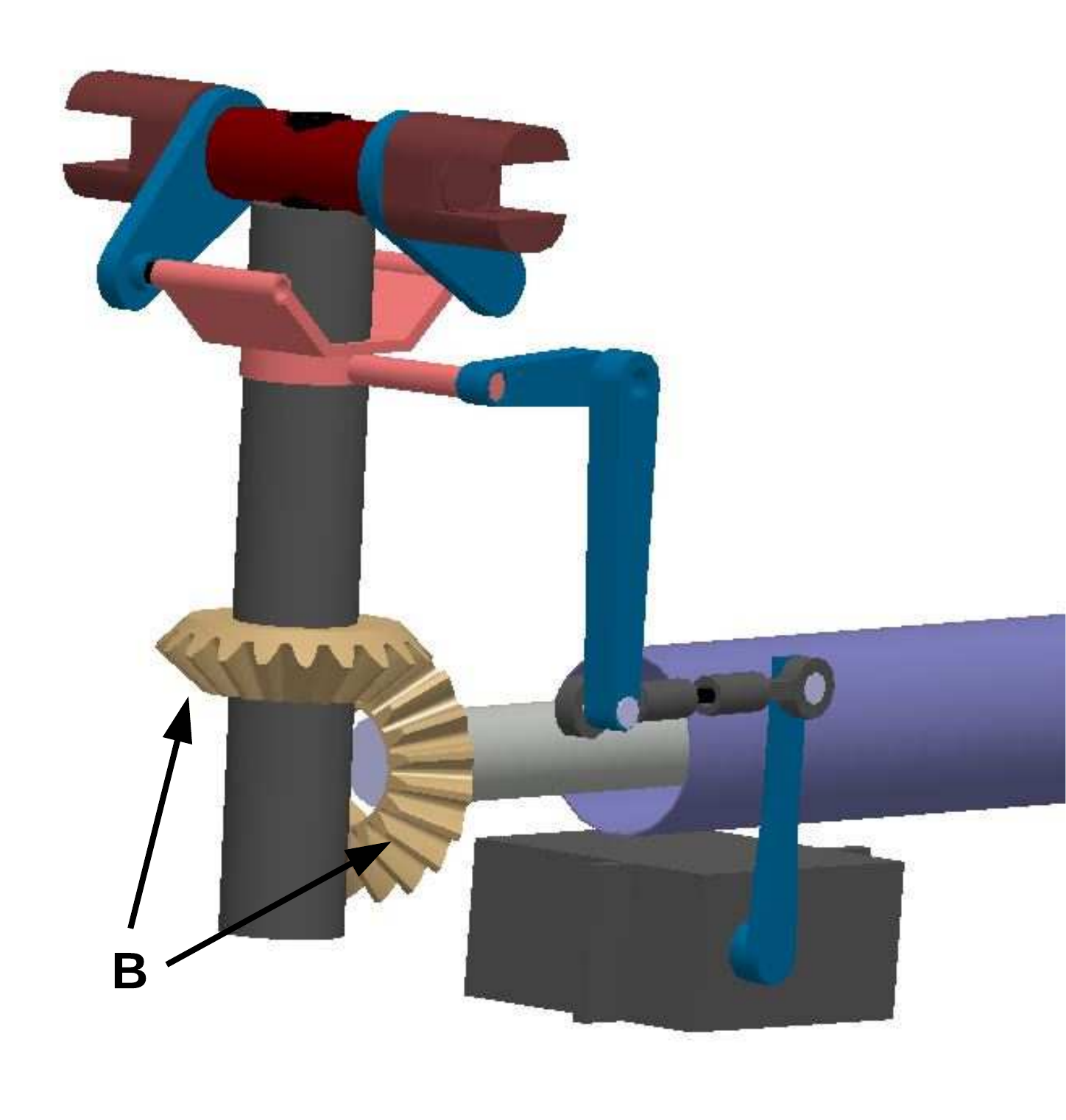}}
 	\caption{Schematic of transmission mechanism and gearbox}
 	\label{fig:Transmission}
 \end{figure}

Transmission mechanism is shown in Fig. \ref{fig:Transmission} (a). The entire transmission system that transfers power from engine to a propellers consists of three parts. 
\begin{enumerate}
\item[1)] Part 1 (Gearbox): To transmit power from the engine to central shafts, gearbox has been used, shown in Fig. \ref{fig:Transmission} (b). The gearbox contains 5 spur gears denoted by E, M1, M2, S1, S2. E is the gear attached to the engine shaft, M1 and M2 are gears in the middle, S1, S2 are gears attached to central shafts transferring power to the rotors. Two stage gearbox design has been adopted here to have a smooth transfer of power from the engine shaft to the central shafts. Power requirement during hover is more and hence this power is going to dictate the gear design. The engine will be operating at about 12800 RPM during hover and the rotors are to operate at 3200 RPM, which means RPM has to be reduced by a factor of 4. For the two stage gearbox design, the RPM will be reduced to half in first stage and to one forth in second stage. This requires the gears to have gear ratios as - $\text{E:M1:M2} = 1:2:2$ and $\text{M1:S1:S2}= 1:2:2$. For a preliminary design purpose, based on standard design practices, pressure angle ($\phi_0$) of 20 degree, addendum ($a_t$) of 1$m$ and dedendum ($d_t$) of 1.2$m$ is chosen for design calculations, where $m$ is module. Diameter and number of teeth are determined considering the space constraints as well as the interference between the teeth using Eq.~(\ref{eqn:teeth_interfer}) (refer \cite{AGMA}). Face widths ($b_t$) are determined using American Gear Manufacturers Association (AGMA) strength equation, Eq.~(\ref{eqn:AGMA}). 
\begin{equation}\label{eqn:teeth_interfer}
T_1 \geq \frac{2a_t\frac{1}{T_2}P_d}{\sqrt{1+\frac{1}{T_2}(\frac{1}{T_2}+2)sin^2\phi_0}-1}
\end{equation}
\begin{equation}\label{eqn:AGMA}
\sigma_b = \frac{F_t P_d}{b_tY}K_v K_o K_m K_f
\end{equation}
here $T_1$ and $T_2$ are number of teeth on gear 1 and 2 under consideration. $P_d$ is pitch circle diameter. $\sigma_b$ is maximum bending stress. $F_t$ is the transmitted load. $Y$ is Lewis factor. $K_v\; K_0,\;K_m,\;K_f$ factors for dynamic loading, overload, load distribution on the gear, stress concentration effect respectively. Table \ref{tab:Gearbox_Specs} lists out all the specifications for the gears in the gearbox.

\item[2)] Part 2 (Transfer bevel gears): For transferring power from central shafts to respective rotor shafts, there are two transfer bevel gears systems which consists of 3 bevel gears (B) each, shown in Fig. \ref{fig:Transmission} (c). One of these three gears is connected to one of the central shafts (CS1, CS2) and transfer power to other two bevel gears kept at $90^{\circ}$ and attached to the corresponding rotor shafts (RS1-RS4). All of these three gears are identical, or in other words have gear ratio of 1:1, as the rotational speed being transferred is same.

\item [3)] Part 3 (Rotor bevel gears): For every rotor, there are two identical bevel gears kept at $90^{\circ}$ to transfer power from rotor shafts to the rotor hub, shown in Fig. \ref{fig:Transmission} (d), (e). The position of the bevel gear attached to rotor hub is such that the adjacent rotors rotate in opposite direction while diagonally opposite rotors rotate in the same direction.
\end{enumerate}

\begin{table}
	\centering
	\caption{Specifications of gears in the transmission}
	\label{tab:Gearbox_Specs}
	\begin{tabular}{c@{\hskip 0.2in}  c@{\hskip 0.2in}  c@{\hskip 0.2in}  c@{\hskip 0.2in}  c }  
		\hline \hline
		\textbf{Parameter} & E & M1,M2 & S1,S2 & B \\
		\hline
		Diameter(in cm) & 2 & 4&8& 3.6\\
		No. of teeth & 17 & 34 & 68 &20\\
		Face width(in cm) & 3.6 & 3.6&3.6& 0.8\\
		Module, m (in mm) & 1.2& 1.2&1.2& 1.8\\
		Adendum (in mm)& 1.2& 1.2&1.2&1.8\\
		Dedundum (in mm) & 1.44 & 1.44&1.44&4.1\\			
		\hline	\hline	
	\end{tabular}
\end{table}

\begin{figure}[h]
  \centering
  \subfigure[CAD drawing]{\includegraphics[width = 0.47\textwidth]{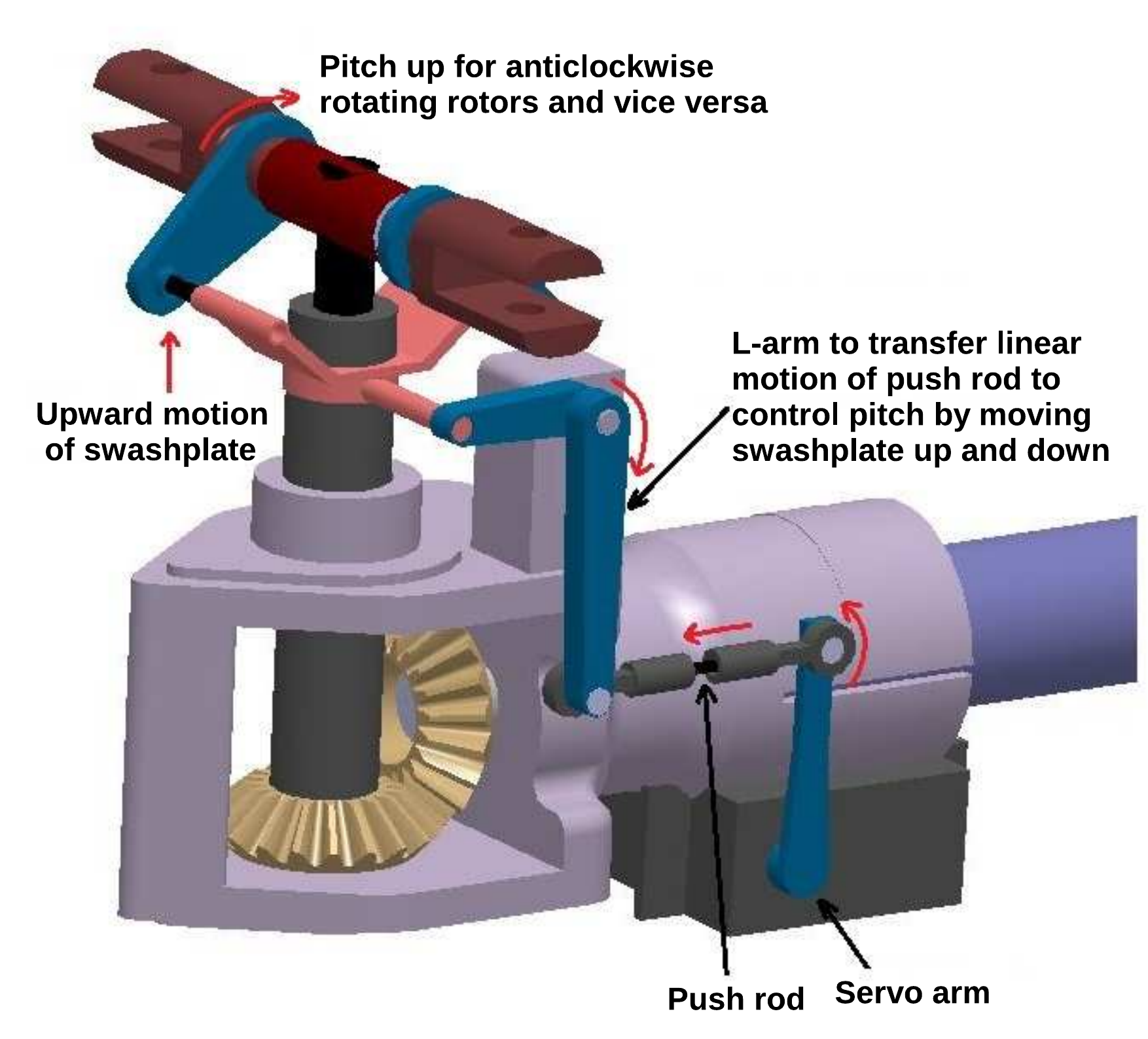}}
    \subfigure[Actual mechanism]{\includegraphics[width = 0.47\textwidth]{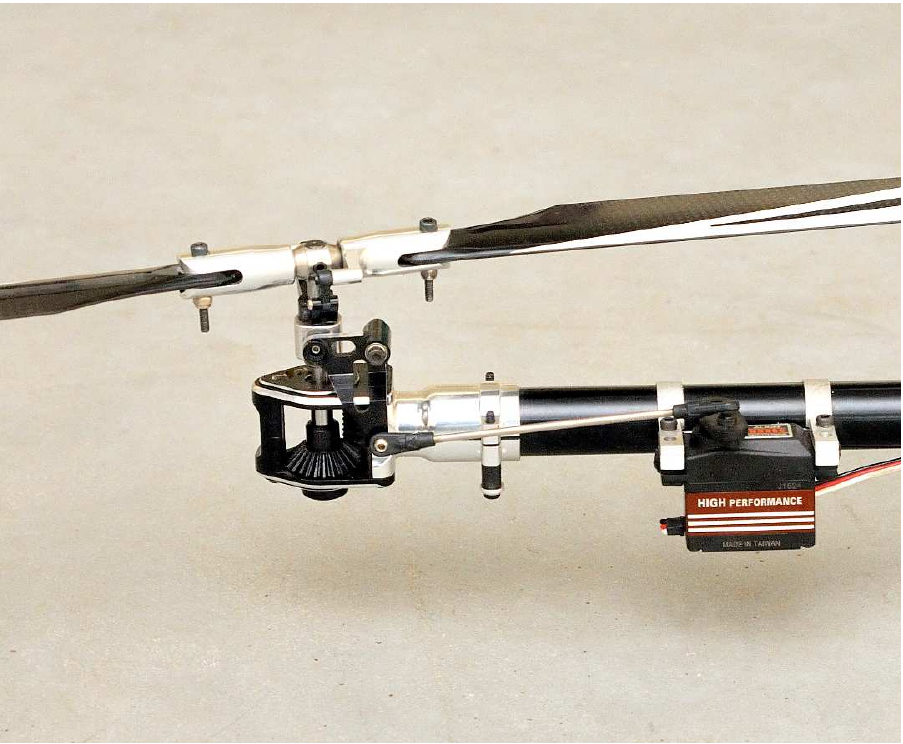}}
 	\caption{Schematic of pitch change mechanism}
	\label{fig:Pitch_change}
\end{figure}

Figure \ref{fig:Pitch_change} shows pitch changing mechanism in detail. The pitch of the blade is changed by moving swashplate, having only heave degree of freedom, up or down by using a servo motor mounted for each rotor on the shaft joining the rotors. For example, as the servo arm is rotated anticlockwise the L-arm rotates clockwise which pushes the swashplate in upward direction which increases the pitch of the anticlockwise rotating blades connected to the swashplate. 

The rotational speed of the rotors needs to be changed from that during hover for the forward flight. For the current design, this will be achieved by using throttle to change the operating speed of the engine to achieve desired rotational speed and power.

\section{Preliminary structural analysis}
To decide upon the dimensions of basic supporting frame and internal wing structure, preliminary structural analysis is performed using CATIA software. First, the parameters are roughly chosen for given loading condition and then static finite element analysis is performed using CATIA software to assess if the structure has adequate strength.

\begin{figure}
 	\centering
\includegraphics[scale=.7]{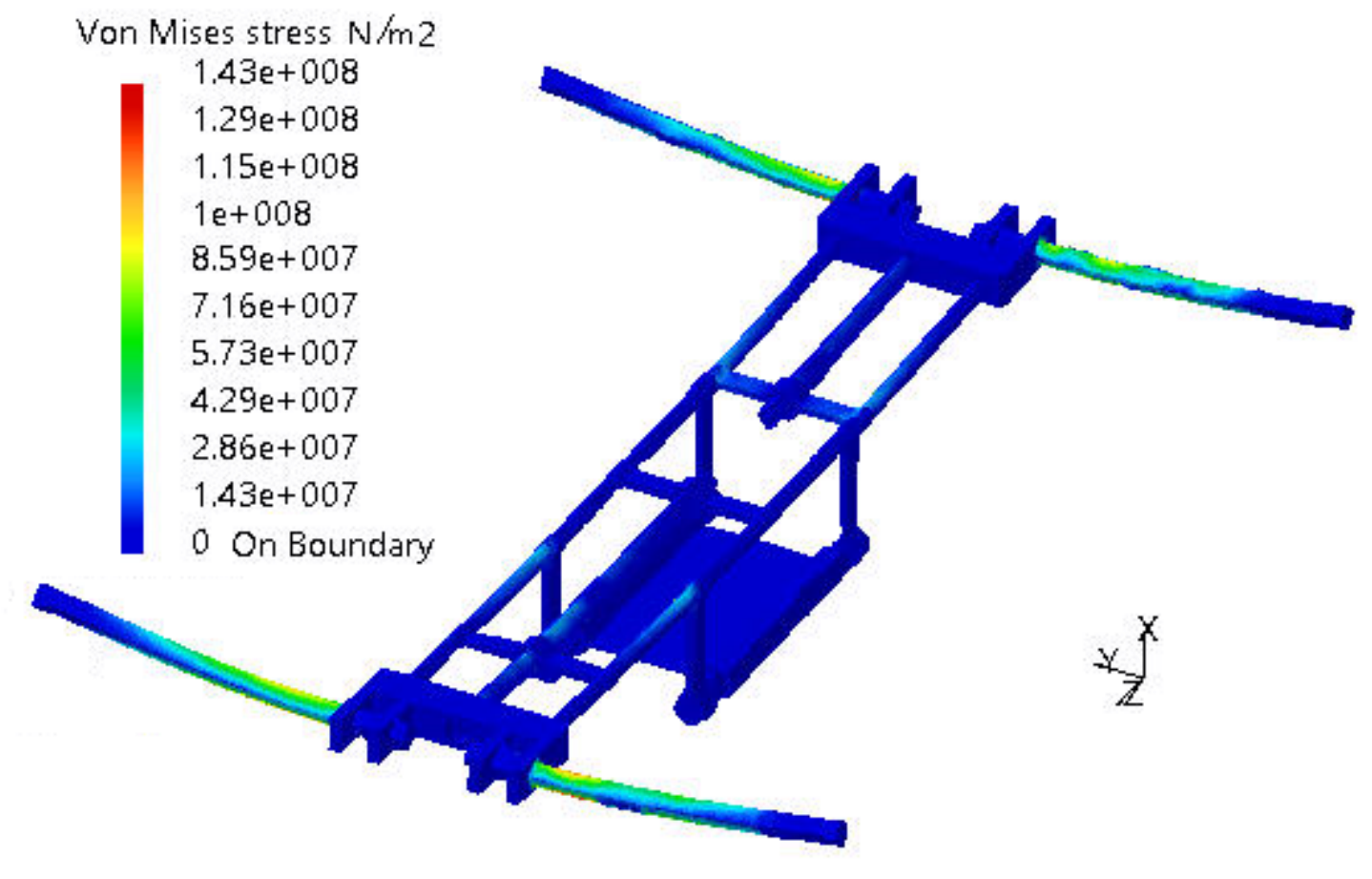}
  	\caption{Structural analysis of quadrotor frame}
 	\label{fig:Struct_quad}
 \end{figure}

For the main quadrotor frame, the expected rotor thrust during hover condition is applied as point loads at the end of the rotor shaft and the central portion of the frame is fixed to mimic the effect of weight of the vehicle. Figure~\ref{fig:Struct_quad} shows stress variation over the frame. With the chosen dimensions for frame structure, the factor of safety is around 1.9 which is slightly on the higher side. Since, the objective is not to optimize the structural strength, the chosen dimensions are retained. 

\begin{figure}
	\centering
\includegraphics[scale=.7]{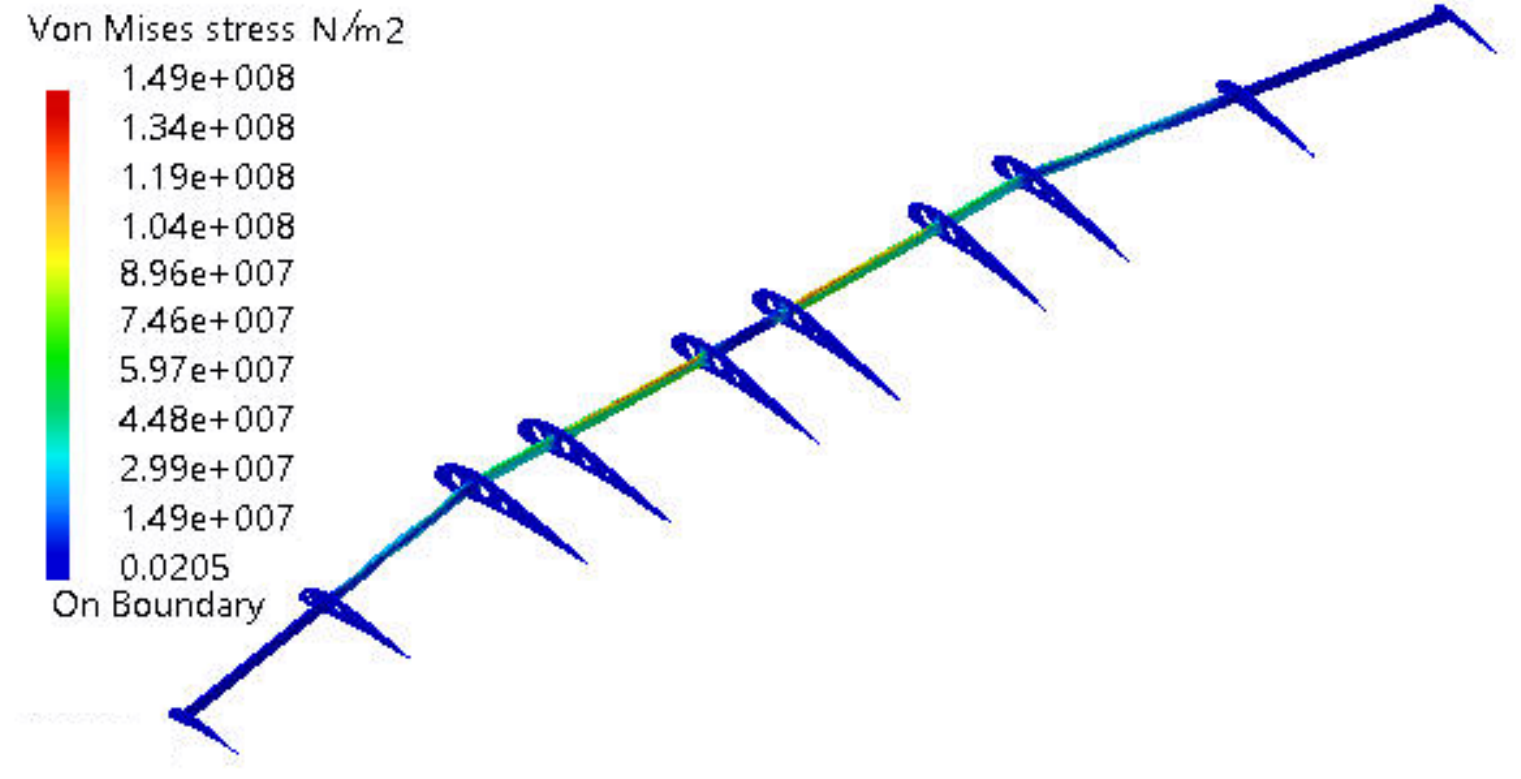}
	\caption{Structural analysis of wing structure}
	\label{fig:Struct_wing}
\end{figure}

For the wing structure, consisting of spars and ribs, the main load carrying members of the wing, distributed loads are applied. Section-wise constant uniformly distributed load is applied having resultant load equal to weight for a single wing. Similarly, drag is also applied, assuming its value to be 10\% of the lift. The load distribution and the stress variation for this loading is shown in Fig.~\ref{fig:Struct_wing}. The spar has rectangular cross-section of 18 mm X 6 mm. The tip ribs have 2 mm thickness while all other ribs are 3 mm thick. For the given dimensions of spar, the factor of safety is around 1.85 which gives adequate margin for preliminary design.

\section{Weight estimate and center of gravity location}
\begin{figure}
	\centering
	\subfigure[Front view]{\includegraphics[scale=.35]{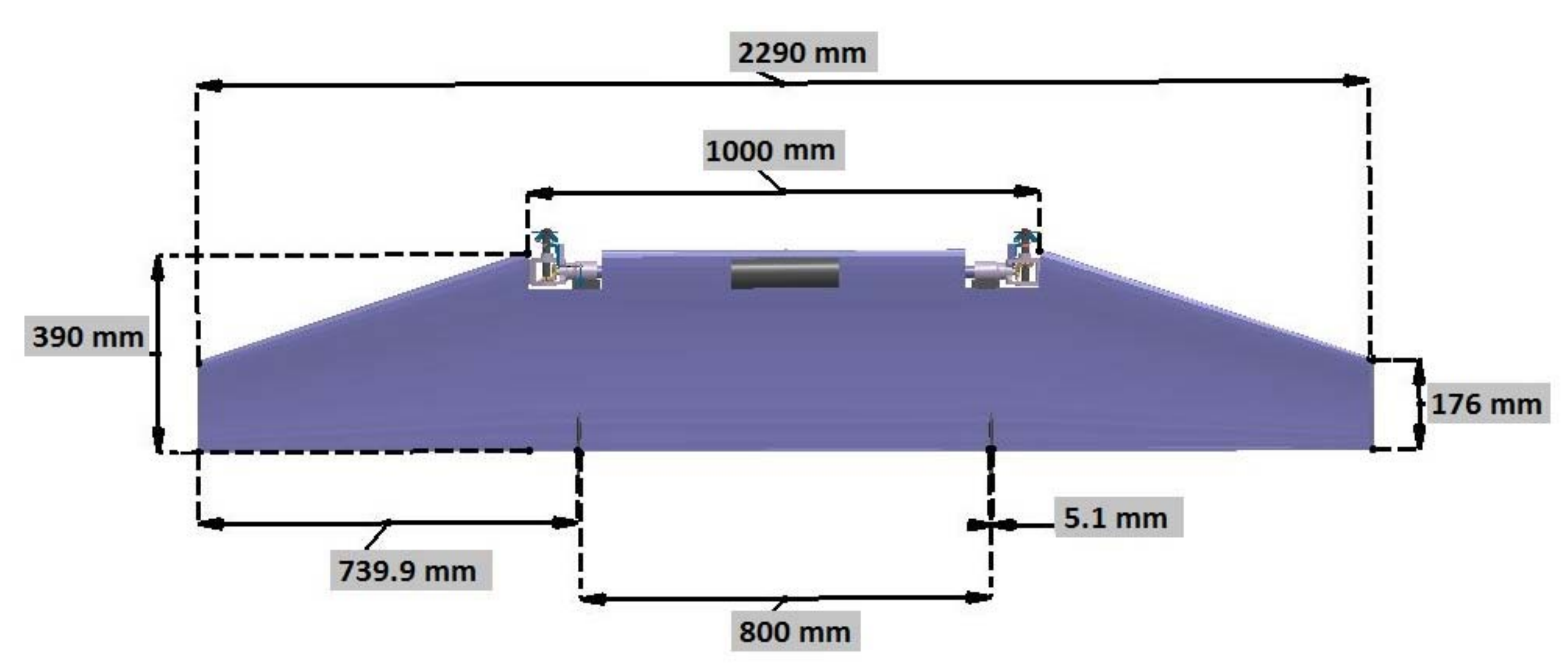}}\\
	\subfigure[Top view]{\includegraphics[scale=0.35]{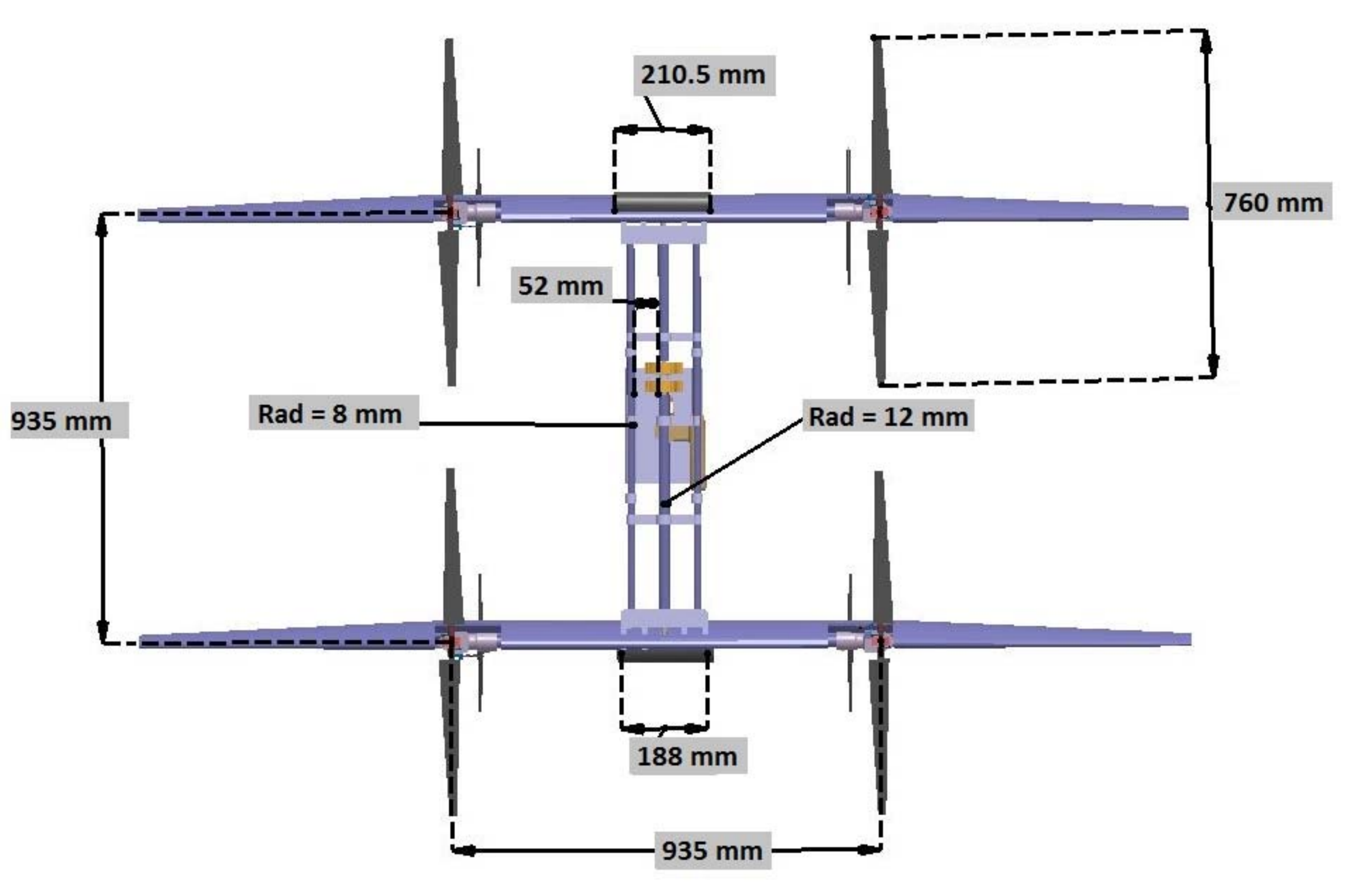}}\\	
	\subfigure[Side view]{\includegraphics[scale=.30]{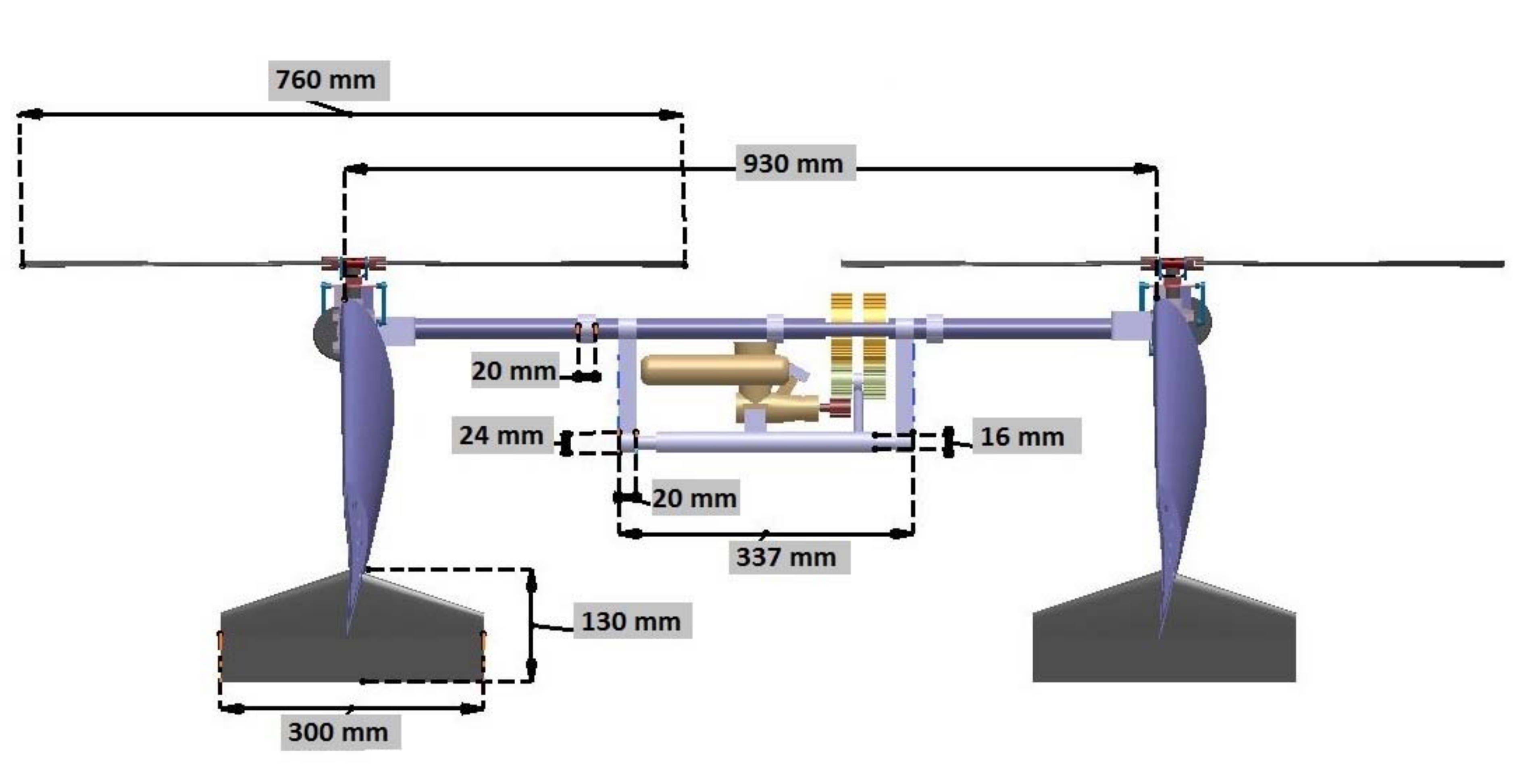}}
	\caption{Three view drawing of designed UAV}
	\label{fig:3D}
\end{figure}

The detailed drawing of the vehicle is made in CATIA having the dimensions of rotor and wing as designed above and respective materials are assigned to each and every part. The software gives the total weight and that of the individual parts. The weights are listed in Table \ref{tab:Weights}. For initial calculations the gross weight of the vehicle was assumed to be 15 kg but after initial design and detailed drawing it was realized that the weight of the vehicle without payload is more than 11 kg. Therefore, the calculations were repeated for different initial guess until the empty and design weight including payload were in agreement. The table shows weights after final iteration. Center of gravity, because of almost symmetrical design, lies at the center of the quadrotor frame with offset of 15.7 cm towards the trailing edge from leading edge of the wing. The three view drawing of the final design is shown in Fig.~\ref{fig:3D}.

\begin{table}
	\centering
	\caption{Weights of the individual parts}
	\label{tab:Weights}
	\begin{tabular}{ c@{\hskip 0.2in} c@{\hskip 0.2in} c@{\hskip 0.2in} c}  
		\hline \hline
		\textbf{Component} & \textbf{Weight (Kg)} & \textbf{Quantity} & \textbf{Total weight  (Kg)} \\
		\hline
		Wing assembly & 1.985& 2 & 3.970\\
		Rotor assembly & 0.396 & 4 & 1.584 \\
	        Engine assembly + Gearbox & 2.477 & 1 &2.477 \\
		Rest frame structure & 3.326 & 1 & 3.326\\
		Payload & 6 & 1 & 6\\
		Fuel &1.15 & 1 & 1.15\\
		\hline
		\multicolumn{3}{c}{\textbf{Total weight of the vehicle (Kg)}} &18.5\\
		\hline
	\end{tabular}
\end{table}

\section{Avionics and telemetry}

The proposed design is inherently unstable and therefore an on-board autopilot is required to stabilize the vehicle continuously in various modes of flight. The open source PixHawk autopilot board with on-board ARM Cortex M4 processor (STM32f407) based micro-controller with a clock speed of 168 MHz is used for stabilization and control for proof-of-concept demonstration of hovering flight discussed in the next section. It has nine degrees-of-freedom IMU (3 axis digital MEMS gyro with range of $\pm$2000 deg/s, 3 axis digital MEMS accelerometer with measuring range of $\pm$16 g and 3 axis digital MEMS Magnetometer with range of $\pm$12 Gauss), pressure sensor and GPS. The autopilot loop on the PixHawk runs at 200 Hz. 

The autopilot software uses open source Mavlink communication protocol for all RTOS inter-process communications as well as ground link telemetry communications related to vehicle health, flight modes, alerts, current position, autopilot arm / disarm status, way-point transfer / switching and sensor calibration. The Mavlink communication protocol also offers cyclic redundancy check (CRC) checksum to reduce data corruption during communication between the ground station and the vehicle. The Linux version of open-source QGroundControl ground station software is used for real-time monitoring and logging of telemetry data. The telemetry module pair operating at 433 MHz is used for all vehicle to ground station communication.

\section{Proof-of-concept prototype flight testing}
\begin{figure}
	\centering
	\includegraphics[scale=.6]{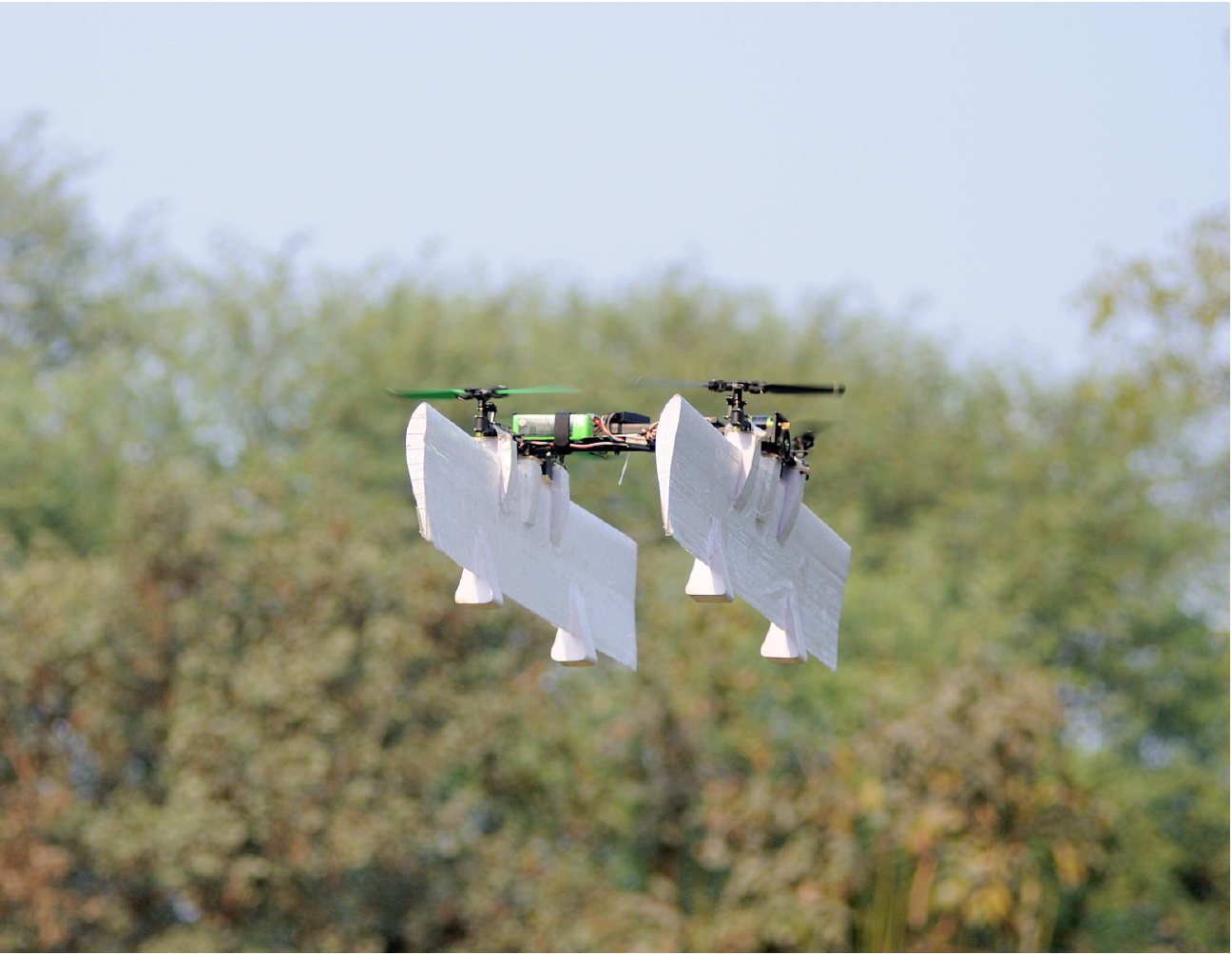}
	\caption{Variable pitch quadrotor biplane UAV prototype}
	\label{fig:Quad_Biplane}
\end{figure}

Before, initiating the fabrication of the designed vehicle, a proof-of-concept scaled vehicle is developed based on this design to demonstrate the concept. The baseline quadrotor frame is fabricated using off-the-shelf components to fabricate a variable pitch quadrotor frame in which rotors are powered using a single brushless DC motor and transmission system. The wings fabricated using foam are then attached to the quadrotor frame with the help four aluminum clamps, balsa wood and an adhesive. The PixHawk autopilot board is mounted on the vehicle to enable attitude stablization for hovering flight. The bare variable pitch quadrotor airframe weighs 1000 g, both the wings combined weigh 160 g (are 1 m long each) and the battery is 250 g which amounts to a gross takeoff weight of 1400 g. The vehicle thrust is measured experimentally using a load cell and it is established that the prototype vehicle produced a maximum thrust of around 18 N. The assembled quadrotor biplane prototype in hovering flight is shown in Fig. \ref{fig:Quad_Biplane}. The proof-of-concept demonstration of hovering flight required the implementation of a PID controller for the variable pitch quadrotor system which is discussed below. 

\subsection{PID control for the variable pitch quadrotor}
For attitude stabilization and tracking a simple PID controller is developed and implemented onboard the PixHawk autopilot board. The development of the entire flight envelope autopilot is not the focus of the present paper and would be taken up in future work. 
\begin{figure}[h]
	\centering
	\includegraphics[scale=0.5]{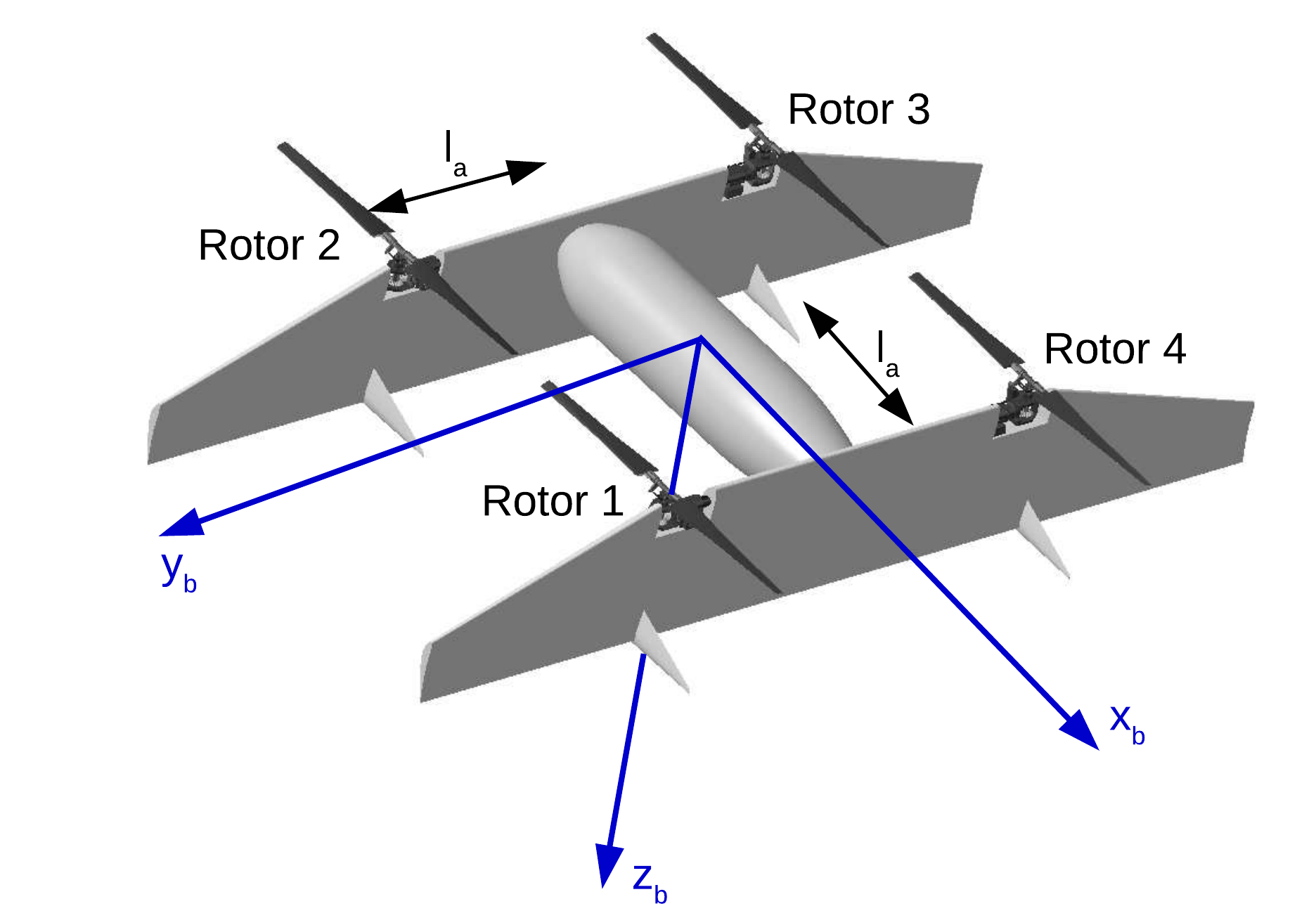}
	\caption{Coordinate frame}
	\label{fig:frame}
\end{figure}

\textbf{Attitude control}:
To achieve the desired orientation $\mathbf{E}_d=[\phi_d\quad \theta_d\quad  \psi_d]'$, moments are generated proportional to the Euler angles' errors and their rates using the following expression.
\begin{equation}\label{eqn:PID_att}
\mathbf{M} = - \mathbf{K}_p^a \mathbf{e}_a -
\mathbf{K}_i^a \int_{0}^{t} \mathbf{e}_a(\tau)d\tau - \mathbf{K}_d^a \mathbf{\dot{e}}_a
\end{equation}
where $ \mathbf{e}_a=\mathbf{E-E}_d $ and $\mathbf{M}=[l \quad m \quad n]'$
\\
\textbf{Position control}: 
To track position $\mathbf{r}_d=[x_d\quad y_d\quad z_d]'$, feedback translational accelerations~\cite{Cutler}, in all the 3 directions, are produced by using the position error and its rate as given below.
\begin{equation}
\mathbf{\ddot{r}}_{fb} = - \mathbf{K}_p^p \mathbf{e}_p -
\mathbf{K}_i^p \int_{0}^{t} \mathbf{e}_p(\tau)d\tau - \mathbf{K}_d^p \mathbf{\dot{e}}_p
\end{equation}
where  $ \mathbf{e}_p=\mathbf{r-r}_d$. This gives us total accelerations that are required to be applied to the vehicle as follows,
\begin{equation}
\mathbf{a}=M(\mathbf{\ddot{r}}_d+\mathbf{\ddot{r}}_{
	fb}-\mathbf{g})
\end{equation}
where $\mathbf{g}=[0 \quad 0 \quad g]'; \quad \mathbf{a}=[a_x \quad a_y \quad a_z]' $. From this accelerations total thrust, desired pitch and roll angles can be calculated as follows~\cite{Gupta}.
\begin{equation}\label{eqn:PID_thrust}
\left.
\begin{array}{l}
T = M\norm{\mathbf{a}}\\
\phi_d = sin^-{}^1(u_xsin(\psi_d) - u_ycos(\psi_d))\\
\theta_d = sin^-{}^1(\frac{u_xcos(\psi_d) + u_ysin(\psi_d)}{cos(\phi_d)})
\end{array}
\right \}
\end{equation}
where \quad $ u_x = \frac{-M a_x}{T_d}; \quad
u_y = \frac{-M a_y}{T_d} $

\textbf{Control allocation}: In case of variable pitch quadrotor, the thrust and torque of each rotor are dependent on the the trust coefficients of the rotors which in turn depend on their blade pitch. The total thrust and moments are related to thrust coefficients, obtained using blade element momentum theory, are given below~\cite{Gupta}.

\begin{equation}\label{eqn:thrust_moment}
\left.
\begin{array}{l}
T=K_F (C_{T_{1}}+C_{T_{2}}+C_{T_{3}}+C_{T_{4}})\\
l=K_F l_a(-C_{T_{1}}-C_{T_{2}}+C_{T_{3}}+C_{T_{4}})\\
m=K_F l_a(C_{T_{1}}-C_{T_{2}}-C_{T_{3}}+C_{T_{4}})\\
n=\frac{K_FR}{\sqrt{2}}(-|C_{T_{1}}|^\frac{3}{2}+|C_{T_{2}}|^\frac{3}{2}-|C_{T_{3}}|^\frac{3}{2}+|C_{T_{4}}|^\frac{3}{2})
\end{array}
\right \}
\end{equation}
Here $K_F$ is rotor force constant and subscript 1 to 4 denotes the rotor number. Since the yaw moment is non-linearly related to thrust coefficients, for simplicity it is linearized around the hover condition giving expression for yaw moment as in Eq.~\ref{eqn:yaw}.
\begin{equation}\label{eqn:yaw}
n=K_FR\sqrt{\frac{C_{T_{h}}}{2}}(-|C_{T_{1}}|+|C_{T_{2}}|-|C_{T_{3}}|+|C_{T_{4}}|)
\end{equation}
Where $C_{T_{h}}$ is thrust coefficient of each rotor during hover. Equation \ref{eqn:thrust_moment} and \ref{eqn:yaw} are used to obtain thrust coefficients of the each rotor to apply thrust and moments as required from Eqs. (\ref{eqn:PID_att}) and (\ref{eqn:PID_thrust}). The correlation between thrust coefficients and blade pitch for each rotor is obtained through experiments which is used to generate PWM signal for each of the servo motors that is controlling the blade pitch of the rotors. 

\begin{figure}
	\centering
	\subfigure[Attitude with time]{\includegraphics[scale=0.5]{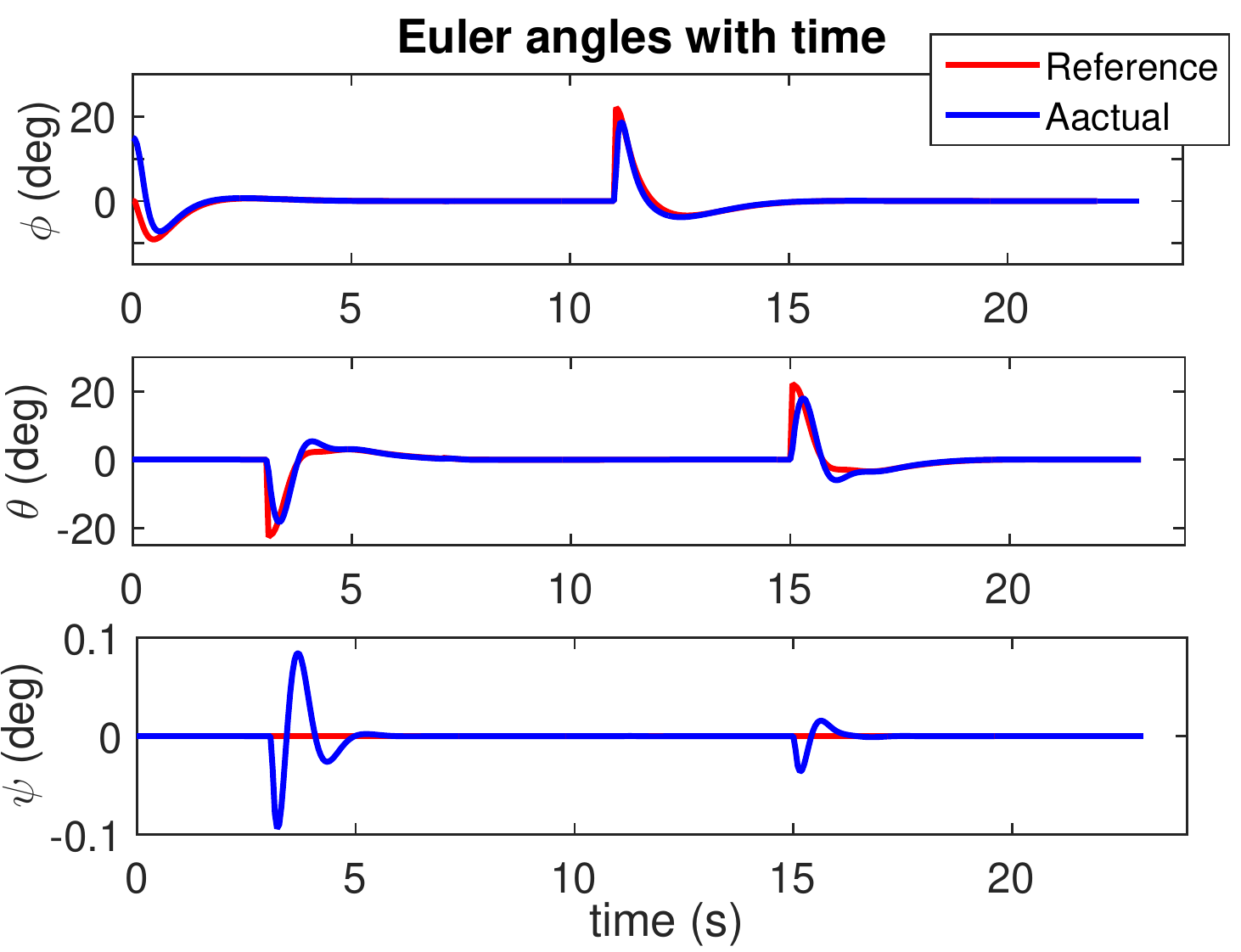}}
	\subfigure[Position with time]{\includegraphics[scale=0.5]{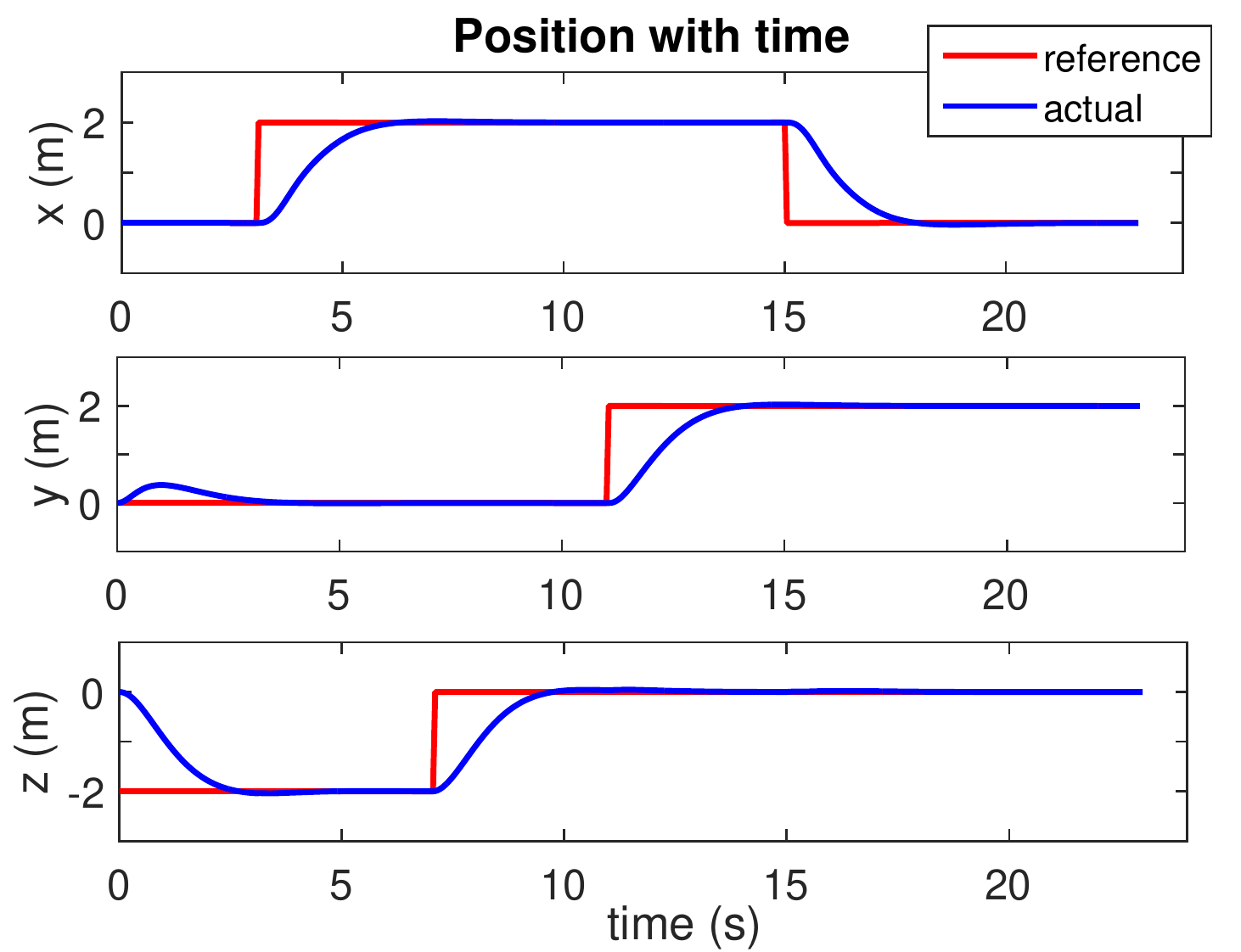}}\\
	\subfigure[Position in 3d with time]{\includegraphics[scale=0.5]{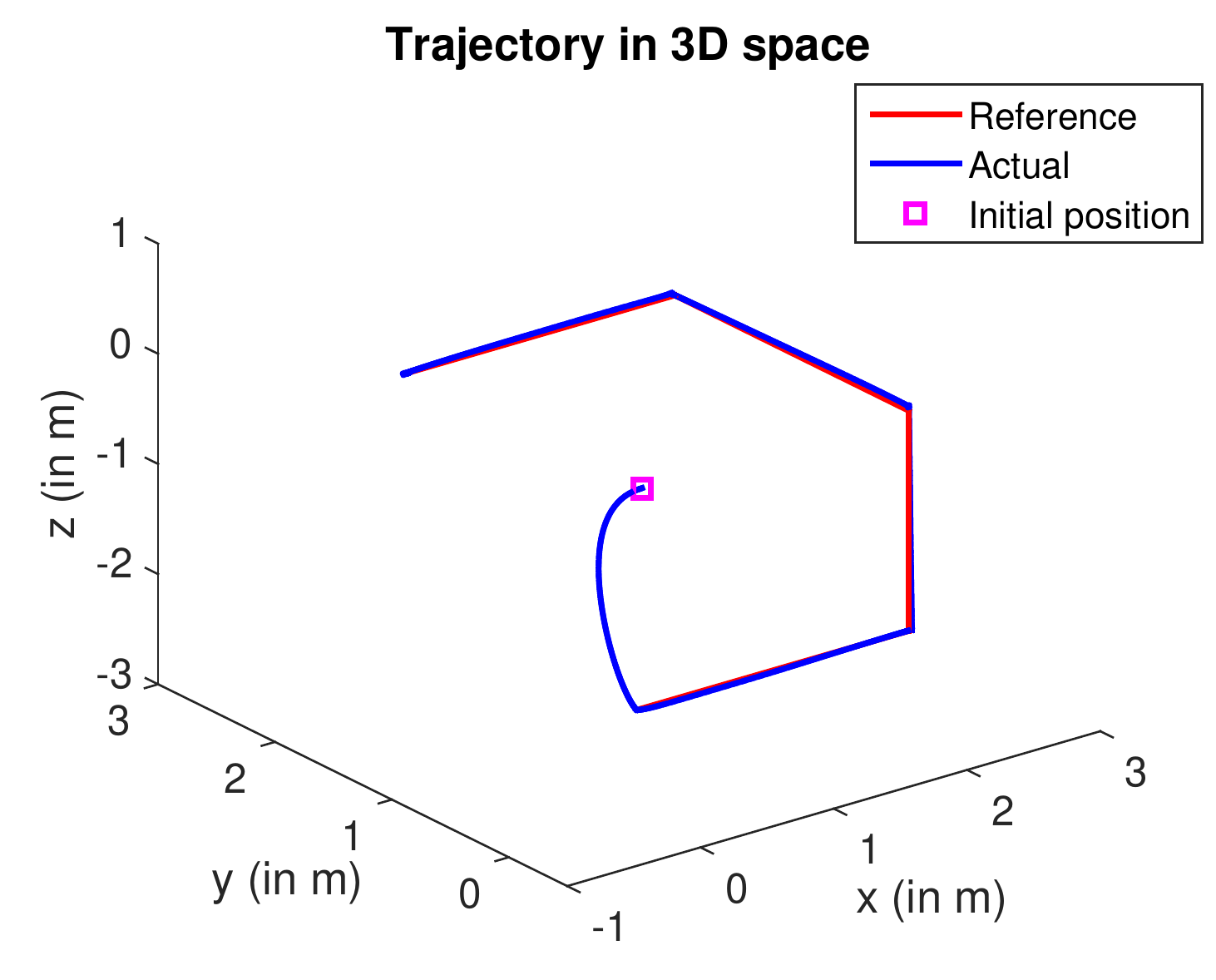}}

	\caption{Simulation for way point tracking}
	\label{fig:Simulation}
\end{figure}

\textbf{Simulation results}: Before the implementation of the controller on the actual vehicle, simulations are carried out for way points tracking. The quadrotor is commanded to track $(x_d,y_d,z_d,\psi_d)=\{(0,0,-2,0),\thinspace (2,0,-2,0),\thinspace (2,0,0,0),\thinspace (2,2,0,0),\thinspace (0,2,0,0)\}$. As observed from Fig. \ref{fig:Simulation} (b) and (c), quadrotor tracks the given way points. Similarly, Fig \ref{fig:Simulation} (c) shows that the required attitude angles generated for position control are also tracked.  

\begin{figure}
	\centering
	\subfigure[Roll attitude]{\includegraphics[scale=0.5]{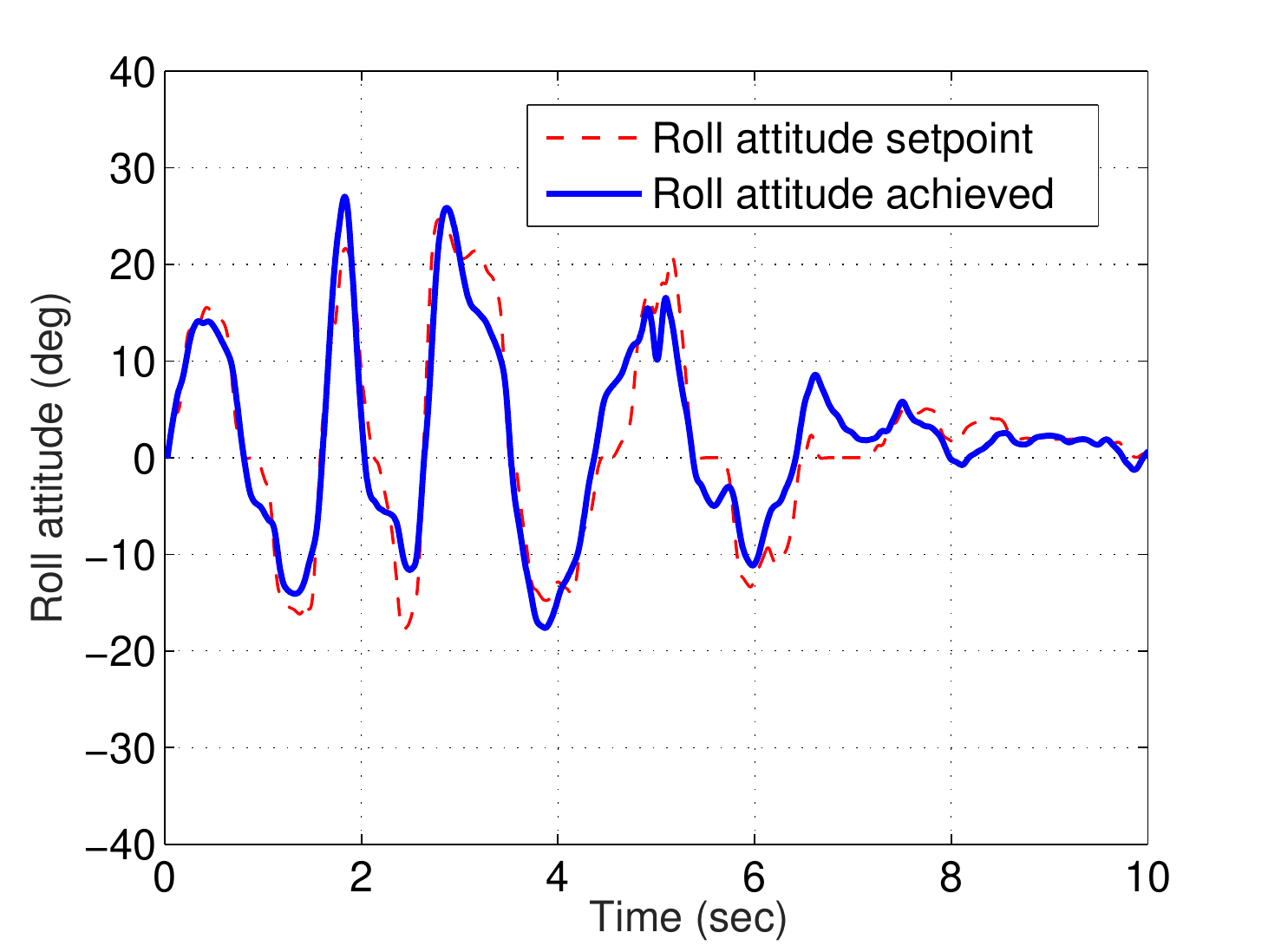}}
	\subfigure[Pitch attitude]{\includegraphics[scale=0.5]{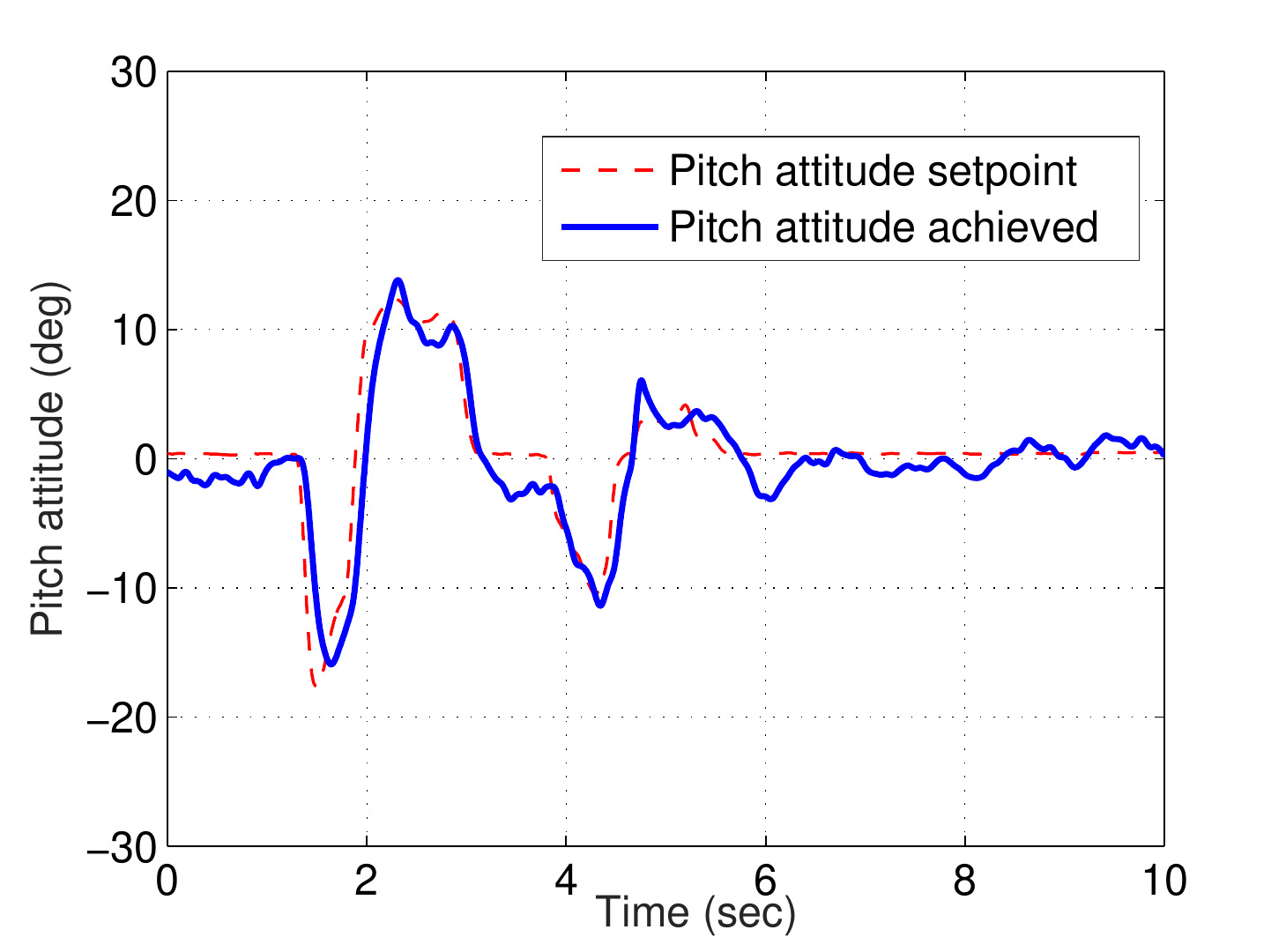}}\\
	\subfigure[Yaw attitude]{\includegraphics[scale=0.5]{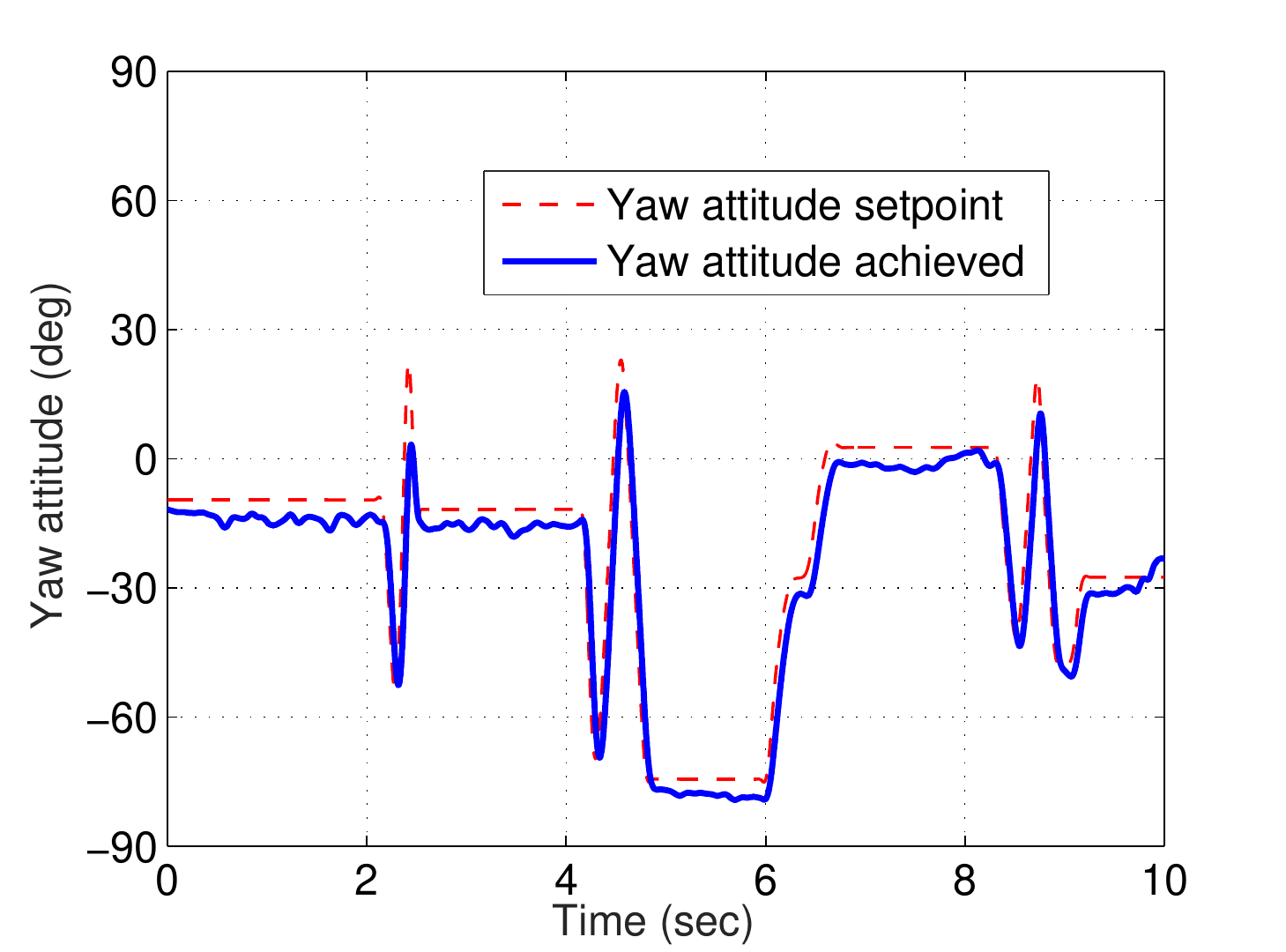}}
	\caption{Experimental validation of controller for attitude tracking of the scaled proof-of-concept prototype}
	\label{fig:Experimental}
\end{figure}

\textbf{Experimental results}: The PID algorithm is implemented on PixHawk autopilot board and used to stabilize and control the prototype during hovering flight. Figure~\ref{fig:Experimental} shows the attitude tracking performance of the prototype vehicle in manual flight. It is observed that the controller tracks roll, pitch and yaw attitude satisfactorily for the vehicle to enable stable hovering flight shown in Fig.~\ref{fig:Quad_Biplane}.  

\section{Concluding remarks}
Conceptual design, of various systems of a novel Quadrotor Biplane VTOL UAV concept and flight test demonstration of proof-of-concept prototype is presented in this paper. Ingenious use of various technologies such as variable pitch quadrotor, biplane, variable tip speed, and engine as power source makes the vehicle unique in itself at the same time a disruptive technology for today's market. The proposed design is expected to be robust and compact. It is highly maneuverable and capable of performing aggressive maneuvers due to its variable pitch proprotors. It has long range due to addition of the biplane wings configuration in its structure and can fly for significantly longer duration due to use of fuel engine. Its landing gears also serve as vertical fins, improving the yaw damping for directional stability. It is designed to land in remote unprepared terrain due to its hovering and VTOL capabilities. 

The design of proprotors, wings, transmission and structural components is systematically undertaken. The aerodynamic design of the proprotors is carried out using modified BEMT. The modified BEMT considers swirl and ignores the small inflow angle assumption. The BEMT predictions are validated against experimental results for helicopter rotor as well as propeller during hover. The proprotor design highlights the conflict between the requirements for hover and forward flight. Final proprotor design with 24$^\circ$ preset angle and -24$^\circ$ twist is arrived at by giving 70\% weightage to forward flight and 30\% weightage to hovering flight.  The operating RPM of the proprotors is reduced from 3200 during hover to 2000 during forward flight to ensure optimal performance during cruise flight. This is required to decrease profile power loss during forward flight. The estimated power consumption during forward flight mode is 64\% less than that required for hover, which justifies the need for hybrid UAVs as that studied in this paper. 

Aerodynamic design of the wings is obtained using a typical monoplane wing design approach while optimizing the benefits of the biplane wing configuration. The gears used in transmission mechanism are designed using the standard gear design approach with the help of AGMA equation. The engine is selected as per the power requirement and ensuring high power to weight ratio. Preliminary structural analysis is performed on the base quadrotor frame as well as on the wings using CATIA software. A proof-of-concept vehicle prototype is fabricated and a PID controller is developed to demonstrate stable hovering flight and attitude tracking.   

\section*{Future work}
As a part of future work, first the designed configuration would be fabricated. The controller for the entire flight envelope of the vehicle would be developed and tested, first on the proof-of-concept vehicle and then on the designed prototype. The vehicle would be tested under various flight conditions and its performance would be systematically evaluated.

%

\section*{Acknowledgments}
Authors would like to thank Mr. Ramdas Gadekar, a Ph.D. research scholar in Aerospace Engineering department at IIT Kanpur for his valuable assistance during the proof-of-concept prototype flight testing. Authors would also like to thank Ms. Kirti Bhatnagar, Senior Project Associate for her help in CATIA drawing. This research did not receive any specific grant from funding agencies in the public, commercial, or not-for-profit sectors. 


\end{document}